\documentclass[aps,prd,twocolumn,amsmath,amssymb,citeautoscript,superscriptaddress]{revtex4-1}
\usepackage{graphicx}
\usepackage{dcolumn}
\usepackage{bm}
\usepackage{accents}

\newcommand{\aap}{Astron.\ Astrophys.}
\newcommand{\mnras}{Mon.\ Not.\ R.\ Astron.\ Soc.}
\newcommand{\apjl}{Astrophys.\ J.\ Lett.}
\newcommand{\apjs}{Astrophys.\ J.\ Suppl.\ Ser.}
\newcommand{\araa}{Ann.\ Rev.\ Astron.\ Astrophys.}

\usepackage{ulem}
\usepackage{xcolor}

\begin{document}

\title{Probing the Black Hole Metric. I. Black Hole Shadows and Binary Black-Hole Inspirals}

\author{Dimitrios Psaltis}
\affiliation{Steward Observatory and Department of Astronomy, University of Arizona, 933 N. Cherry Ave., Tucson, AZ 85721, USA}

\author{Colm Talbot}
\affiliation{LIGO Laboratory, California Institute of Technology, Pasadena, CA 91125, USA}

\author{Ethan Payne}
\affiliation{School of Physics and Astronomy, Monash University, Clayton VIC 3800, Australia}
\affiliation{OzGrav: The ARC Centre of Excellence for Gravitational Wave Discovery, Clayton VIC 3800, Australia}

\author{Ilya Mandel}
\affiliation{School of Physics and Astronomy, Monash University, Clayton VIC 3800, Australia}
\affiliation{OzGrav: The ARC Centre of Excellence for Gravitational Wave Discovery, Clayton VIC 3800, Australia}
\affiliation{Birmingham Institute for Gravitational Wave Astronomy and School of Physics and Astronomy, University of Birmingham, Birmingham, B15 2TT, United Kingdom}


\begin{abstract}
In General Relativity, the spacetimes of black holes have three fundamental properties: (i) they are the same, to lowest order in spin, as the metrics of stellar objects; (ii) they are independent of mass, when expressed in geometric units; and (iii) they are described by the Kerr metric. In this paper, we quantify the upper bounds on potential black-hole metric deviations imposed by observations of black-hole shadows and of binary black-hole inspirals in order to explore the current experimental limits on possible violations of the last two predictions. We find that both types of experiments provide correlated constraints on deviation parameters that are primarily in the $tt$-components of the spacetimes, when expressed in areal coordinates. We conclude that, currently, there is no evidence for a deviations from the Kerr metric across the 8 orders of magnitudes in masses and 16 orders in curvatures spanned by the two types of black holes.  Moreover, because of the particular masses of black holes in the current sample of gravitational-wave sources, the correlations imposed by the two experiments are aligned and of similar magnitudes when expressed in terms of the far field, post-Newtonian predictions of the metrics. If a future coalescing black-hole binary with two low-mass (e.g.,  $\sim 3 M_\odot$) components is discovered, the degeneracy between the deviation parameters can be broken by combining the inspiral constraints with those from the black-hole shadow measurements. 
\end{abstract}


\maketitle

\section{Introduction}

Over the past century, numerous predictions of the theory of General Relativity (GR) have been tested against multiple experiments and astrophysical observations~\cite{Will1993}. Even though all these investigations aim to test the same theory, they nevertheless address different combinations of its ingredients, different aspects of its predictions, and are performed on widely different scales. A very stringent constraint imposed in one setting does not necessarily preclude the importance of testing different predictions of the theory in a different setting. 

At one level, it is common to distinguish gravitational tests between those that test Einstein's equivalence principle and those that test the field equation~\cite{Will1993}. The former search for violations of the weak equivalence principle, of the local Lorentz invariance, and of the local position invariance. The latter assume the validity of Einstein's equivalence principle and, therefore, that spacetime is endowed by a metric and that test particles and photons follow geodesics in this metric. These tests then use observations to map the metric of an object and test whether its parameters are consistent with GR predictions. 

Among metric tests, there is a critical distinction between those that measure parameters of time-independent metrics and those that probe the dynamics of the theory. For example, all solar-system tests and many of the tests involving binary pulsars explore equilibrium metrics~\cite{Will2014}, whereas gravitational-wave tests with pulsar timing~\cite{Wex2014}, direct gravitational-wave observations~\cite{Abbott2016,Abbott2019,*Ilsi2019,GWTC2:GR}, and cosmological tests~\cite{Ferreira2019} explore dynamical metrics. A different distinction between metric tests separates those that involve vacuum metrics from the ones in which the coupling of matter with the gravitational field plays an important role. The latter category includes not only the cosmological tests, for which the feedback of matter to the gravitational field determines the evolution of structure formation, but also, e.g., tests of scalar-field gravity with pulsars, in which the presence of matter in the strong gravitational fields of the neutrons stars gives rise to dipole radiation (which has not been observed)~\cite{Wagoner1970,Damour1993,*Damour1996}. 

Even though the breadth of gravitational tests explores many aspects of the underlying theory and its predictions, it is usually impossible to assess the relative significance of any one particular upper limit on possible deviations. This is true because there is no compelling alternative to GR that is based on fundamental physics arguments and that leaves observable signatures at the scales of most experiments. In other words, without a plausible alternative to the theory to guide the tests and our thinking,  it is not viable to ask how large a deviation one would expect from any test. Moreover, it is impossible to guess at what scale would such deviations become observable. 

For these reasons, a wide variety of gravitational tests have been performed with objects and in settings that probe a vast range of scales: from sub-mm length scales to the size of the observable Universe, from the $GM/Rc^2\sim 10^{-9}$ potentials~\footnote{Here, $M$ denotes the mass scale of the test, $R$ the length scale, $G$ is the gravitational constant, and $c$ is the speed of light.} of terrestrial experiments to the order-unity potentials of black holes, and from the $GM/R^3c^2\sim 10^{-12}$~cm$^{-2}$ curvature scales of neutron stars to the $\sim 10^{-57}$~cm$^{-2}$ curvature scale of the cosmological constant~\cite{Baker2015}. 

The differences in the qualitative character between the gravitational tests, the observables used, and the tools employed hamper our ability to cross compare the resulting constraints of any deviations from GR. If a specific modification to the GR field equations is considered, then the complete theory can be used to make predictions for and be compared against all types of observations and astrophysical systems. However, translating empirical constraints on general modifications from one setting to another poses serious challenges. For example, it is very hard to ask in a theory-independent way how the results of cosmological tests affect the predictions for the gravitational-wave emission from coalescing neutron stars. Perhaps more importantly for the design of future experiments, it is often impossible to understand whether existing constraints on modifications from GR already preclude the detection of beyond-GR phenomena in a previously unexplored setting.

In recent years, a new set of gravitational tests has emerged that probe a previously unexplored regime: that of the near-field regions of astrophysical black holes. The LIGO/Virgo detection of gravitational waves from coalescing black holes and neutron stars has led to tests of the dynamics of GR with stellar-mass objects, during the inspiral and ringdown stages of the events~\cite{Abbott2016,Abbott2019,*Ilsi2019,GWTC2:GR}\footnote{Using the gravitational-wave events as sirens, additional tests have been performed on the propagation of gravitational waves across the large distances from the sources to the Earth, in order to constrain the mass of the graviton~\cite{Abbott2016} or the speed of gravitational waves that is allowed to be different from the speed of light in some modified gravity theories inspired by cosmological observations~\cite{Baker2017}}. The constraints from gravitational-wave tests are typically expressed in terms of upper limits on deviations from the GR predictions on a set of parametric post-Newtonian terms in the waveform expression. These terms amalgamate both potential deviations in the metrics of the individual black holes from the Kerr solution (when they are at large distances) as well as potential deviations in the strength, polarization, and angular distribution of the radiated gravitational waves, i.e., the dynamics of the theory.

Moving to the larger masses and smaller curvatures of the supermassive black holes in the centers of galaxies, the monitoring of the orbits of stars within a few thousand Schwarzschild radii from the black hole in the center of the Milky Way, Sgr~A*, has led to application of two classical GR tests to this black-hole environment: the measurement of the gravitational redshift (which leads to a test of the equivalence principle)~\cite{Gravity2019,Do2019} and the detection of orbital precession of the nearest star~\cite{Gravity2020}. As in the case of the precession of Mercury in the Solar System, the constraints from the latter test are expressed in terms of the usual coefficients of the parametric post-Newtonian (PPN) framework~\cite{Will2014}. Additional constraints can be imposed on equivalence principle violations via potential changes in the fine-structure constant~\cite{Hees2020} or on the Yuwaka strength and scale of a putative fifth force~\cite{Hees2017}.

More recently, the Event Horizon Telescope (EHT) has detected the shadow of the black hole in the high-resolution image obtained from the center of the M87 galaxy~\cite{PaperI,*PaperVI}. Comparing the observed shadow size to that predicted for the mass of the black hole that was known {\it a priori} from stellar dynamics has led to constraints on the possible deviation of the black-hole spacetime from the Kerr metric~\cite{Psaltis2020}. These constraints were expressed in terms of upper limits on parameters of metrics that have been designed to be different from Kerr, while ensuring that no pathologies are present outside the horizons~\cite{Johannsen2011,Vigeland2011,Johannsen2013,Johannsen2013b}. They were further translated into constraints on the post-Newtonian expansions of these metrics, in order to compare them with earlier, weak-field tests.

A common denominator among many of these avenues of testing GR with black holes or other stellar objects is the set of constraints they impose on the equilibrium metrics of objects with different compositions and masses. However, as discussed above, these constraints are expressed in different ways that are specific to each test because, e.g.,  they are merged with parameters that quantify the dynamics of the theory (as is the case of the gravitational-wave tests) or employ complexity that is necessary to avoid pathologies (as is the case of the shadow tests). 

The aim of this series of papers is to combine all existing tests of metrics of astrophysical objects in order to test three important GR predictions for the metrics of black holes, i.e., that: {\em (i)\/} the metric of a black hole, expanded to first order in spin, is identical (when expressed in geometric units) to that of a slowly spinning star, i.e., it is the Schwarzschild spacetime with the first-order frame-dragging terms; {\em (ii)\/} all black holes, independent of mass or curvature, are described by the same metric; and {\em (iii)\/} the black-hole spacetime is described by the Kerr metric. 

The approach we will follow here is to use the constraints that are imposed with each type of test and calculate their implications for the values of the various terms in a parametric post-Newtonian expansion of the equilibrium black-hole metrics. It is important to emphasize here that, in many cases involving black-hole tests, we will not be testing post-Newtonian expansions of the metrics. Instead, we will be constraining the parameters for regular, well behaved metrics that deviate from the GR predictions and then use these constraints to place bounds on deviations of the corresponding post-Newtonian parameters of these metrics.

Albeit not comprehensive, this approach allows for a comparison between the results of various tests and identifies the unique aspects of equilibrium metrics that each test is sensitive to. Moreover, this approach facilitates the comparison of the new constraints to those imposed by previous Solar System, pulsar, and cosmological tests, which we will address in forthcoming papers.

In this first paper, we will focus on the tests that involve the black-hole shadow and the inspiral phase of coalescing black holes and aim to address the latter two GR predictions discussed above. Section \S II introduces the formalism and coordinate system for the parametric post-Newtonian equilibrium metric used throughout this series of papers. Section \S III presents the black-hole shadow constraints in terms of several parametric metrics that deviate from Kerr without introducing pathologies and translates them into the parametrization of the post-Newtonian expansion of the equilibrium metrics. Section \S IV follows the same approach for tests that use LIGO/Virgo observations of gravitational waves emitted during the inspiral phase of compact binary coalescence. In Section \S V, we discuss the key results and future prospects.

\smallskip

\section{Parametric Metrics of Isolated Static Objects}

As discussed in Section \S I, the goal of this work is to translate the constraints imposed by various gravitational tests into bounds on deviations from the GR solution for the equilibrium metric of an isolated object. This first paper will focus on the results of two types of observations: of the black-hole shadow images obtained with the EHT and of the gravitational waves from coalescing black holes detected with LIGO/Virgo.

The shadow of a Kerr black hole is highly circularly symmetric, up to near-extremal values of the spin, and has a diameter that depends very weakly on the spin of the black hole, for all observer inclinations. This is a consequence of a fortuitous near cancellation of the effects of frame dragging and of the quadrupole mass moment of the Kerr spacetime~\cite{Johannsen2010b}. The extremely weak dependence of the shadow diameters on spin was also shown to be preserved in several metrics that are parametrically different from Kerr~\cite{Johannsen2013c,Medeiros2020}. Moreover, the constraints on deviations from the Kerr metric imposed by the diameter measurement of the shadow in M87 were shown to depend very weakly on the assumed spin~\cite{Psaltis2020}. 

In the case of the black hole binaries observed with LIGO/Virgo, the predicted inspiral waveforms do depend on the spins of the black holes, starting at the 1.5 Post-Newtonian (PN) order (see, e.g., Ref.~\cite{Khan2016}).  However, the majority of coalescing black holes observed to date are consistent with having low or moderate spins, $\chi \lesssim$~few~tenths~\cite{Abbott2019,GWTC2,*GWTC2:pop}. Because both sets of measurements provide very weak bounds on the spins of the black holes, the focus of this work will be on their implications for the equilibrium spacetimes of non-spinning, isolated objects.

\begin{figure}[t]
\includegraphics[width=0.48\textwidth]{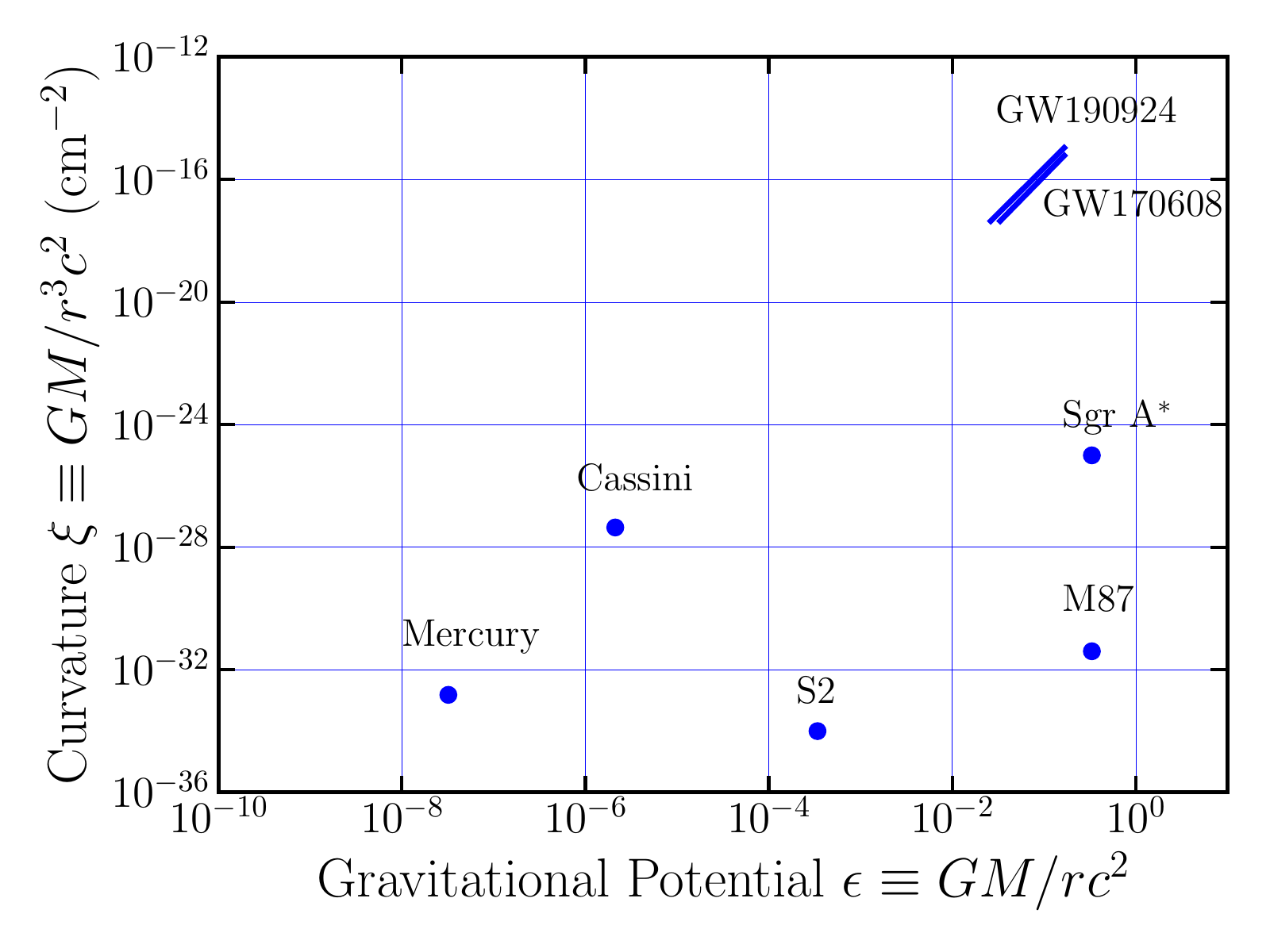}
\caption{\label{fig:EHTprobes} Characteristic gravitational potential and curvature probed by two Solar-System tests (the perihelion precession of Mercury and the light deflection measured with Cassini) as well as the same quantities probed by the detection of gravitational waves from the LIGO/Virgo events GW170608 and GW190924, by the detection of periapsis precession in the S2 star around Sgr~A$^*$, and by the observation of the black-hole shadow in the M87 and Milky Way galaxies (after~\cite{Baker2015}).}
\end{figure}

In GR, Birkhoff's theorem ensures that the external spacetime of a spherically symmetric, isolated object is unique and is described by the Schwarzschild metric. The theorem is not necessarily valid in other metric theories of gravity; spherically symmetric isolated objects may be described by different metrics depending on the gravitational potential, the curvature, or even the nature of the object. For example, in scalar-tensor modifications to gravity with quadratic couplings~\cite{Damour1993,Damour1996} or with Chameleon screening~\cite{Joyce2015,Burrage2018}, the external spacetime of a stellar object depends on the density of matter in its interior. In the first class of theories, the modifications are important at the strong fields of neutron stars, whereas in the second class of theories, the modifications become significant in the low-density envelopes of giant stars.  However, in all such theories, the unique external spacetime of a non-rotating black hole remains the Schwarzschild solution~\cite{Thorne1971,*Scheel1995,*Psaltis2008b,*Sotiriou2012}. For this reason, the external spacetimes of objects of different composition, mass, or nature (i.e., stellar objects vs.\ black holes) may not be the same. Indeed, demonstrating conclusively that the external spacetimes of two different spherically symmetric static objects, when expressed in gravitational units, are not the same would serve as a direct evidence that GR will have to be modified.

The external spacetime of a spherically-symmetric compact object can be written in many forms and coordinate systems. In order to facilitate comparison with earlier constraints, especially those at weak gravitational fields, we will translate our results into predictions for a general post-Newtonian parametrization of the metric of a spherically symmetric object. In doing so, we will be implicitly assuming that the values of the various parameters depend on the mass, gravitational potential, gravitational curvature, or nature of the object under consideration.

\begin{table*}[t]
\caption{Fields Probed by Different Gravitational Tests}
\label{tab:systems}
\begin{ruledtabular}
\begin{tabular}{lcccc}
\textrm{Test}&
\textrm{Mass} &
\textrm{Distance} &
\textrm{Potential $\epsilon$} &
\textrm{Curvature $\xi$} \\
&
\textrm{$(M_\odot)$} &
\textrm{(cm)} &
&
\textrm{(cm$^{-2})$} \\
\colrule
Cassini (Shapiro) & 1 & $7\times 10^{10}$ & $2\times 10^{-6}$ & $4\times 10^{-28}$\\
Mercury (perihelion) & 1 & $5\times 10^{12}$ & $3\times 10^{-8}$ & $2\times 10^{-33}$\\
GW170608 & 19 & $\sim(2-9)\times 10^7$ & 1/30-1/6 & $\sim 5\times 10^{-18}-6\times 10^{-16}$\\
GW190924\_021846 & 14 & $\sim(1-8)\times 10^7$ & 1/38-1/6 & $\sim 5\times 10^{-18}-1\times 10^{-15}$\\
S2 Star (periapsis) & $4\times 10^6$ & $2\times 10^{15}$ & $3\times 10^{-4}$ & $1\times 10^{-34}$\\
Sgr~A$^*$ (shadow) & $4\times 10^6$ & $2\times 10^{12}$ & 1/3 & $9\times 10^{-26}$\\
M87 (shadow) & $6.5\times 10^9$ & $3\times 10^{15}$ & 1/3 & $4\times 10^{-32}$\\
\end{tabular}
\end{ruledtabular}
\end{table*}

For comparison with other GR tests, the gravitational potential
\begin{equation}
    \epsilon\equiv \frac{GM}{rc^2}
\end{equation}
and curvature
\begin{equation}
    \xi\equiv \frac{GM}{r^3c^2}
\end{equation}
probed by various gravity tests are shown in Table~\ref{tab:systems} and Fig.~\ref{fig:EHTprobes}. These include the black-hole shadow tests of M87 and, in the near future, of Sgr~A* (evaluated at $r=3GM/c^2$), the inspiral test of GW170608 and GW190924 (evaluated at the range of separations that are detectable by the LIGO/VIRGO detectors, see below), the periapsis precession test with the S2 star around Sgr~A*~\cite{Gravity2020} (evaluated at the periapsis distance), and for two classical Solar-System tests with Cassini and the perihelion precession of Mercury~\cite{Will2014}

\subsection{Parametric Post-Newtonian Metrics for non-Spinning Objects}

The traditional GR tests in the Solar System with the Parametric Post-Newtonian (PPN) framework have been performed with a metric written in terms of isotropic coordinates~\cite{Will1993}, i.e.,
\begin{equation}
ds^2=g_{tt,I} dt^2+g_{rr,I} \left(dr_I^2+r_I^2 d\Omega\right)\;,
\end{equation}
where the subscript $I$ denotes that the coordinates are isotropic. 

Following the PPN approach, and denoting the traditional 1PN parameters as $\beta_1$ and $\gamma_1$, the non-zero metric components in isotropic coordinates become
\begin{eqnarray}
g_{tt,I}&=&-1+\frac{2}{r_I}-\frac{2 \beta_1}{r_I^2}+\frac{3 \beta_2 }{2 r_I^3}+{\cal O}(r^{-4})\nonumber\\
g_{rr,I}&= &1+\frac{2 \gamma_1 }{r_I}+\frac{3 \gamma_2 }{2 r_I^2}+{\cal O}(r^{-3})\;,
\label{eq:2PNI}
\end{eqnarray}
where $G=c=M=1$. In GR, the values of all PPN parameters for the Schwarzschild spacetime, as defined here, are equal to unity.

In contrast, testing GR with black-hole shadows or inspiral waveforms is more naturally performed using parametric metrics in areal coordinates, i.e.,
\begin{equation}
ds^2=g_{tt,S} dt^2+g_{rr,S} dr_S^2+r_S^2d\Omega\;,
\end{equation}
where the subscript $S$ denotes the fact that the area of a closed surface at coordinate $r_S$ is always equal to $4\pi r_S^2$. This is because it has been recently shown that, in areal coordinates, the diameter of the shadow of a non-spinning black hole and the inspiral waveform during a coalesence event depend only on the $tt-$components of the metrics~\cite{Psaltis2020,Carson2020}.

Converting between isotropic and areal radial coordinates is, nevertheless, trivial using the transformation 
\begin{equation}
r_I= r_S-\gamma_1  +\frac{ \left(2 \gamma_1 ^2-3 \gamma_2 \right)}{4 r_S}+{\cal O}(r_S^{-2})\;.
\end{equation}
For the parametrization shown in Eq.~(\ref{eq:2PNI}), the result is
\begin{eqnarray}
g_{tt,S}&=& -1+\frac{2}{r_S}-\frac{2 (\beta_1 -\gamma_1)}{r_S^2}  
+\frac{2}{r_S^3}\left\{\frac{1}{4} \left[2 \left(\gamma_1 ^2-1\right)\right.\right.
  \nonumber\\
&&\left.\left.-8 (\beta_1  \gamma_1 -1)+3 (\beta_2 -1)+3 (\gamma_2    -1)\right]\right\}\nonumber\\
&&+{\cal O}(r^{-4})\
\label{eq:2PN_gtt}
\end{eqnarray}
and
\begin{equation}
g_{rr,S}=1+\frac{2 \gamma_1 }{r_S}+\frac{\gamma_1 ^2+3 \gamma_2 }{r_S^2}
+{\cal O}(r^{-3})\;.
\label{eq:2PN_grr}
\end{equation}
If, as in Ref.~\cite{Psaltis2020}, the  various PN parameters were introduced instead in the expansion of the metric components written in areal coordinates as
\begin{equation}
g_{tt,S}=-1+\frac{2}{r_S}+2\sum_{i=1}^N \left(-1\right)^{i} \frac{\zeta_i}{r_S^{i+1}}\;,
\label{eq:PNzeta}
\end{equation}
then the coefficients of the various order become
\begin{eqnarray}
\zeta_1&=&\beta_1-\gamma_1\nonumber\\
\zeta_2&=&\frac{1}{4} \left[2 \left(\gamma_1 ^2-1\right)-8 (\beta_1  \gamma_1 -1)\right.\nonumber\\
&&\qquad\qquad\left.+3 (\beta_2 -1)+3 (\gamma_2    -1)\right]\;,
\end{eqnarray}
etc, when expressed in terms of the present parametrization.

In the Solar System, the PPN parameter $\gamma_1$ is constrained via measurements of the deflection of light and of the Shapiro delay for signals that graze the solar surface~\cite{Will2014}. The most stringent limit to date has been achieved with the Cassini mission and is $\vert\gamma_1-1\vert<2.3\times 10^{-5}$. On the other hand, the PPN parameter $\beta_1$ is constrained, in combination with $\gamma_1$, via measurements of the perihelion precession of Mercury and of the Nordtvedt effect in Lunar Ranging~\cite{Will2014}. The most stringent limit to date is $\vert\beta_1-1\vert<8\times 10^{-5}$.

Albeit proven to be very effective~\cite{Will2011}, the expansion of the metric in a series over inverse powers of the coordinate radius is, of course, formally valid when the series converges. This is expected to be true, as long as there are no pathologies in the metric down to the radius probed by a particular test and the expected deviations from GR increase in strength as the radius from the central object decreases (i.e., not when searching for Yukawa-type corrections). However, even if formally converging, interpreting different astrophysical settings requires a careful assessment of the truncation errors in comparison to the measurement uncertainties. We will address these convergence issues for the black-hole shadow tests in \S III and for the inspiral tests in \S IV.

\subsection{Expected Magnitude of Deviations}

Because of the lack of compelling alternatives to GR that arise from fundamental physics arguments and lead to astrophysically relevant effects, there is no first-principles approach to estimate the expected magnitudes of corrections to the various PPN parameters. This is especially true for tests with black holes since the Kerr metric is a solution to many simple modifications of the Einstein field equations supplemented with an additional field characterized by constant coupling coefficients~\cite{Thorne1971,*Scheel1995,*Psaltis2008b,*Sotiriou2012}. Obtaining non-Kerr black-hole solutions for modifications to the field equations arising from, e.g., the addition of quadratic terms in the Riemann tensor, requires couplings that are described by dynamical fields. One example of such non-Kerr black holes occur in the so-called Einstein-Dilaton-Gauss-Bonnet [EDGB] theories, were originally obtained in Ref.~\cite{Kanti1996,*Kanti1998}, and were further studied in~\cite{Yunes2011,Yagi2012,Ayzenberg2014,*McNees2016,Antoniou2018a,*Antoniou2018b,Silva2018,*Doneva2018}.

One might naively expect deviations from the Schwarzschild/Kerr metrics only in black holes with masses that are comparable to the scale at which the Einstein-Hilbert action of GR is modified. It is straightforward to show, however, that this is not necessarily the case. As a proof of principle, we will use the non-spinning black-hole solution for EDGB gravity, when the coupling coefficients are linear functions of the dynamical fields~\cite{Yunes2011}. The Lagrangian action of this theory is
\begin{eqnarray}
S&=& \int d^4x \sqrt{-g}\left\{\kappa R + a_1 \theta R^2+a_2\theta R_{ab}R^{ab}\right.\nonumber\\
&&+a_3\theta R_{abcd}R^{abcd}+a_4\theta R_{abcd}^*R^{abcd}\nonumber\\
&&\left. -\frac{\beta}{2}\left[\nabla_a\theta\nabla^a\theta+2V(\theta)\right]\right\}\;,
\end{eqnarray}
where $R$, $R_{ab}$, and $R_{abcd}$ are the Ricci scalar, Ricci tensor, and Riemann tensor, respectively, $V(\theta)\simeq (1/2)m_\theta \theta^2$ is the potential of the dynamical scalar field, $a_1,...,a_4$ are the coupling coefficients of the Gauss-Bonnet terms, and $\beta$ is the coupling coefficient of the dynamical scalar field. Hereafter, we set the latter to unity, as it can be reabsorbed in a redefinition of the scalar field~\cite{Yagi2012}.

The presence of the dynamical field forces the non-spinning black-hole solution to deviate from Schwarzschild. When expressed in areal coordinates, the two non-trivial components of the metric for this solution are~\cite{Yunes2011}
\begin{equation}
g_{tt,S}=-1 + \frac{2}{r_S}-\frac{\lambda}{3 r_S^3}+{\cal O}(r^{-4})
\end{equation}
and
\begin{equation}
g_{rr,S}=1+\frac{2}{r_S}+\frac{4-\lambda}{r_S^2}+{\cal O}(r^{-3})\;,
\end{equation}
where $\lambda=a_3^2/(\kappa M^4)$ and we have explicitly shown the dependence on the mass of the black hole $M$ that can be measured by, e.g., monitoring the orbits of stars at large distances. Comparing this solution to the PPN expansions~(\ref{eq:2PN_gtt})-(\ref{eq:2PN_grr}) leads, for this theory, to
\begin{eqnarray}
\beta_1&=&1\nonumber\\
\gamma_1&=&1\nonumber\\
\beta_2&=&\frac{1}{9}\left(9+\frac{a_3^2}{\kappa M^4}\right)\nonumber\\
\gamma_2&=&\frac{1}{3}\left(3-\frac{a_3^2}{\kappa M^4}\right).
\end{eqnarray}
In other words, this particular modification to GR leads to non-spinning black-hole spacetimes that have the same 1PN expansion as the Kerr metric but differ only at the 2PN and higher orders. Moreover, the deviation depends only on the dimensionless ratio $\lambda$. 

In principle, the $a_3$ coupling of the quadratic term can be substantially different than the mass of the black hole. However, as long as there is substantial scale separation between the couplings of the quadratic field and the Planck mass, then the metrics of black holes with masses $M\sim (a_3^2/\kappa)^{1/4}$ will show order unity deviations from the Schwarzschild solution at the 2PN (and higher) orders. This argument was discussed explicitly in Ref.~\cite{Stein2014}, where an expression was derived for the effective cut-off scale $\Lambda$ of the EDGB gravity. Deviations from GR reach order unity for a black of mass $M$, when te cut-off scale is
\begin{equation}
    \Lambda\sim 7 \left(\frac{M}{M_\odot}\right)^{-2/5}~{\rm TeV}\;.
\end{equation}

It is important to emphasize here that, among all gravitational tests with black holes, the ones with the smallest masses will lead to the tightest constraints in this theory.  This would lead to the conclusion that the stellar-mass black holes involved in gravitational-wave tests generate the most stringent limits on any possible deviations. This conclusion, however, is an artifact of the assumption intrinsic to this modification that the various couplings are proportional to the dynamical field $\theta$. Different (and perhaps more complex) coupling functions (see, e.g., \cite{Kanti1996,*Kanti1998,Antoniou2018a,*Antoniou2018b}) could easily reverse this trend, as is the case with many modifications of gravity that involve various screening mechanisms (see, e.g., \cite{Joyce2015}).

Finally, even if corrections to the fundamental theory appear at scales that are very different than the masses of astrophysical objects, the presence of horizons surrounding black holes might give rise to horizon-scale classical metric perturbations~\cite{Giddings2016a}. The consequence of such perturbations would be the presence of time-variable, order unity stochastic deviations of black-hole metric parameters, with observable effects both in gravitational-wave emission~\cite{Giddings2016b} and in the images of black-hole shadows~\cite{Giddings2018}. If the timescale of variability of such deviations is longer than the $\sim 10$~h it takes for the EHT to obtain a single snapshot image, then these deviation parameters will be frozen to some arbitrary combination of values.

\section{Black Hole Shadow Tests}

The EHT  generated a high-resolution image of the center of the M87 galaxy~\cite{PaperI} that is characterized by a deep brightness depression surrounded by a ring of emission. This image has been interpreted as the shadow of the central supermassive black hole, cast on the emission from the surrounding plasma. When measured from the reconstructed images, the fractional width of the ring of emission was constrained to be comparable to the nominal resolution of the array~\cite{PaperIV}. On the other hand, when inferred by fitting phenomenological emission models directly to the interferometric visibility data, the fractional width of the ring was constrained to $\lesssim 0.2$, at least for one of the four days of observations~\cite{PaperVI}. 

This property of the emission ring, i.e., that it is very narrow and is truncated at the black-hole shadow, makes it possible to use it as a proxy for measuring the size of the shadow itself. The bias between the diameter of the brightest emission in the ring and that of the shadow was estimated using a large suite of synthetic EHT data based on General Relativistic MagnetoHydroDynamic (GRMHD) simulations and the uncertainty in this bias was found to be comparable to the measured size of the ring itself, as expected. Finally, the inferred size of the shadow was found to be within $\sim 17$\% of the value predicted by the Kerr metric~\cite{PaperVI} using the {\it a priori} known mass-to-distance ratio of the M87 black hole that was based on observations of the motions of stars in its vicinity~\cite{Gebhardt2011}. As proposed in Ref.~\cite{Psaltis2015}, this constitutes a null-hypothesis test of the various assumptions that enter this inference, i.e., that the brightness depression in the image is indeed the black-hole shadow, that the analysis of stellar dynamics provides an accurate measurement of the black-hole mass~\footnote{Had we instead used the mass smaller black-hole mass measurements based on gas dynamics~\cite{Walsh2013}, we would have concluded that the black hole spacetime is not described by the Kerr metric, a situation that was assigned a negligible prior.}, and that the black-hole spacetime is described by the Kerr metric.

Ref.~\cite{Psaltis2020} used the inferred size of the M87 shadow to perform a {\it bona fide} metric test: how much could one deform the Kerr metric and still be consistent with the measurement? Using a variety of parametrically deformed metrics, they showed that, as in the case of Kerr, the shadow size depends very weakly on the spin of the spacetime (defined as the specific angular momentum measured at infinity, divided by the mass of the black hole). Moreover, they demonstrated that, for spherically symmetric spacetimes, the radius of the shadow measured by a distant observer depends only on the $tt-$component of the metric, when expressed in areal coordinates, i.e.,
\begin{equation}
r_{\rm sh}=\frac{r_{\rm ph}}{\sqrt{-g_{tt}(r_{\rm ph})}}\;,
\label{eq:rsh}
\end{equation}
where 
\begin{eqnarray}
    r_{\rm ph}&=&\sqrt{-g_{tt}(r_{\rm ph})} \left(\left.\frac{d\sqrt{-g_{tt}}}{dr}\right\vert_{r_{\rm ph}}\right)^{-1}\nonumber\\
    &=& 2g_{tt}(r_{\rm ph}) \left(\left.\frac{d g_{tt}}{dr}\right\vert_{r_{\rm ph}}\right)^{-1}
\label{eq:rph}
\end{eqnarray}
is the coordinate radius of the photon orbit. Finally, they calculated the constraints imposed by the existing shadow-size measurements on deformed spacetimes that show no deviations at the first PN order (as is the case for, e.g., the black-hole spacetimes in the modified gravity theory discussed in \S2) and are, therefore, fully consistent with all Solar-System tests. Because, as discussed above, there is no guarantee in a general modified gravity theory that the Solar-System constraints are applicable to black-hole spacetimes, this was offered as a proof of principle that shadow observations with current capabilities can be used to impose new and tighter constraints on potential deviations from the Kerr metric for supermassive black holes.

In this paper, we extend the work of Ref.~\cite{Psaltis2020} to explore general forms of deviations from Kerr within multiple phenomenological descriptions without imposing the Solar-System bounds and translate the resulting bounds into limits on the PPN parameters of equilibrium black-hole metrics. 

\subsection{Deformed Metrics without Pathologies}

The outline of a black-hole shadow is the locus of the photon trajectories on the screen of a distant observer that, when traced backwards, become tangent to the surfaces of spherical photon orbits hovering just above the black-hole horizon~\cite{Bardeen1973,*Chandra1983,*Tao2003,*Takahashi2004}. The coordinate radius of the photon orbit for a Schwarzschild spacetime is at $r_S=3M$ and is reduced to $r_S=M$ (i.e., the coordinate radius of the horizon) for a prograde photon orbit around a maximally spinning black hole. Because the Kerr spacetime is regular everywhere outside the horizon and the photon orbits always lie outside the latter, Kerr black holes of all spins are characterized by shadows that can be well defined, at least mathematically (the same is not true for super-spinning Kerr black holes~\cite{Bambi2009}).

However, this is not true in general for deformed Kerr spacetimes. Introducing any naive parametric deformation violates, by construction, the no-hair theorem, which states that the only asymptotically flat vacuum spacetime that is Ricci flat, is free of singularities outside the horizon, and is free of closed time-like loops is the one described by the Kerr metric (we do not consider here charged black holes)~\cite{Israel1967,*Israel1968,*Carter1971,*Hawking1972,*Robinson1975}. Indeed, early attempts to deform the Kerr metric while keeping it Ricci flat led to spacetimes with significant pathologies~\cite{Johannsen2013}. Calculations of black-hole shadows with these deformed spacetimes required excising in an {\it ad hoc} manner the regions with pathologies and limited the solutions to slowly spinning black holes, such that the radii of photon orbits remained outside the pathologies~\cite{Johannsen2010a,Johannsen2010b}. Moreover, the presence of these pathologies effectively precluded any GRMHD simulations of accretion in such spacetimes.

The only way to deform the Kerr metric while removing any pathologies outside its horizon (at least for a broad range of deformation parameters) is to ignore explicitly the requirement that the metric is Ricci flat. In recent years, this has led to a number of parametrically deformed metrics that are free of pathologies but allow for deformations to be dialed in with different phenomenological parameters: the Johannsen-Psaltis metric (hereafter JP)~\cite{Johannsen2011,Johannsen2013} and its extensions by Cardoso, Pani, \& Rico~\cite{Cardoso2014} and Carson \& Yagi~\cite{Carson2020b}, the Modified Gravity Bumpy Kerr metric of Vigeland, Yunes, and Stein~\cite{Vigeland2011} (hereafter MGBK), etc. In a different approach, a general metric can be written in terms of polynomial or rational functions with free coefficients such that, when a particular discrete set of coefficients is chosen, the metric approximates non-Kerr solutions to various modified-gravity field equations; if the non-Kerr solution is free of pathologies, so is the polynomial or rational expansion~\footnote{Albeit useful in numerical studies of known metrics, such parametrizations do not guarantee the absence of pathologies when the various coefficients are chosen outside the discrete sets that are known to describe regular metrics}. This is the approach followed in the Rezzolla-Zhidenko metric (hereafter RZ)~\cite{Rezzolla2014,Konoplya2016}, which has found some use in numerical explorations of black-hole shadows from known non-Kerr metrics~\cite{Younsi2016,Mizuno2018}.

\subsection{Post-Newtonian Expansions of Deformed Metrics}

When written in areal coordinates, the $tt$-component of the JP metric of a non-spinning compact object is~\cite{Johannsen2013}
\begin{equation}
g_{tt}^{\rm JP}=-\left(1-\frac{2}{r_S}\right)\left(1+\sum_{i=2}^\infty\frac{\alpha_{1i}}{r_S^{i}}\right)^{-2}\;,
\label{eq:JPfull}
\end{equation}
where $a_{1i}$ is an infinite sequence of deformation parameters, which are equal to zero for the Schwarzschild metric. In Ref.~\cite{Johannsen2013}, the coefficient $\alpha_{12}$ was set to zero in order to force $\beta_1=1$. In this paper, for reasons discussed above, we will allow potential deviations even at the 1PN order. Written in terms of the PN parametrization discussed in \S2, the result is
\begin{eqnarray}
\zeta_1^{\rm JP}&=&-\alpha_{12}\nonumber\\
\zeta_2^{\rm JP}&=&\alpha_{13}-2\alpha_{12}\nonumber\\
\zeta_3^{\rm JP}&=&-\alpha_{14}+2\alpha_{13}+\frac{3}{2}\alpha_{12}^2\;,
\label{eq:JPPN}
\end{eqnarray}
etc.

Written in areal coordinates, the $tt-$component of the MGBK metric for a non-spinning compact object is~\cite{Vigeland2011}
\begin{equation}
g_{tt}^{\rm MGBK}=-\left(1-\frac{2}{r_S}\right)\left[1-\gamma_1(r_S)-2\gamma_4(r_S)\left(1-\frac{2}{r_S}\right)\right]\;.
\label{eq:MGBKfull}
\end{equation}
This expansion appears to terminate at the $r_S^{-2}$ order because the deformed Kerr metric was designed such that it is characterized by an approximate Killing tensor of that same order. However, the two functions $\gamma_1(r)$ and $\gamma_4(r)$ are arbitrary and can be expanded in series, e.g.,
\begin{equation}
\gamma_{\rm A}=\sum_{n=2}^\infty \frac{\gamma_{{\rm A},n} }{r_S^n}\;,
\end{equation}
where A~$=1$ or 4 and $\gamma_{{\rm A},n}$ are infinite sequences of dimensionless deformation parameters and the $n<2$ terms are equal to zero in order for the metric to be asymptotically flat and have the correct Newtonian limit. Keeping only a few lower-order coefficients, the metric becomes
\begin{eqnarray}
g_{tt}^{MGBK}&=&-1+\frac{2}{r_S}+\frac{\gamma_{1,2}+2\gamma_{4,2}}{r_S^2}\nonumber\\
&&+\frac{-2\gamma_{1,2}+\gamma_{1,3}-8\gamma_{4,2}+2\gamma_{4,3}}{r_S^3}+{\cal O}(r^{-4})\;,\nonumber\\
\end{eqnarray}
such that the PN parameters in areal coordinates are
\begin{eqnarray}
\zeta_1^{\rm MGBK}&=&-\gamma_{1,2}-2\gamma_{4,2}\nonumber\\
\zeta_2^{\rm MGBK}&=&-\gamma_{1,2}+\frac{1}{2}\gamma_{1,3}-4\gamma_{4,2}+\gamma_{4,3}
\label{eq:MGBKPN}
\end{eqnarray}
etc.

Finally, the $tt$-component of the RZ metric for a non-spinning compact object, written in areal coordinates, is~\cite{Rezzolla2014}
\begin{eqnarray}
g_{tt}^{\rm RZ}&=&-\left(1-\frac{r_0}{r_S}\right)\left[1-\epsilon(1-x)\right.\nonumber\\
&&\qquad \left.+(a_0-\epsilon)(1-x)^2+\tilde{A}(x)(1-x)^3\right]\;,
\label{eq:RZfull}
\end{eqnarray}
where
\begin{equation}
x\equiv 1-\frac{r_0}{r_S}\;,
\end{equation}
\begin{equation}
\tilde{A}(x)=\frac{a_1}{1+\frac{a_2 x}{1+\frac{a_3 x}{...}}}\;,
\end{equation}
$r_0$ is the coordinate radius of the infinite redshift surface (heuristically identified with the horizon, if no pathologies exist at larger radii), and $\epsilon$, $a_1$, $a_2$, ... are a sequence of deformation parameters. Writing all radii in terms of the mass of the black hole, as measured at infinity, fixes one of the parameters to
\begin{equation}
\epsilon=-\left(1-\frac{2}{r_0}\right)\;.
\end{equation}
Under these assumptions, the PN parameters of the RZ metric become
\begin{eqnarray}
\zeta_1^{\rm RZ}&=&\frac{1}{2}a_0 r_0^2\nonumber\\
\zeta_2^{\rm RZ}&=&\frac{1}{2}\left[1-\frac{2}{r_0}+a_0 - \frac{a_1}{1+\frac{a_2}{1+\frac{a_3}{...}}}\right] r_0^3\;,
\label{eq:RZPN}
\end{eqnarray}
etc.

\subsection{Photon Orbits and Shadows of Deformed Metrics}

The three different parametrizations~(\ref{eq:JPfull}), (\ref{eq:MGBKfull}), and (\ref{eq:RZfull}) for the deformed metric of a non-spinning object share a common characteristic: they all involve expansions in power-series after the factor $(1-2/r_S)$ has been removed from the $tt$-component of the metric. This ensures that the metric has a surface of infinite redshift (a ``horizon'') for a large range of deformation parameters, which helps in hiding the pathologies introduced by the deformations from the observable universe. However, because of this factoring, a single PN parameter $\zeta_i$ in areal coordinates corresponds to either a finite combination (for the JP and MGBK metrics) or an infinite complex function (for the RZ metric) of the deformation parameters of each metric. Nevertheless, in each case, there is a trivial transformation between the PN parameters $\zeta_i$ and the deformation parameters of the corresponding metric. For this reason, we will use the JP parametrization given by Eq.~(\ref{eq:JPPN}) to place constraints on plausible metric deviations from the measurement of the black-hole shadow diameter. One can then use transformations~(\ref{eq:JPPN}), (\ref{eq:MGBKPN}), or (\ref{eq:RZPN}) to convert them into constraints on the particular deviation parameters for the other metric parametrizations. 

Using Eq.~(\ref{eq:JPfull}) with Eq.~(\ref{eq:rph}) yields an expression for the coordinate radius of photon orbits, i.e.,
\begin{equation}
r_{\rm ph}^{\rm JP}=3+\frac{2}{9}\alpha_{12}+\frac{1}{9}\alpha_{13}+\frac{4}{81}\alpha_{14}+...\;,
\label{eq:rphJP}
\end{equation}
where only linear terms in the deformation parameters have been retained. This is consistent with the power-series expansion of the metric shown in equation~(\ref{eq:JPfull}). Moreover, terms that involve higher powers in the deformation parameters are negligible compared to the corrections introduced by the black-hole spin, which are not measurable with current data and are neglected here.  For example, the first non-linear term is $-2\alpha_{12}^2/243\simeq 0.008 \alpha_{12}$. Evaluating the $tt$-component of the metric at the radius of the photon orbit gives
\begin{equation}
g_{tt}(r_{\rm ph})=-\frac{1}{3}+\frac{2}{81}\alpha_{12}+\frac{10}{729}\alpha_{14}+...\;,
\end{equation}
which remains regular unless the deformation parameters take extremely large values. Note that the $\alpha_{13}$ contribution to this expression is vanishing at linear order.

Expressed in terms of the PN parameters in areal coordinates, the radius of the photon orbit of the JP metric becomes
\begin{equation}
r_{\rm ph}^{\rm JP}=3-\frac{52}{81}\zeta_1 + \frac{17}{81}\zeta_2 - \frac{4}{81}\zeta_3+...
\end{equation}

Finally, inserting equation~(\ref{eq:rphJP}) into the general expression~(\ref{eq:rsh}) for the shadow radius gives
\begin{equation}
r_{\rm sh}^{\rm JP}=3\sqrt{3}\left(1+\frac{1}{9}\alpha_{12}+\frac{1}{27}\alpha_{13}+\frac{1}{81}\alpha_{14}+....\right)\;.
\label{eq:rshJP}
\end{equation}
Expressed in terms of the PN parameters in areal coordinates, the shadow radius becomes
\begin{equation}
r_{\rm sh}^{\rm JP}=3\sqrt{3}\left(1-\frac{19}{81}\zeta_1+\frac{5}{81}\zeta_2-\frac{1}{81}\zeta_3+...\right)\;.
\label{eq:rshJPPN}
\end{equation}

It is important to emphasize that the shadow tests performed here and in Ref.~\cite{Psaltis2020} do {\em not\/} employ a parametric post-Newtonian metric. Indeed, the coefficients of the various terms in Eq.~(\ref{eq:rshJPPN}) are different from the expression obtained if calculating the size of a black-hole shadow using directly the PN metric~(\ref{eq:PNzeta}). Instead, the shadow tests are performed using metrics that remain regular all the way down to their horizons and constraints are imposed on the parameters of these regular metrics. However, in order to compare the shadow tests to those of earlier, weak-field tests, these constraints are then translated into equivalent constraints on the post-Newtonian parameters of these metrics.

The convergence properties of the series~(\ref{eq:rshJP}) and (\ref{eq:rshJPPN}) are difficult to explore formally. Nevertheless, the coefficients of the terms in Eq.~(\ref{eq:rshJP}) are decreasing by successive powers of 3 and those in Eq.~(\ref{eq:rshJPPN}) by successive powers of $\sim 4-5$. This is expected given that they are all the result of power series expansions in $1/r_S$ and the radius of the photon orbit, which determines primarily the size of the shadow, is $r_S=3$ for the Schwarzschild metric. In order for higher-order terms to have a significant impact on the size of the shadow, they need to be successively increasing by corresponding powers of $\sim 3-5$.

\subsection{Metric Constraints from the Measured Size of the M87 Shadow}

Our goal is to place constraints on possible deviations from the Kerr metric using the inferred size of the black-hole shadow in M87, given the mass of the black hole measured at large distances, in the Newtonian limit, as prior information. However, as discussed earlier, the EHT imaging observations of M87 do not directly measure the size of the shadow but rather the size of the bright ring of emission that surrounds it. 

In order to connect the two, we use the model described in~Ref.~\cite{PaperVI} that incorporates a number of steps in order to convert the prior mass measurement to a prediction of the size of the bright ring. In particular, we quantify the prior in terms of the angular size in the sky of one gravitational radius for a black hole of mass $M$ at a distance $D$, i.e., $\theta_{\rm g}\equiv GM/(c^2D)$. We introduce the correction factor $a$ between the angular diameter $\hat{d}_{\rm m}$ of the peak emission and the angular diameter of the shadow $2r_{\rm sh}$ such that
\begin{equation}
    \hat{d}_{\rm m}= 2 a r_{\rm sh}\;.
\end{equation}
Finally, we calculate the angular size of the shadow for a given prior mass-to-distance ratio for the black hole, while allowing for a possible fractional deviation $\delta$ in the prediction of the Kerr metric such that
\begin{equation}
    \hat{d}_{\rm m}= 2 a (1+\delta) r_{\rm sh}=2 a (1+\delta) 3\sqrt{3} \left(\frac{GM}{Dc^2}\right)\;.
    \label{eq:dmodel}
\end{equation}
Even though $\delta$ represents any possible deviation, for the particular model discussed above, it is equivalent to
\begin{equation}
    \delta=\frac{19}{81}\left(-\zeta_1+\frac{5}{19}\zeta_2-\frac{1}{19}\zeta_3+...\right)\;.
    \label{eq:delta}
\end{equation}
Eq.~(\ref{eq:dmodel}) allows us to infer or constrain the deviation $\delta$ from the Kerr metric predictions, given the prior information on the mass-to-distance ratio $M/D$ of the black hole, the measurement of the diameter $\hat{d}_{\rm m}$ of the bright ring of emission with the EHT, and a model for the correction factor $a$.

The prior $P(\theta_{\rm g})$ on the angular size in the sky of one gravitational radius for the black hole in M87 has been measured using stellar dynamics in Ref.~\cite{Gebhardt2011} and quantified in Ref.~\cite{PaperIV}. Here we use the full numerical information on the prior, which peaks at $\theta_{\rm g,0}\sim 3.62~\mu$as and is asymmetric (it can be represented approximately as $\theta_{\rm g}=3.62^{+0.60}_{-0.34}~\mu$as).

The correction factor $a$ has been calibrated using $\sim 100$ synthetic images from accretion-flow simulations that span different black-hole spins, magnetic field configurations in the accretion flows, and models for the plasma physics. Even though the particular simulations used in the calibration were performed for the Kerr metric, the primary source of the error budget arises from the thermodynamic properties of the plasma in the inner accretion flow; the properties of the metric enter predominantly in the imprint of gravitational lensing on the image, which is the size and shape of the black-hole shadow. We use here a Gaussian distribution for $a$ with a mean value $6\sqrt{3}a_0=11.35$ and a standard deviation of $\sigma_a/a_0=11.4$\% (see Table~4  and Fig.~26 of Ref.~\cite{PaperVI}). Note that the width of this distribution incorporates the small ($\pm 4$\%) spread in the predicted shadow size due to the unknown black-hole spin and observer inclination~\cite{Johannsen2010b}.

\begin{figure}[t]
\includegraphics[width=0.48\textwidth]{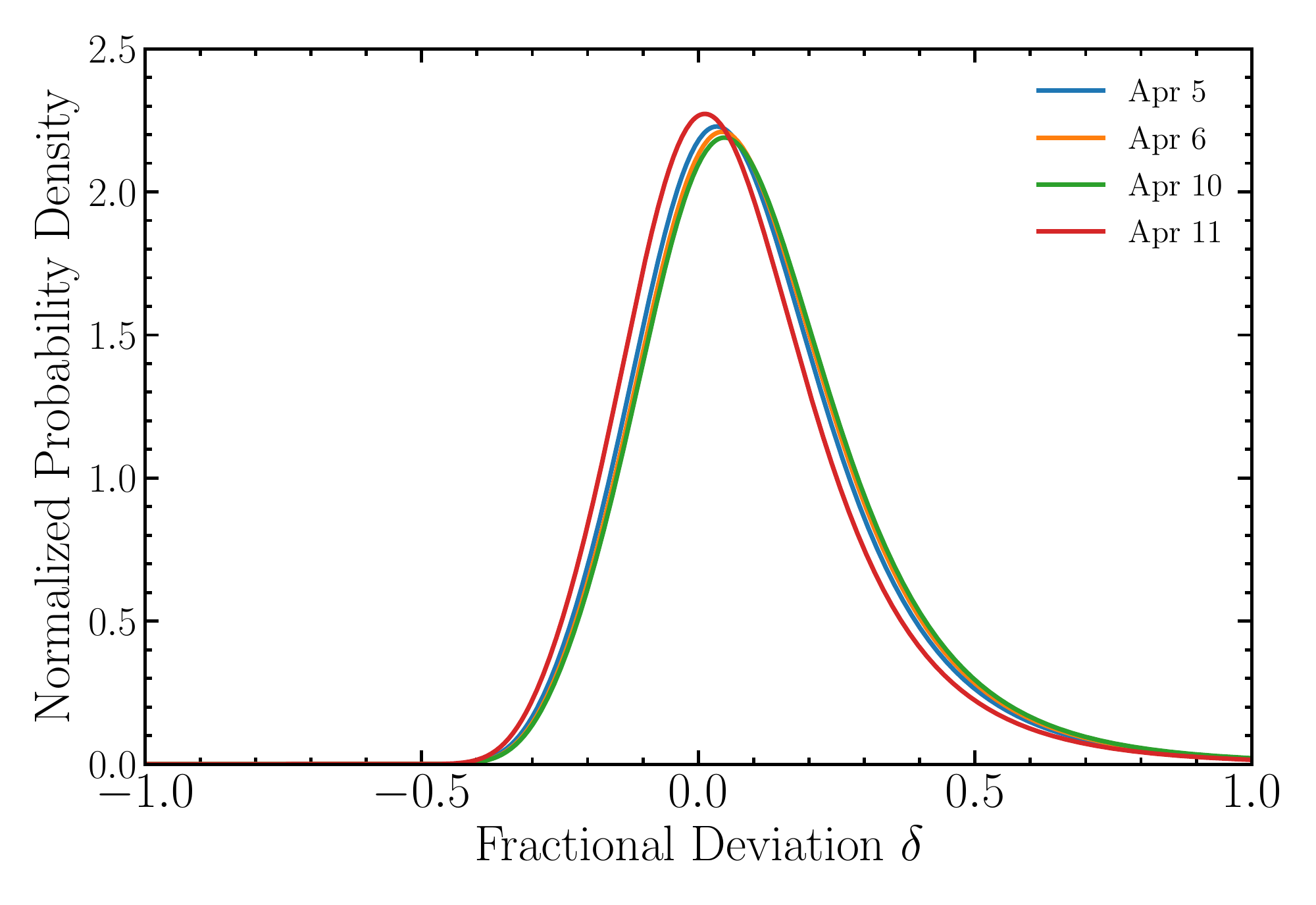}
\caption{\label{fig:EHTpost} The posterior over the fractional deviation $\delta$ between the size of the shadow predicted using the Kerr metric for the M87 black hole and the size inferred for the 4 days of the 2017 EHT observations. These posteriors are based on the stellar dynamics measurement of the mass-to-distance ratio for the black hole taken as a prior and incorporate the uncertainty introduced by the difference between the size of the black-hole shadow and that of the bright ring of emission, as well as the uncertainty due to the unknown black-hole spin and observer inclination. For the Apr~5  observations, the result is $\delta=0.03^{+0.20}_{-0.16}$ (68\% credible level) and is approximately the same for the remaining days.
}
\end{figure}

The EHT observed the black hole in the center of M87 for four days across the span of a week in April~2017. Interferometric data were collected in two frequency bands (HI and LO) and images were generated for each day and each frequency band separately~\cite{PaperIV}. Moreover, two different geometric model images as well as images generated from GRMHD simulations were fit directly to the interferometric visibility data, in order to infer the geometric parameters of the emission rings~\cite{PaperVI}. Both the image-domain and visibility-domain analyses led inferred image sizes that are consistent with each other and across the different days of observation and different frequency bands. Here, we use the posteriors for the measurement of the diameter of the ring of emission inferred using the \texttt{xs-ringgauss} model for the combined HI+LO datasets, in all four days of observations. We denote the most likely value of each measurement by $\hat{d}$ and its uncertainty by $\sigma_{\rm d}$ (see Table~3 of Ref.~\cite{PaperVI}).

The posterior over the deviation parameter $\delta$ is given by
\begin{eqnarray}
    P(\delta\vert \hat{d},\sigma_d)&=&C \int d\delta\int da \int d\theta_{\rm g} \;{\cal L}[\hat{d},\sigma_d|\theta_{\rm g},a,\delta] \nonumber\\
    &&\qquad\qquad\times P(\delta) P(\theta_{\rm g}) P(a)\;,
\end{eqnarray}
where $C$ is an appropriate normalizaton constant and ${\cal L}[\hat{d},\sigma_{d}|\theta_{\rm g},a,\delta]$ is the likelihood of measuring a ring of size $\hat{d}$ given the model parameters $\theta_{\rm g}$, $a$, and $\delta$. Hereafter, we will assume a flat prior in the fractional deviation $\delta$, with limits that are much larger than unity.  Assuming a Gaussian distribution for both the likelihood function and the model correction factor $a$, we perform two of the integrals analytically such that
\begin{eqnarray}
    &&P(\delta\vert \hat{d},\sigma_d)=\int d\theta_{\rm g}P(\theta_{\rm g})\frac{1}{\sqrt{2\pi} \Sigma}\nonumber\\
    &&
    \exp\left[-\frac{108 a_0^2 (1+\delta)^2 \theta_{\rm g}^2 -12\sqrt{3} a_0
    (1+\delta)\theta_{\rm g}\hat{d}+\hat{d}^2}{2\Sigma^2}\right]\;,
\end{eqnarray}
where 
\begin{equation}
    \Sigma=\sqrt{108(1+\delta)^2\theta_{\rm g}^2 \sigma_a^2+\sigma_d^2}\;.
\end{equation}
After folding in the prior over $\theta_{\rm g}$ and performing the last integral numerically, we obtain, for each of the four days of observations, the posteriors shown in Fig.~\ref{fig:EHTpost}.

For the Apr~5~2017 observations, the result is $\delta=0.03^{+0.20}_{-0.16}$ (68\% credible level) and is approximately the same for the remaining days. This implies that the limit on the deviation parameters becomes
\begin{equation}
-0.55\le -\zeta_1+\frac{5}{19}\zeta_2-\frac{1}{19}\zeta_3+... \le 0.98\;.
\label{eq:EHTlimits}
\end{equation}

It is possible to use inequalities~(\ref{eq:EHTlimits}) in order to constrain any individual parameter $\zeta_i$, allowing for only that parameter to attain a non-zero value, as was done, e.g., in Ref.~\cite{Psaltis2020}.   However, because the shadow for a non-spinning compact object is circularly symmetric, its image provides only a single data point: its radius. For a general parametric extension of the metric with an infinite sequence of deviation parameters, it is evident from the above inequality that this single data point can only constrain the linear combination~(\ref{eq:EHTlimits}) of the infinite series of parameters. Because of this complete degeneracy, it is impossible to quantify credible levels for each of the deviation parameters separately by marginalizing over the remaining parameters.

\begin{figure}[t]
\includegraphics[width=0.48\textwidth]{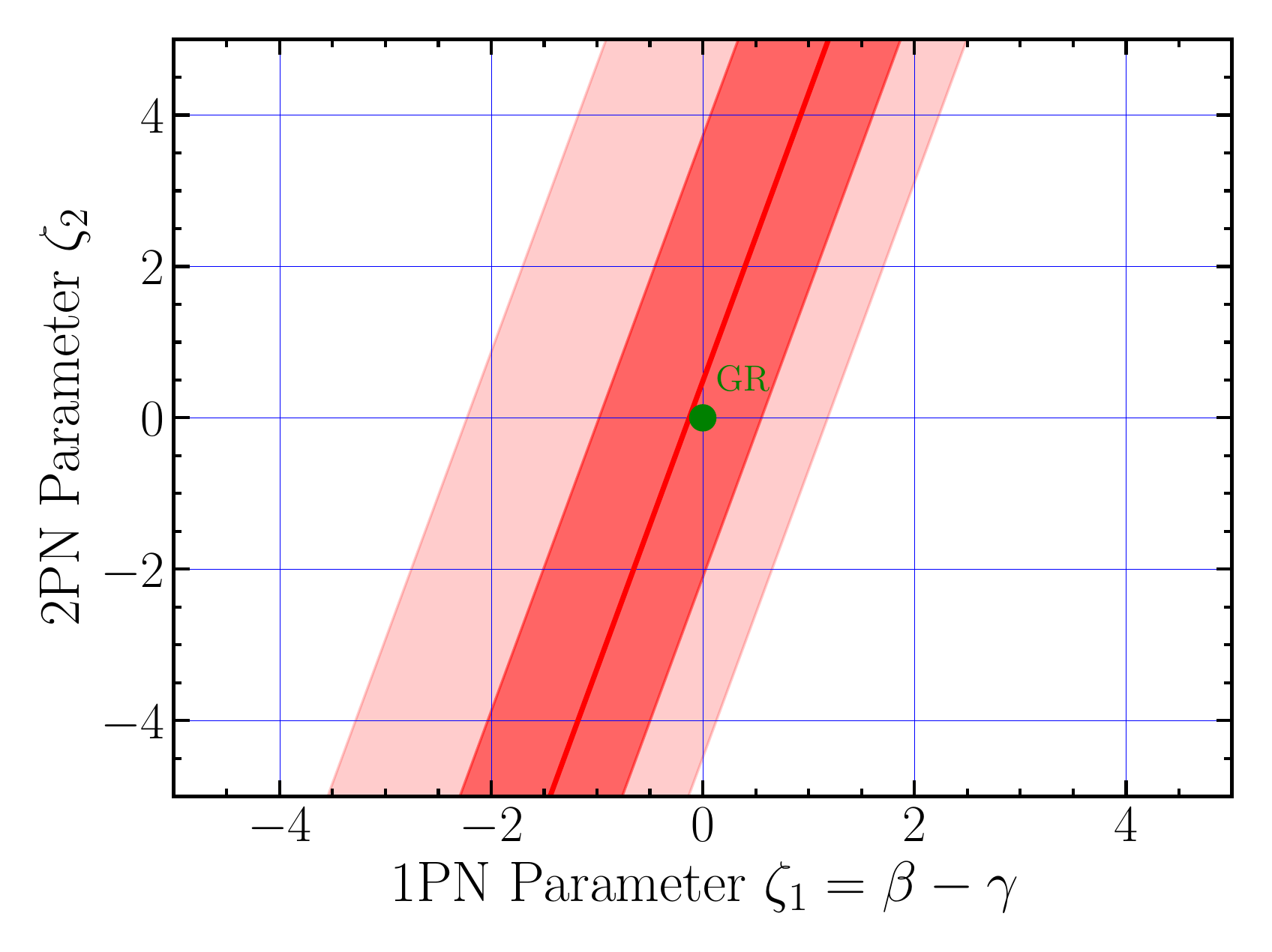}
\caption{\label{fig:EHTlimits} Correlated bounds on the 1PN and 2PN parameters $\zeta_1$ and $\zeta_2$ imposed by the measurement of the size of the black-hole shadow in M87. The bounds were calculated with the JP parametrization of deformations from the Kerr metric and the shaded regions show the regions of the parameter space with deviation $\delta$ that is within the 68\% and 95\% credible intervals for the April~5 observations. The bounds assume that all higher-order deviation parameters provide negligible corrections to the predicted shadow size, which is true when $\vert\zeta_i\vert\lesssim 1$ for $i\ge 3$.}
\end{figure}

Nevertheless, the coefficients of each parameter in the sum of Eq.~(\ref{eq:EHTlimits}) appear to be decreasing by a factor of $\sim 4-5$ between successive orders. As a result, if the deviation parameters are comparable to each other or are also decreasing with increasing order, then the series will be converging within a very small number of terms. In fact, even if the 3PN parameter $\zeta_3$ is of order unity, the resulting correction to the predicted shadow size will be at the $\sim 1$\% level, i.e., smaller than the measurement uncertainty of $\lesssim 20$\%. Assuming that only the first two PN parameters, $\zeta_1$ and $\zeta_2$, provide contributions that are measurable with the current observations, inequality~(\ref{eq:EHTlimits}) results in the correlated upper limits shown in Fig.~\ref{fig:EHTlimits}. 

\section{Black Hole Inspiral Tests}

The LIGO/Virgo detectors~\cite{AdLIGO, AdVirgo} have observed gravitational waves emitted during the inspiral, coalescence, and ringdown phases of several black-hole binaries and used them to place constraints on potential deviations from various GR predictions for black-hole spacetimes~\cite{Abbott2016,Abbott2019,GWTC2:GR}. The tests performed to date can be broadly categorized into three large groups: (i) those involving the inspiral phase, when the coalescing black holes are at distances larger than their horizons; (ii) those involving the ringdown phase of the remnant black hole that is shedding its short-lived gravitational hair, and (iii) the tests involving the polarization and propagation of gravitational waves from the binary to the Earth. The various categories are, of course, not independent from each other and substantial information can be obtained by exploring, e.g., whether the black-hole parameters inferred from the inspiral phase of an event with GR waveforms are consistent with those inferred from the ringdown phase.

In this paper, we will focus only on the tests involving the inspiral phase for three reasons. First, the majority of the individual black holes coalescing in the various detected events appear to have small spins (see~\cite{GWTC2,*GWTC2:pop} for a detailed discussion of the spin distribution and the evidence for non-zero spins in some of the observed systems), in contrast to the remnant black holes ringing down, for which the inferred spins are $\sim 0.7$. Second, during the inspiral phase, the coalescing black holes are at separations larger than the effective radius of the innermost stable circular orbit in the system, so that their individual spacetimes are mildly perturbed away from the equilibrium solutions. Moreover, the relatively large separations allow for useful constraints to be obtained even with a post-Newtonian expansion of the waveforms. Finally, modeling the inspiral phases allows us to place constraints on deviations from the GR predictions for the equilibrium black-hole metrics that can be directly compared to those obtained using the EHT observations, as discussed in the previous section. 

To date, the LIGO Scientific Collaboration and Virgo Collaboration have confirmed the observation of gravitational waves from 47 compact binary coalescences with a false alarm rate below 1 per year, including 44 binary black hole systems, two binary neutron star systems, and GW190814, which is likely a binary black hole system~\cite{GWTC2,*GWTC2:pop}. In this paper, we reanalyze the LIGO/Virgo data for three sources, focusing exclusively on tests of GR during the inspiral phase of three events: {\em (i)\/} GW150914, the first and highest signal-to-noise ratio binary black hole merger observed to date; {\em (ii)\/} GW170608, the low-mass system with the strongest constraints on GR deviations to date; {\em (iii)\/} GW190924\_021846, the lowest mass binary black hole system observed to date. We analyze these three out of the large number of available systems in order explore the mass dependence of the constraints.

\begin{figure}[t]
\includegraphics[width=0.48\textwidth]{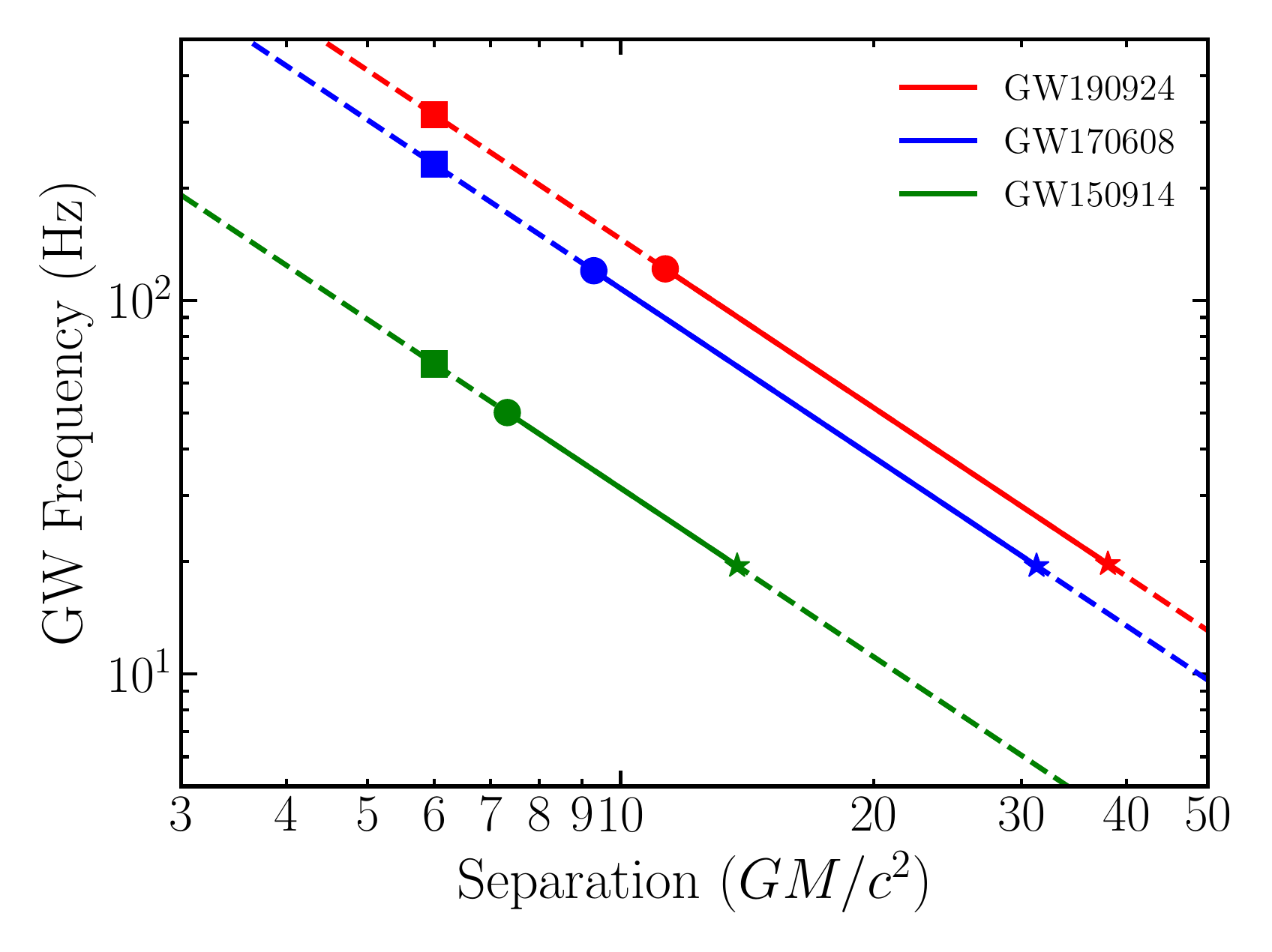}
\caption{\label{fig:LIGOsources} The frequencies of the emitted gravitational waves as a function of the binary separation, for the three black-hole binary events that we explore here. The stars denote the lowest frequency analyzed, in each case this is 20 Hz; the filled squares denote the frequency at which the separation is $\sim 6 GM/c^2$; the filled circles denote the frequencies at which we terminate the analyses of the early inspiral phases of the events. The solid lines identify the range of frequencies and corresponding binary separations that contribute  to the constraints on deviations from the GR predictions explored here.}
\end{figure}

Figure~\ref{fig:LIGOsources} shows the frequency of the emitted gravitational waves as a function of the binary separation for the three sources we analyze here. For the purposes of this figure we set the total mass of GW150914 to $M=65 M_\odot$, of GW170608 to $M=19 M_\odot$, and of GW190924\_021846 to $14 M_\odot$~\cite{GWTC2}. The stars denote the gravitational-wave frequencies ($\sim 20$~Hz) at which the signal amplitudes become discernible above the noise of the detectors. The filled squares represent the frequencies at which the orbital separation is $6 GM/c^2$, i.e.,
\begin{equation}
   f_6=\frac{1}{6\sqrt{6}\pi}\left(\frac{GM}{c^3}\right)^{-1}\;.
\end{equation}
Finally, the filled circles represent the frequencies at which we  terminate  the  analyses of the early inspiral phases of the events in order to ensure that our constraints are driven solely by the inspiral regimes (see below). The solid lines identify the range of frequencies and corresponding binary separations that primarily inform the tests of deviations from GR predictions based on the inspiral phases.

\subsection{Post-Newtonian Tests of Black-Hole Binary Inspirals}

In the study of potential deviations from the GR waveforms during the inspiral phases, the LIGO/Virgo collaboration has been employing, among others, the \texttt{IMRPhenomPv2} waveform model~\cite{IMRPhenomPv2}, which is a precessing modification to the aligned-spin waveform model \texttt{IMRPhenomD}~\cite{Khan2016}.
For this model, the phase evolution during the inspiral phase up to the third PN order is exactly the \texttt{TaylorF2} post-Newtonian model
\begin{eqnarray}
\Phi_{\rm GR}(f)&=&2\pi f t_c - \phi_c - \pi/4\nonumber\\
&+&\frac{3}{128\eta}\left(\pi f M\right)^{-5/3} \sum_{i=0}^{7} \phi_i \left(\pi f M\right)^{i/3}\;,
\label{eq:Phemom}
\end{eqnarray}
where $i=2p$ is twice the PN order (in the counting system used here) and $t_c$ and $\phi_c$ are the time and phase at coalescence. Here $M=m_1+m_2$ is the total mass of the system and $\eta\equiv m_1 m_2/M^2$ is the symmetric mass ratio. Beyond this, there are additional phenomenological correction terms at higher order.  Neglecting spin corrections, the first few $\phi_i$ terms in GR are
\begin{eqnarray}
\phi_0&=&1\nonumber\\
\phi_1&=&0\nonumber\\
\phi_2&=&\frac{3715}{756}+\frac{55}{9}\eta\nonumber\\
\phi_3&=&-16\pi\nonumber\\
\phi_4&=&\frac{15293365}{508032}+\frac{27145}{504}\eta+\frac{3085}{72}\eta^2\;.
\label{eq:GRphi}
\end{eqnarray}
The deviation from the GR predictions are then usually parametrized as a fractional shift $\delta \hat{\phi}_i$ in $\phi_i$ as described in~\cite{Agathos2014}. Note that, throughout this section, we have reverted to showing explicitly the dependence of the various terms on the mass of the binary, because of the impact of the actual value of the mass on the correlations between the inferred parameters, as we describe below.

\begin{figure}[t]
\includegraphics[width=0.48\textwidth]{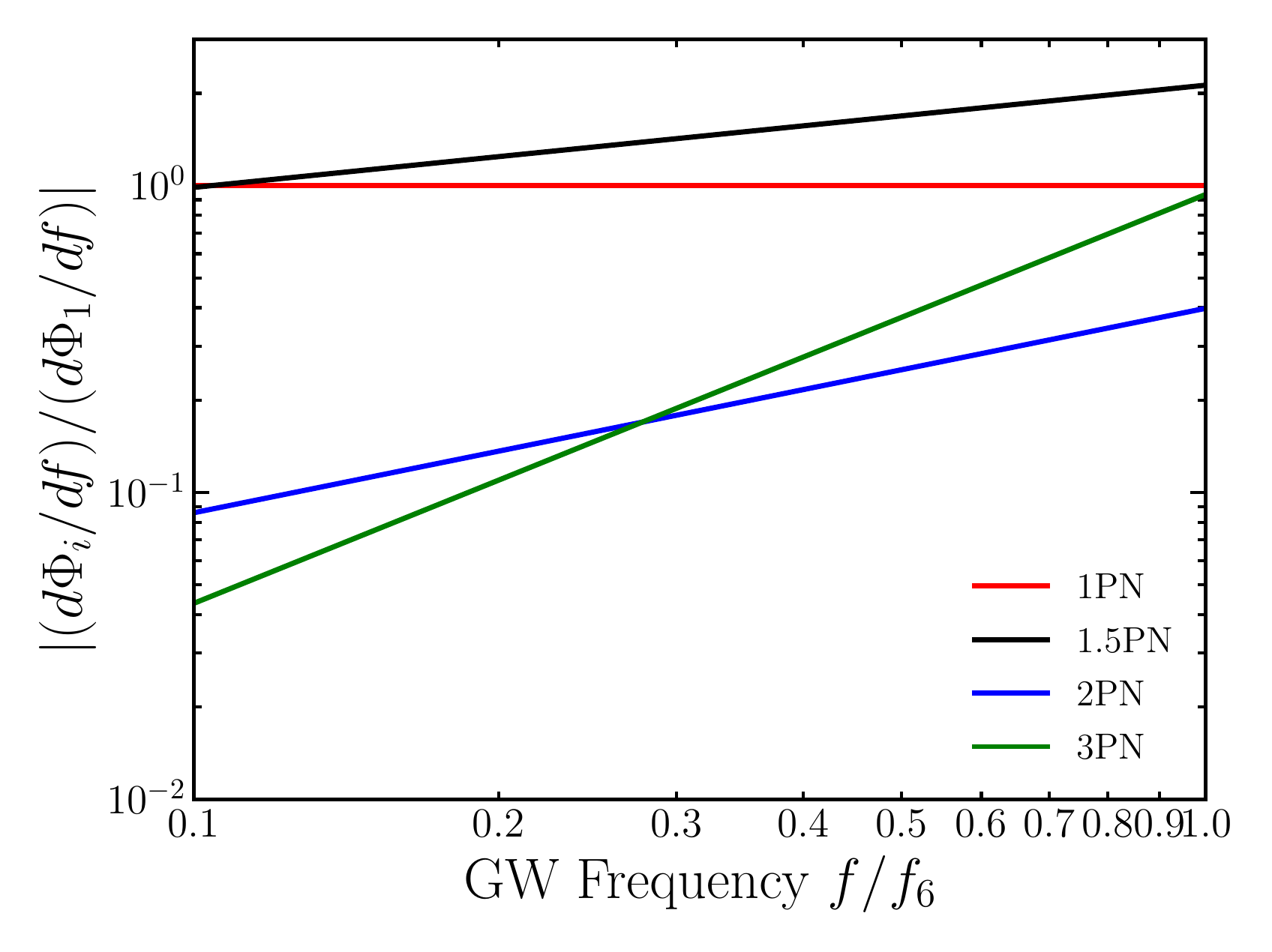}
\caption{\label{fig:LIGO_conv} The relative importance of different PN orders in the calculation of the frequency derivative of the phase of the waveform, as a function of the wave frequency. The PN order is $p=i/2$, as in Eq.~(\ref{eq:Phider}), and all contributions are normalized to the 1PN order. The wave frequency is expressed in units of $f_6$, the frequency at a separation equal to $6GM/c^2$, at which point the inspiral phase ends. All terms have been calculated for a binary with equal masses, i.e., $\eta=1/4$ , and zero spins.}
\end{figure}

The particular values of the phase are, of course, arbitrary, and can be masked by an adjustment in the phase $\phi_c$. Instead, the inference of the model parameters is driven primarily by the evolution of the phase of the gravitational wave with frequency. In the Fourier domain, where the \texttt{IMRPhenomPv2} model is defined, this is captured by the first derivative of the phase of the waveform with respect to frequency, i.e.,
\begin{equation}
\label{eq:Phider}
\frac{d\Phi(f)}{df}=\sum_{i=0}^{7}\frac{d\Phi_i(f)}{df} \;,
\end{equation}
with
\begin{equation}
\frac{d\Phi_i(f)}{df}=\frac{M}{128\eta} (i-5)\pi^{(i-5)/3}  \left(f M\right)^{(i-8)/3}\phi_i(1+\delta\hat{\phi}_i)\;,
\end{equation}
where again $i=2p$ is twice the PN order. 

In order to explore the convergence of this series, we first focus on the GR case, i.e., set $\delta \hat{\phi}_i=0$, and express the frequency in terms of $f_6$ by setting $\hat{f}\equiv f/f_6$. Figure~\ref{fig:LIGO_conv} shows the relative importance of the various PN orders with respect to the 1PN order. At the lowest frequencies accessible with LIGO/Virgo, e.g., $\hat{f}\sim 0.1$ for GW170608, the relative contribution of the 2PN and 3PN orders are comparable to each other and $\sim 10$\% with respect to the 1PN order. As expected, their relative contribution increases (to $\sim 1/2$) as the event approaches the end of the inspiral phase and the PN expansion becomes less accurate.

Incorporating potential deviations from the GR waveforms leads to (showing only the integer PN orders, which describe potential deviations from Kerr of the equilibrium black-hole metrics; see below)
\begin{equation}
\frac{d\Phi(f)}{d \hat{f}} \sim   -\frac{8.9}{\hat{f} ^2} (1+\delta \hat{\phi}_2)-\frac{3.5}{\hat{f}^{4/3}}
   (1+\delta \hat{\phi}_4)-\frac{8.3}{\hat{f}^{2/3}}(1+\delta \hat{\phi}_6)\;,
\end{equation}
where, for clarity, we have set $\eta=1/4$. If we evaluate this expression at $f=(1/10)f_6$, i.e., approximately when the LIGO and Virgo interferometers start detecting the signal from GW170608, we obtain
\begin{eqnarray}
    \left.\frac{d\Phi(f)}{d \hat{f}} \right\vert_{\hat{f}=1/10}&\simeq & -1002\left(1 + 0.89 \delta\hat{\phi}_2 + 0.08\delta\hat{\phi}_4\right.\nonumber\\
    &&\qquad\qquad\left.+ 
 0.04\delta\hat{\phi}_6 + ...\right)\;.
 \label{eq:PhinonGR}
\end{eqnarray}
The convergence becomes increasingly slower as the binary separation decreases towards $6GM/c^2$ and the frequency of the wave increases towards $f_6$. 

We show later in this section that the constraints on the various deviation parameters are actually driven by the second derivative of $\Phi$ with respect to frequency, because changes in the first derivative can be masked by an adjustment in the coalescence time $t_c$. Had we considered the convergence of the second derivative with frequency, we would have reached a similar conclusion.

The rate of convergence for the waveforms of coalescing binary black holes (see, e.g., Fig.~\ref{fig:LIGO_conv}) is similar to the rate of convergence of the series~(\ref{eq:rshJPPN}) used in tests involving the size of the black-hole shadow. This might appear counter intuitive, since the separations of the binaries in the inspiral phase are in the range $\sim (6-40) GM/c^2$, whereas the radius of the photon orbit is at $3GM/c^2$. The reason lies in the fact that the coefficients of the various PN orders in the phase evolution of the waveforms are increasing rapidly with PN order: for a binary system with equal masses, Eq.~(\ref{eq:GRphi}) gives $(\phi_1,\phi_2, \phi_3, \phi_4, \phi_5, \phi_6, ...)\simeq(0,6, -50,46,154,-652,...)$.

Previous tests of the compact binary phase evolution have mostly allowed for just one parameter to deviate at a time (but see \cite{Abbott2016} for the analysis of GW150914 in which multiple parameters were varied simultaneously). This is done in order to improve the constraints at specific post-Newtonian orders and reduce computational challenges. When all deviation parameters are allowed to vary, their posteriors are highly correlated and, for the 1PN and higher terms, span a range comparable to that imposed by the priors (see~\cite{Abbott2016}).  In fact, if all post-Newtonian coefficients are allowed to deviate from the GR values, the problem is under-determined.  The number of free parameters governing the intrinsic phasing of the inspiral, which includes the masses, the spins, and the deviation terms terms $\delta\hat{\phi}_i$, is greater than the number of post-Newtonian coefficients.  Therefore, even if all coefficients were to be measured perfectly in the absence of noise, the physical parameters and deviation terms can only be constrained to lie on a degenerate hypersurface whose dimensionality is given by the number of excess parameters relative to the number of measured post-Newtonian terms.  However, in order to compare the LIGO/Virgo constraints to those imposed by the measurement of the black-hole shadow in M87, we are interested specifically in the correlations between the various deviation parameters.

Another complication in using the inspiral gravitational waveforms to constrain the equilibrium spacetimes of black holes, which is our aim here, is the need to have a model for the radiative (or dissipative) properties of the underlying theory. In fact, it is well known that several modifications to GR give rise to altered waveforms, even though the black-hole metrics in these modified theories remain identical to the Kerr solution~\cite{Barausse2008}. This is the reason why the post-Newtonian constraints imposed by the LIGO/Virgo events on, e.g., the phase evolution of the detected gravitational waves amalgamate potential deviations of both the equilibrium spacetimes (i.e., the conservative properties of the theory) and of the gravitational-wave emission and propagation (the dissipative and radiative properties of the theory).

In order to break the complete degeneracy between varying the physical parameters of the black-hole binaries and the deviation terms of the metrics, we explore a different cross section of the parameter space of possibilities compared to previous work (e.g.,~\cite{GWTC2:GR}). In accordance with our goal to use gravitational-wave observations to constrain the equilibrium spacetimes of black holes, we will consider the case where the radiative aspects of the theory are the same as in GR and allow only for the black-hole spacetimes to deviate from the GR solutions. 
The emission of gravitational waves, which drives the phase evolution of the waveforms, will again start at the 2.5PN order, as in GR, and the  gravitational-wave amplitude will be proportional to the second time derivative of the quadrupole moment of the spacetime. However, this derivative will be determined by the time-dependent relative positions of the two black holes in the orbit, which in turn depend on the equilibrium spacetimes of the individual black holes. This is how the PN parameters that we are concerned with in this study enter the calculation of the waveforms and can be constrained with observations~\footnote{We do not explore the possibility that back-scattering of gravitational waves off the background -- so-called tails of tails terms -- could contribute to higher non-integer deviations in the gravitational-wave signature: the 2.5PN term corresponds to a fixed phase offset that is marginalized over, and we restrict our analysis to terms at or below 3PN.}.

\subsection{Modeling Inspiral Data with ppE Waveforms}

In accordance with the reasoning described above, we will follow the procedure outlined in Refs.~\cite{Carson2020,Cardenas2020,Tahura2018} for calculating the waveforms of gravitational waves during the inspiral phase based on the parametric post-Einstein (ppE) phenomenological approach~\cite{Yunes2009} and using effective one-body dynamics~\cite{Buonanno1999}\footnote{Even though the effective one-body approach was proven to be valid only for the radiative properties of GR, this is consistent with the approach in this work of allowing only for the equilibrium spacetimes to be different than GR}. This enables us to compare model waveforms to LIGO/Virgo data by allowing for simultaneous deviations at more than a single PN term while reducing the extent of correlations discussed above. Because we will be using the ppE formalism, we are also able to incorporate amplitude and phase information in our model.

As shown in Ref.~\cite{Carson2020}, the waveform evolution during the inspiral phase depends only on the $tt$-components of the black-hole metrics as written in areal coordinates. In the ppE formalism, the waveform during the inspiral phase of a binary, written in the frequency domain, is given by~\cite{Yunes2009}
\begin{eqnarray}
\tilde{h}_{\rm ppE}(f)&=& {\cal A}_{\rm GR}(f)\left(1+\alpha_{\rm ppE} u^{a_{\rm ppE}}\right)\nonumber\\
&&\qquad\qquad\exp\left\{i\left[\Phi_{\rm GR}(f)+\beta_{\rm ppE} u^{b_{\rm ppE}}\right]\right\} \;,
\label{eq:hppE}
\end{eqnarray}
where $A_{\rm GR}$ and $\Phi_{\rm GR}(f)$ are the GR predictions for the amplitude and phase of the waveform, $u\equiv (\pi {\cal M} f)^{1/3}$, ${\cal M}=M \eta^{3/5}$ is the chirp mass, $M=m_1+m_2$ is the total mass, $m_1$ and $m_2$ are the masses of the two black holes, and $\eta=m_1 m_2/M^2$ is the symmetric mass ratio, as before. The various parameters with the ``ppE'' subscripts describe potential deviations from the GR predictions. In order to incorporate the individual PN orders, we write
\begin{eqnarray}
\tilde{h}_{\rm ppE}(f)&=& {\cal A}_{\rm GR}(f) e^{i\Phi_{\rm GR}(f)} {\displaystyle \prod_{p=1}^{3}} {\cal A}_{{\rm ppE}, p} e^{i\Phi_{{\rm ppE}, p}} \\
{\cal A}_{{\rm ppE}, p} &=& \left(1+\alpha_{{\rm ppE}, p} u^{a_{{\rm ppE}, p}}\right) \nonumber\\
\Phi_{{\rm ppE}, p} &=& \beta_{{\rm ppE}, p} u^{b_{{\rm ppE}, p}}.
\end{eqnarray}

The two ppE parameters that affect the phase evolution of the gravitational wave, which we will use here, are the amplitude $\beta_{\rm ppE}$ and the power-law index $b_{\rm ppE}$. The latter is fixed for each successive post-Newtonian order $p$ by~\cite{Carson2020}
\begin{equation}
b_{{\rm ppE},p}=2p-5\;.
\label{eq:bppE}
\end{equation}
The former depends on the particular parametrization of the PPN metric in areal coordinates. For the metric~(\ref{eq:2PN_gtt}), application of Eq.~(11) of~\cite{Carson2020} for the $p-$th PN order gives
\begin{equation}
    \beta_{{\rm ppE},p}=(-1)^{p+1}\frac{5(p+1)(2p+1)}{8(2p-8)(2p-5)\eta^{2p/5}}\zeta_p\;.
    \label{eq:betappE}
\end{equation}
Applied to the 1PN and 2PN orders, we obtain
\begin{eqnarray}
\label{eq:betaPPEPN}
\beta_{\rm ppE,1}&=&\frac{5\zeta_1}{24\eta^{2/5}}=\frac{5(\beta_1-\gamma_1)}{24\eta^{2/5}}\nonumber\\
\beta_{\rm ppE,2}&=&-\frac{75\zeta_2}{32\eta^{4/5}}\nonumber\\
&=&-\frac{75}{128\eta^{4/5}} \left[2 \left(\gamma_1 ^2-1\right)-8 (\beta_1  \gamma_1 -1)\right.\nonumber\\
&&\qquad\qquad\left.+3 (\beta_2 -1)+3 (\gamma_2    -1)\right]\;,
\end{eqnarray}
Note that there is one important difference between the phase evolution during the inspiral phase of the ppE waveform~(\ref{eq:hppE}) and the \texttt{IMRPhenomPv2} model~(\ref{eq:Phemom}) that arises when terms at the 2.5PN and higher order are considered: the \texttt{IMRPhenomPv2} model involves  logarithmic correction terms in frequency, whereas the ppE waveforms do not. Nevertheless, the magnitudes of these logarithmic corrections are subdominant for the cases we consider here and can be neglected.

The amplitude terms are directly related to the phase terms via~\cite{Carson2020}
\begin{equation}
    a_{\rm ppE}=2p
\end{equation}
and
\begin{equation}
    \alpha_{\rm ppE}=\frac{(-1)^{p+1}(p+1)(2p-1)}{3}\frac{\zeta_p}{\eta^{2p/5}}\;.
\end{equation}

Using Eqs.~(\ref{eq:hppE})-(\ref{eq:betaPPEPN}), it is also straightforward to connect the fractional deviations $\delta\hat{\phi}_i$, inferred from the LIGO/Virgo measurements~\cite{GWTC2:GR} to the ppE coefficients and to the various PN parameters of the metric~(\ref{eq:2PN_gtt}). The result for the 1PN and 2PN orders is
\begin{equation}
\zeta_1\equiv \beta_1-\gamma_1=\frac{(743+924\eta)}{1344}\delta\hat{\phi}_2
\end{equation}
and
\begin{equation}
\zeta_2=-\frac{(3058673+5472432\eta+4353552\eta^2)}{10160640}
\delta\hat{\phi}_4\;.
\end{equation}

\begin{figure*}[t]
\includegraphics[width=0.75\textwidth]{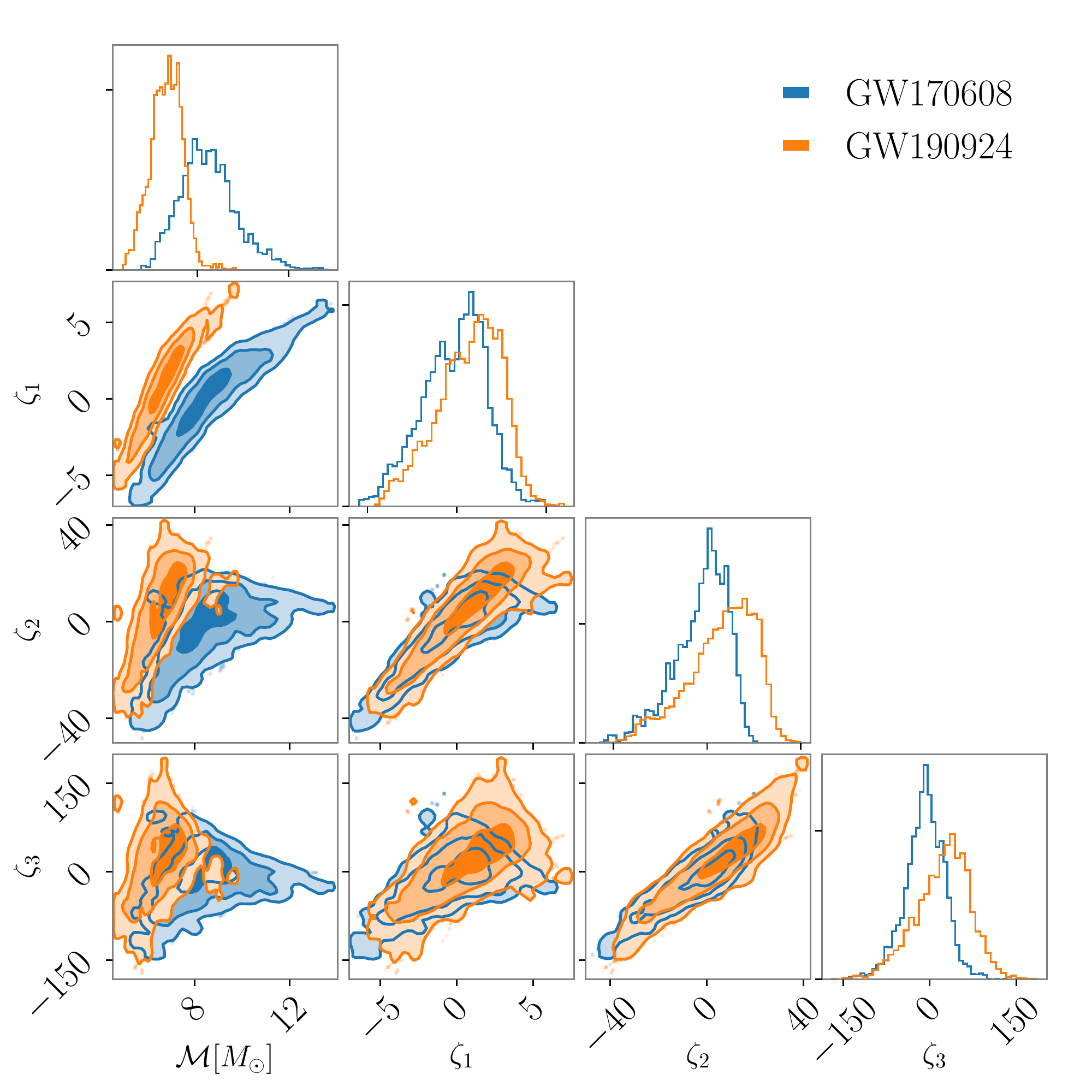}
\caption{\label{fig:LIGO_zeta}
Correlated uncertainties between the chirp mass $\cal{M}$ and the parameters $\zeta_p$ that describe deviations from the GR metrics at various post-Newtonian orders. These results were obtained for two relatively low-mass binary black hole mergers, GW170608 and GW190924, which have total masses of $\sim 19 M_{\odot}$ and $\sim 14 M_{\odot}$ respectively. There are tight mass-dependent correlations between the ppE parameters. The mass dependence of these correlations creates the wedge structure in the two-dimensional posterior distributions.
}
\end{figure*}

\begin{figure*}[t]
\includegraphics[width=0.75\textwidth]{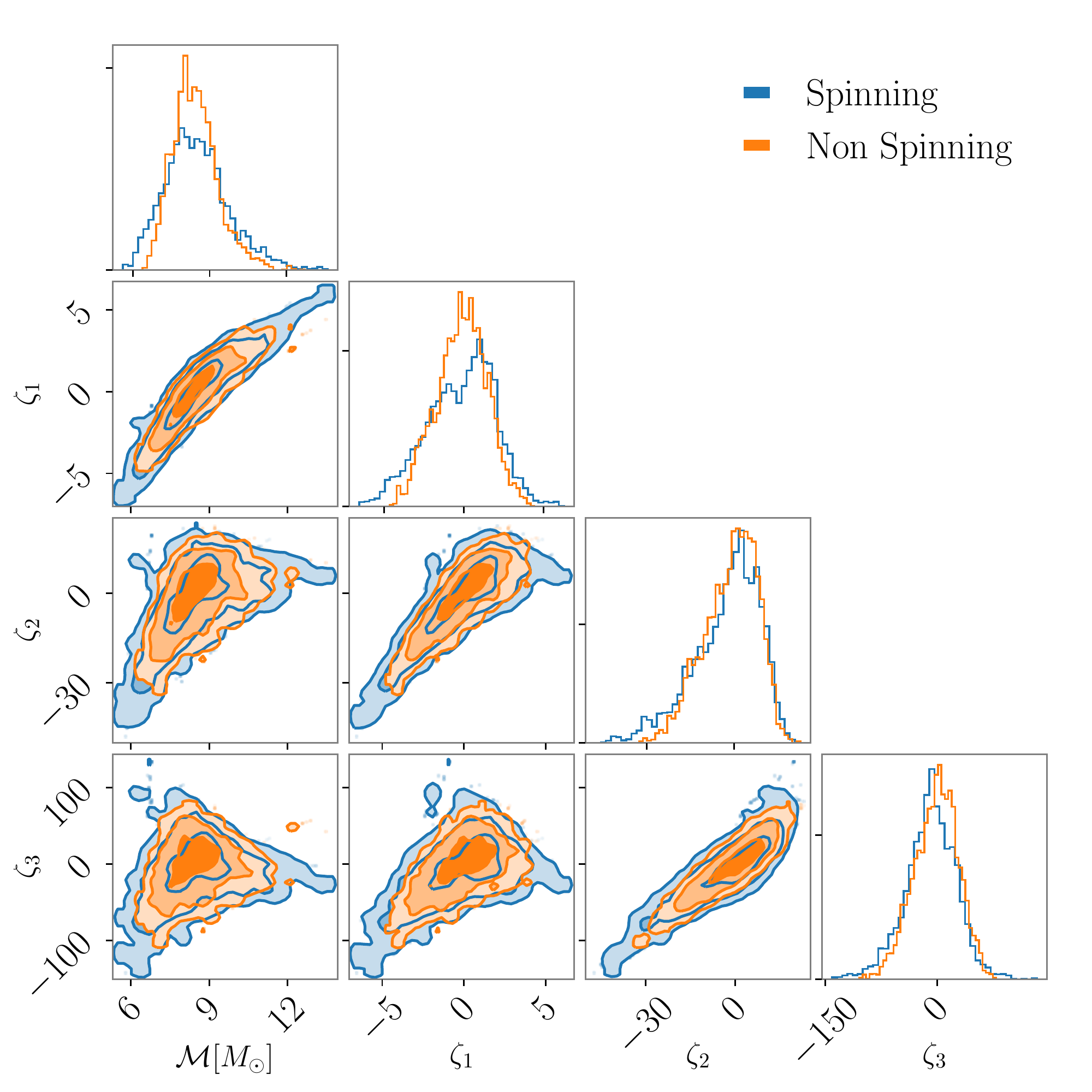}
\caption{\label{fig:LIGO_zeta_zero_spin}
Comparing correlated uncertainties between the chirp mass $\cal{M}$ and the parameters $\zeta_p$ that describe deviations from the GR metric at various post-Newtonian orders for GW170608, when including or excluding spin effects in the GR waveform.  Including spin effects in the GR waveform does not significantly change the posterior distributions for the non-GR parameters.
}
\end{figure*}

We use the waveform~(\ref{eq:hppE}) to explore the correlations in the expanded parameter space through the ensemble Markov chain Monte Carlo sampler \texttt{emcee}~\cite{ForemanMackey2013} as implemented in \texttt{Bilby}~\cite{Ashton2019}. In order to reduce the burn-in time, we choose our starting ensemble by taking samples from the posterior distributions obtained for the case where GR is assumed to be correct~\cite{RomeroShaw2020} and draw starting samples for the non-GR parameters from tightly peaked distributions around zero. We ensure convergence of the MCMC ensemble by visually inspecting the chains.

We employ the same prior distribution on the GR parameters as in~\cite{RomeroShaw2020} and impose uniform priors on the post-Newtonian terms $\zeta_p \in [-1000, 1000]$. We analyze four seconds of data for GW150914 and sixteen seconds of data for GW170608 and GW19092\_021846, with the trigger time placed two seconds before the end of the analyzed data. We neglect the effect of calibration uncertainty in our analysis as it has been shown to be negligible for these events~\cite{Payne2020}. We use noise power spectral density estimates obtained with the \texttt{BayesLine} algorithm~\cite{Abbott2019b, Littenberg2015}. In order to ensure that our constraints are driven solely by the inspiral regime, we impose upper frequency cutoffs of $120$~Hz, $120$~Hz, and $50$~Hz for GW170608, GW190924\_021846, and GW150914 respectively.  

Figure~\ref{fig:LIGO_zeta} shows the one- and two-dimensional marginalized posterior distributions for the $\zeta_p$ PN deviation parameters and the chirp mass for GW170608 and GW190924\_021846. (The results are not shown for GW150914 since too few cycles from the inspiral regime were observed and the deviation parameters are much less constrained.) All deviation parameters are highly correlated with each other and with the chirp mass of the system. Since gravitational-wave interferometers are more sensitive to the phase evolution of the source than to the amplitude, the posteriors here are not significantly more constraining than when we allowed for just the phase to deviate from the GR prediction.

In this analysis, although we allow for the black holes to have non-zero spins, we have neglected the effects of spins on the corrections to the GR waveforms. In order to explore the impact of this assumption on our results, we repeat the analysis while fixing the black hole spins to zero. In Figure~\ref{fig:LIGO_zeta_zero_spin}, we compare the one- and two-dimensional marginalized posterior distributions obtained for GW170608 when including and excluding spin effects in the GR waveform template. Due to degeneracies between black hole mass and spin in the evolution of the gravitational waveform, fixing the spins to zero leads to marginally tighter constraints on other parameters. This translates into slightly narrower credible intervals on the non-GR parameters.

We aim to develop a simple understanding of the correlations using the following arguments. The similarity of two waveforms $\tilde{h}_1(f)$ and $\tilde{h}_2(f)$ is expressed through their normalized match
\begin{equation}
\begin{split}
    \mathcal{O}(\tilde{h}_1, \tilde{h}_2) &= \frac{I_{12}}{\sqrt{I_{11} I_{22}}}\ , \\
     I_{12} &\equiv \Re \int df \, \frac{\tilde{h}_1(f) \tilde{h}^*_2(f)}{S_n(f)} ,
\end{split}
\end{equation}
where $S_n(f)$ is the frequency-dependent noise power spectral density of the interferometer. If the normalized match is close to unity, then the noise-weighted residuals from match-filtering waveform $\tilde{h}_1$ against $\tilde{h}_2$ are low.  In other words, if $\tilde{h}_1$ represents the actual signal in the data and $\tilde{h}_2$ the waveform model used for parameter estimation, the likelihood of observing the data given the model is high.

Assuming that the two waveforms differ only in the phase evolution, as is the dominant effect measured by the interferometers, and expanding the phases around a fiducial frequency $f_0$, we obtain
\begin{eqnarray}
    \tilde{h}_i(f)&=&A \exp\left[i \Phi_i(f)\right]\nonumber\\
    &\sim& A \exp\left[
    i \Phi_i(f_0)+ i \frac{d\Phi_i}{df}\Big|_{f_0}(f-f_0)+\right.\nonumber\\
    &&\qquad\qquad\left.\frac{i}{2}\frac{d^2\Phi_i}{df^2}\Big|_{f_0}(f-f_0)^2+...\right]
\end{eqnarray}
In the stationary phase approximation and, ignoring the overall phase of the integral, which can be incorporated into $\phi_c$ and is marginalized over during the analysis, the overlap integral becomes
\begin{eqnarray}
    I_{12}&=&\int \frac{df}{S_n(f)} \exp\left[i
    \left(\frac{d\Phi_1}{df}-\frac{d\Phi_2}{df}\right)(f-f_0)\right]\nonumber\\
    &+&\int \frac{df}{S_n(f)} \exp\left[\frac{i}{2}
    \left(\frac{d^2\Phi_1}{df^2}-\frac{d^2\Phi_2}{df^2}\right)(f-f_0)^2\right]+...\;.
\end{eqnarray}
The first of these two integrals amounts to a time offset (cf.~the term $2\pi f t_c$ in the waveform phase in Eq.~(\ref{eq:Phemom})) and is also marginalized over.  As a result, to leading order, the likelihood is maximized when the rapidly oscillating term inside the second square bracket is close to zero, i.e., when the second derivatives of the phase functions are similar at the frequencies of interest.

Requiring that the second derivative of the GR waveform at a frequency $f_0$ is equal to the second derivative of the waveform when the deviation parameters $\zeta_1$ and $\zeta_2$ are allowed to take non-zero values results in the anticorrelation 
\begin{eqnarray}\label{eq:correlation}
    \zeta_2&\sim& \frac{2}{5\pi^{2/3} (f_0 M)^{2/3}}\zeta_1\nonumber\\
    &\sim& 4.2 \left(\frac{f_0}{100~{\rm Hz}}\right)^{-2/3}
    \left(\frac{M}{19 M_\odot}\right)^{-2/3} \zeta_1\;,
\end{eqnarray}
where we have evaluated the various expressions for $\eta=1/4$ and chose $f_0=100$ Hz, roughly the frequency at which the LIGO and Virgo noise spectral density is minimal, as the fiducial frequency value. This captures the correlation seen in the results of the comparison of the model to the data (Fig.~\ref{fig:LIGO_zeta}). The mass dependence in this expression also accounts for the wedge shapes of the correlations between the various deviation parameters. A similar understanding of the correlations could be obtained with a traditional Fisher matrix analysis.

Surprisingly, for the masses of the two sources analyzed here, the direction of this correlation is nearly parallel to the one between the PN parameters of the metric, as inferred from the measurement of the size of a black-hole shadow (cf.~Eq.~[\ref{eq:EHTlimits}]). The slope of the correlation, however, depends on the mass of the system.  More massive systems spend too few cycles in the inspiral regime to be useful for probing post-Newtonian coefficients.  However, lower-mass binary black holes are particularly promising in breaking the degeneracy with EHT observations and providing a complementary test.  Binary neutron stars have even lower masses, but neutron stars may exhibit different couplings than black holes in alternate theories of gravity.  Alternatively, a change in the detector's typical sensitive frequency band, represented by $f_0$, would allow future third-generation ground-based detectors and particularly the LISA space instrument to provide constraints with very different correlations.

\section{Discussion}

In this paper, we explored the constraints imposed on potential deviations from the Kerr metric by the observation of the black-hole shadow in the M87 galaxy and the detection of gravitational waves during the inspiral phase of binary black-hole coalescence. There are a number of similarities and differences between these two types of tests of the Kerr metric. 

The shadow observations probe the equilibrium spacetimes of black holes whereas the detections of gravitational waves also probe the dynamics of the theory. For this reason, when comparing the two types of tests, we only consider the constraints placed by gravitational-wave observations on the metrics of the black holes. On the other hand, the shadow observations probe length and mass scales that are 8 orders of magnitude larger and curvature scales that are 16 orders of magnitude smaller than those measured by LIGO/Virgo data. It is therefore conceivable that the stellar-mass black holes probed by gravitational-wave observations and the supermassive black holes probed by shadow observations might not be described by the same metric. In that case, the constraints imposed by the two types of tests cannot be combined, but only offer complementary information.

\begin{figure}[t]
\includegraphics[width=\linewidth]{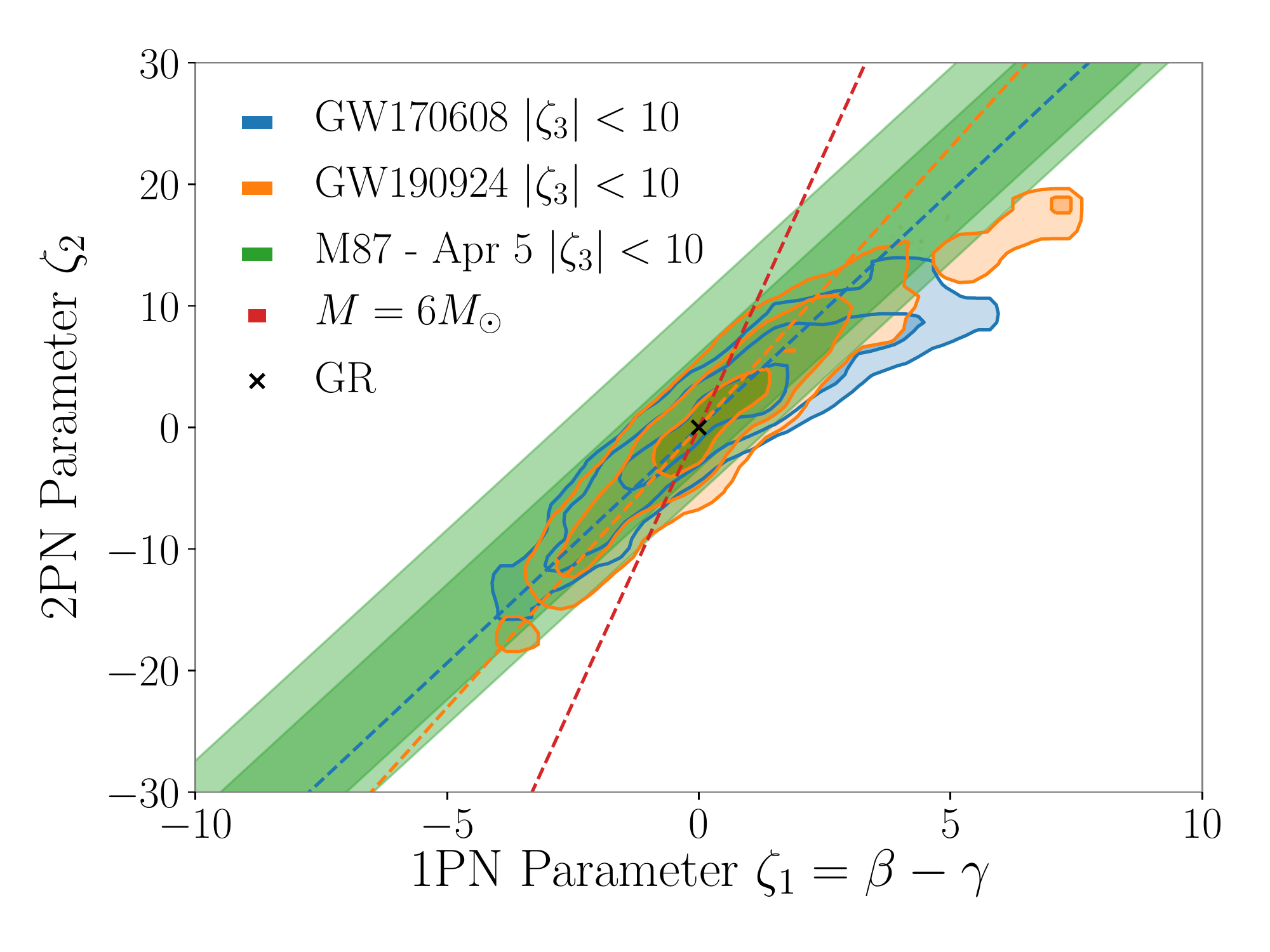}
\caption{\label{fig:LIGO_EHT_corr}
Comparison of the correlated posteriors over the two PN parameters $\zeta_1$ and $\zeta_2$ as inferred from the black-hole shadow test (green slanted contours) and the two coalescing events GW170608  (blue) and  GW190924\_021846 (orange). The blue and orange contours show the 1-, 2-, and 3-$\sigma$ credible regions. For both tests, we assume that the 3PN parameter $|\zeta_3| < 10$ and all higher-order PN parameters introduce negligible corrections. Because of a fortuitous coincidence, the inspiral tests for binary black holes with a total mass of $\sim 20 M_\odot$ and detectors with a peak sensitivity around $\sim 100$~Hz lead to correlated uncertainties that are parallel to those of the black-hole shadow tests. The dashed lines indicate the expected correlation from Eq.~(\ref{eq:correlation}). The red curve indicates the expected correlation for a binary black hole system with a low total mass of $6M_\odot$. The detection of such a binary system can assist in breaking the degeneracies between the deviation parameters.}
\end{figure}

There is limited information available from either type of observation on aspects of the spacetime that are controlled by the black-hole spins. For the case of the shadows, this is a consequence of a fortuitous near cancellation of the effects of frame dragging and of the quadrupole moment of the spacetimes, both of which depend on spin~\cite{Johannsen2010b}. For the gravitational-wave observations, there seems to be a paucity of merging binary black holes in the Universe with substantial spins~\cite{GWTC2,*GWTC2:pop}, although merger products have dimensionless spins of $\sim 0.7$ and their impact on the ringdown portion of the waveform has been explored \cite{GWTC2:GR}. Neglecting the effects of black-hole spins, as we have done here, substantially reduces the complexity of the problem.

\begin{figure}[t]
\includegraphics[width=\linewidth]{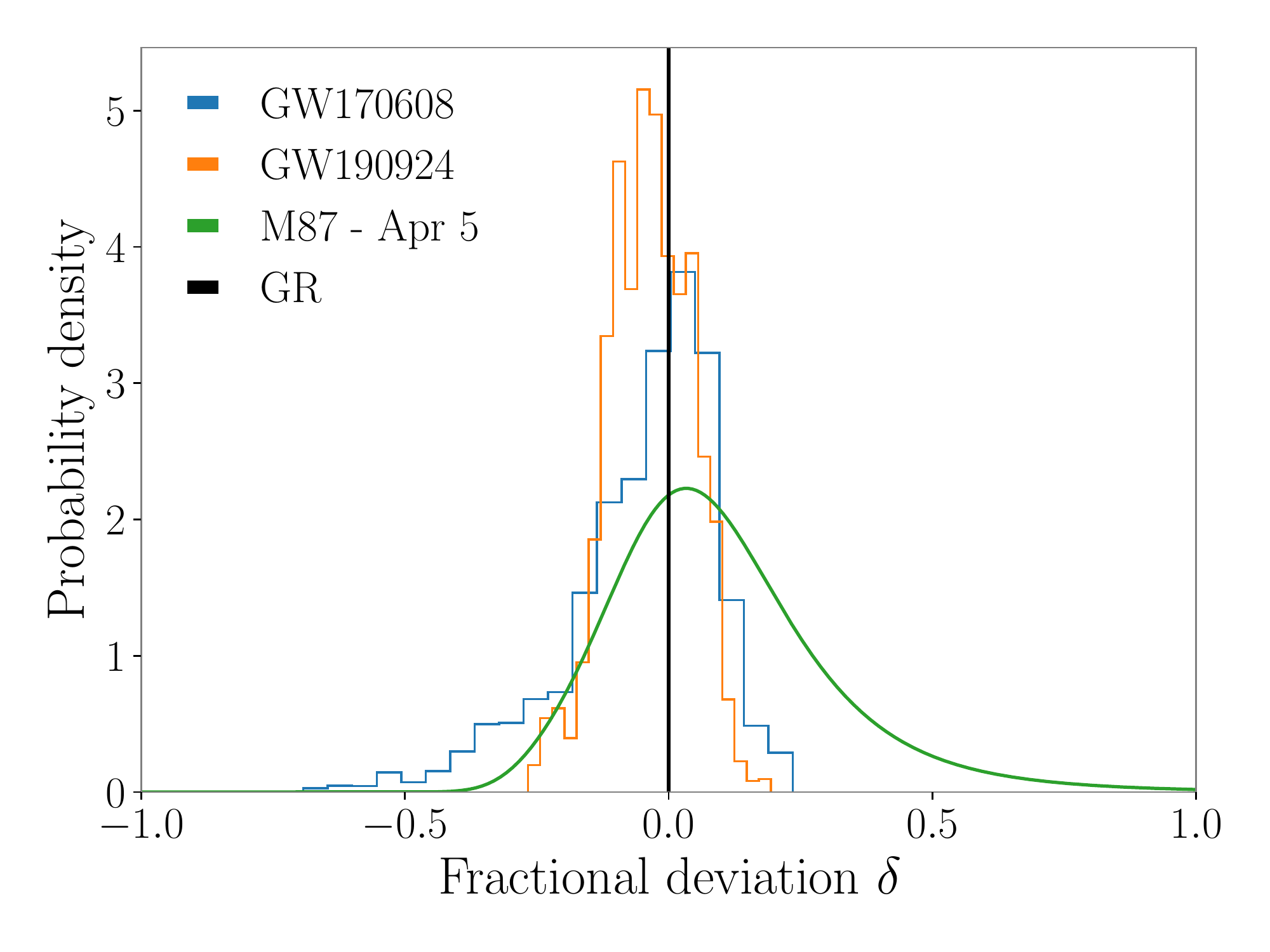}
\caption{\label{fig:LIGO_shadow}
Posterior distributions for the fractional deviation $\delta$ (Eq.~(\ref{eq:delta})) of the $tt-$component of the metric from the GR solution, as imposed by the EHT measurement of the size of the black-hole shadow in M87 (green), and the LIGO/Virgo measurements of the inspiral phases of events GW170608  (blue) and GW190924 (orange). Both EHT and LIGO/Virgo measurements are consistent with no deviations from GR and lead to bounds of comparable magnitude. 
}
\end{figure}

We have chosen here to express the constraints on possible deviations from the Kerr metric in terms of the coefficients of the parametric post-Newtonian expansions of the metric predictions for weak-field tests (even though for the case of the shadow tests we use metrics that are remain regular all the way to the horizons). This might appear to not be warranted by the fact that both tests probe strong gravitational fields, for which the post-Newtonian expansion converges slowly. Indeed, the radius of the photon orbit is at $\sim 3GM/c^2$, the radius of the shadow is $\sim 5 GM/c^2$, and the frequencies detected by LIGO/Virgo probe typical separations $6-30 GM/c^2$ during the inspiral phases. However, the relatively low accuracy of the observations allows us to construct such post-Newtonian expansions without significant concerns. For example, the size of the black-hole shadow in M87 has been measured to be consistent with the Kerr predictions to within an accuracy of $\sim 17\%$~\cite{PaperVI}. If naturalness forces all post-Newtonian terms to have coefficients of the same order of magnitude, then successive terms are a factor of $\sim 4-5$ smaller than the previous ones; at the 3PN order the correction is at the percent level, i.e., below the current observational uncertainty. The same argument can also be made for the post-Newtonian expansion of the gravitational waveforms and their observations, though these coefficients are known to increase at higher PN orders (see \S IVA).  The situation is, of course, starkly different in the case of Solar System tests only because the fractional observational uncertainties there are as small as $10^{-5}$ and therefore require an expansion that converges more rapidly (as the PPN expansion does)~\cite{Will2014}.

When we consider only the constraints from the gravitational-wave observations on the metrics of non-spinning black holes and not on the dynamics, both the shadow size tests and the gravitational-wave tests depend only on the $tt$-component of their metrics written in areal coordinates~\cite{Carson2020,Psaltis2020}. Because of another fortuitous coincidence, the correlations between the deviation parameters constrained from the shadow test are degenerate with the correlations constrained from the gravitational-wave tests, for binary black holes with a total mass of $\sim 20\ M_\odot$ and detectors with a peak sensitivity around $100$ Hz.  This similarity in the degeneracy between the two types of tests is shown explicitly in Fig.~\ref{fig:LIGO_EHT_corr} for the 1PN and 2PN deviation parameters as is the expected degeneracy~(\ref{eq:correlation}) that we derived in the previous section.

The similarity in the degeneracy between the various PN parameters in the two types of tests allows us to project both the EHT and the LIGO/Virgo correlated constraints on a single parameter, i.e., the fractional deviation $\delta$ defined in Eq.~(\ref{eq:delta}), which is directly related to the deviation of the shadow size from the GR prediction. The resulting posterior distributions for this parameter obtained for both the shadow-size and the inspiral tests are shown in Fig.~\ref{fig:LIGO_shadow}. The GR prediction is indicated by the solid black vertical line. This Figure shows that that current imaging and gravitational-wave data provide similar constraints on any deviations of slowly spinning black-hole spacetimes from the Kerr metric across the mass and curvature scales probed by the two experiments. We note that implicit in the analysis that led to Figure~\ref{fig:LIGO_shadow} is that the deviations at fourth and higher post-Newtonian orders are negligible as are deviations in the dynamics of the theory, which are described by the half-integer orders in the gravitational waveforms, and non-linear combinations of the $\zeta_i$. For this Figure, we additionally impose that $|\zeta_i| < 10$ for all non-GR correction terms; not including this constraint approximately doubles the width of the posteriors for the GW events.

A future detection of a low-mass binary black hole, as well as observations with future ground-based and space detectors that are sensitive to a different range of gravitational-wave frequencies, would provide constraints with degeneracies that are not parallel to those of the shadow test. This is shown explicitly in Fig.~\ref{fig:LIGO_EHT_corr}, where the expected line of degeneracy for a $3M_\odot+3M_\odot$ black-hole binary is shown. Under the assumption that the black-hole metrics do not depend on the mass or curvature of the objects involved, such an inspiral test can be combined with the shadow tests in order to break the degeneracies and constrain the individual deviation parameters. 

\bigskip

\begin{acknowledgments}
DP is grateful to F.\ \"Ozel for many discussions and for comments on the manuscript and thanks P.\ Christian, L.\ Medeiros, D.\ Heumann, L.\ Stein, and E.\ Berti for their input. CT and EP thank K.~Chatziioannou, M.~Isi, N.~Johnson-McDaniel, P.~Lasky, E.~Thrane, S.~Vitale, and A.~Weinstein for helpful comments and discussions. DP is supported in part by NSF PIRE award 1743747 and NSF AST-1715061. DP and IM acknowledge the hospitality of the Aspen Center of Physics, where initial discussions that led to this work took place in the summer of 2016.  EP and IM acknowledge support from the Australian Research Council Centre of Excellence for Gravitational Wave Discovery (OzGrav), through project number CE170100004.  IM is a recipient of the Australian Research Council Future Fellowship FT190100574. CT acknowledges support of the National Science Foundation and the LIGO Laboratory. 
LIGO was constructed by the California Institute of Technology and Massachusetts Institute of Technology with funding from the National Science Foundation and operates under cooperative agreement PHY-1764464. This research has made use of data, software and/or web tools obtained from the Gravitational Wave Open Science Center~\cite{Vallisneri2015, OpenData, GWTC1Data, GWTC2Data} (\href{https://www.gw-openscience.org}{https://www.gw-openscience.org}), a service of LIGO Laboratory, the LIGO Scientific Collaboration and the Virgo Collaboration. LIGO is funded by the U.S. National Science Foundation. Virgo is funded by the French Centre National de Recherche Scientifique (CNRS), the Italian Istituto Nazionale della Fisica Nucleare (INFN) and the Dutch Nikhef, with contributions by Polish and Hungarian institutes.
\end{acknowledgments}

\bibliography{grtests.bib}

\begin{thebibliography}{99}%
\makeatletter
\providecommand \@ifxundefined [1]{%
 \@ifx{#1\undefined}
}%
\providecommand \@ifnum [1]{%
 \ifnum #1\expandafter \@firstoftwo
 \else \expandafter \@secondoftwo
 \fi
}%
\providecommand \@ifx [1]{%
 \ifx #1\expandafter \@firstoftwo
 \else \expandafter \@secondoftwo
 \fi
}%
\providecommand \natexlab [1]{#1}%
\providecommand \enquote  [1]{``#1''}%
\providecommand \bibnamefont  [1]{#1}%
\providecommand \bibfnamefont [1]{#1}%
\providecommand \citenamefont [1]{#1}%
\providecommand \href@noop [0]{\@secondoftwo}%
\providecommand \href [0]{\begingroup \@sanitize@url \@href}%
\providecommand \@href[1]{\@@startlink{#1}\@@href}%
\providecommand \@@href[1]{\endgroup#1\@@endlink}%
\providecommand \@sanitize@url [0]{\catcode `\\12\catcode `\$12\catcode
  `\&12\catcode `\#12\catcode `\^12\catcode `\_12\catcode `\%12\relax}%
\providecommand \@@startlink[1]{}%
\providecommand \@@endlink[0]{}%
\providecommand \url  [0]{\begingroup\@sanitize@url \@url }%
\providecommand \@url [1]{\endgroup\@href {#1}{\urlprefix }}%
\providecommand \urlprefix  [0]{URL }%
\providecommand \Eprint [0]{\href }%
\providecommand \doibase [0]{https://doi.org/}%
\providecommand \selectlanguage [0]{\@gobble}%
\providecommand \bibinfo  [0]{\@secondoftwo}%
\providecommand \bibfield  [0]{\@secondoftwo}%
\providecommand \translation [1]{[#1]}%
\providecommand \BibitemOpen [0]{}%
\providecommand \bibitemStop [0]{}%
\providecommand \bibitemNoStop [0]{.\EOS\space}%
\providecommand \EOS [0]{\spacefactor3000\relax}%
\providecommand \BibitemShut  [1]{\csname bibitem#1\endcsname}%
\let\auto@bib@innerbib\@empty
\bibitem [{\citenamefont {{Will}}(1993)}]{Will1993}%
  \BibitemOpen
  \bibfield  {author} {\bibinfo {author} {\bibfnamefont {C.~M.}\ \bibnamefont
  {{Will}}},\ }\href@noop {} {\emph {\bibinfo {title} {{Theory and Experiment
  in Gravitational Physics}}}}\ (\bibinfo {year} {1993})\BibitemShut {NoStop}%
\bibitem [{\citenamefont {{Will}}(2014)}]{Will2014}%
  \BibitemOpen
  \bibfield  {author} {\bibinfo {author} {\bibfnamefont {C.~M.}\ \bibnamefont
  {{Will}}},\ }\href {https://doi.org/10.12942/lrr-2014-4} {\bibfield
  {journal} {\bibinfo  {journal} {Living Reviews in Relativity}\ }\textbf
  {\bibinfo {volume} {17}},\ \bibinfo {eid} {4} (\bibinfo {year}
  {2014})}\BibitemShut {NoStop}%
\bibitem [{\citenamefont {{Wex}}(2014)}]{Wex2014}%
  \BibitemOpen
  \bibfield  {author} {\bibinfo {author} {\bibfnamefont {N.}~\bibnamefont
  {{Wex}}},\ }\href@noop {} {\bibfield  {journal} {\bibinfo  {journal} {arXiv
  e-prints}\ ,\ \bibinfo {eid} {arXiv:1402.5594}} (\bibinfo {year}
  {2014})}\BibitemShut {NoStop}%
\bibitem [{\citenamefont {{Abbott}}\ \emph {et~al.}(2016)\citenamefont
  {{Abbott}}, \citenamefont {{et al}}, \citenamefont {{LIGO Scientific}},\ and\
  \citenamefont {{Virgo Collaborations}}}]{Abbott2016}%
  \BibitemOpen
  \bibfield  {author} {\bibinfo {author} {\bibfnamefont {B.~P.}\ \bibnamefont
  {{Abbott}}}, \bibinfo {author} {\bibnamefont {{et al}}}, \bibinfo {author}
  {\bibnamefont {{LIGO Scientific}}},\ and\ \bibinfo {author} {\bibnamefont
  {{Virgo Collaborations}}},\ }\href
  {https://doi.org/10.1103/PhysRevLett.116.221101} {\bibfield  {journal}
  {\bibinfo  {journal} {\prl}\ }\textbf {\bibinfo {volume} {116}},\ \bibinfo
  {eid} {221101} (\bibinfo {year} {2016})}\BibitemShut {NoStop}%
\bibitem [{\citenamefont {{Abbott}}\ \emph
  {et~al.}(2019{\natexlab{a}})\citenamefont {{Abbott}}, \citenamefont {{et
  al}}, \citenamefont {{LIGO Scientific}},\ and\ \citenamefont {{Virgo
  Collaborations}}}]{Abbott2019}%
  \BibitemOpen
  \bibfield  {author} {\bibinfo {author} {\bibfnamefont {B.~P.}\ \bibnamefont
  {{Abbott}}}, \bibinfo {author} {\bibnamefont {{et al}}}, \bibinfo {author}
  {\bibnamefont {{LIGO Scientific}}},\ and\ \bibinfo {author} {\bibnamefont
  {{Virgo Collaborations}}},\ }\href
  {https://doi.org/10.1103/PhysRevD.100.104036} {\bibfield  {journal} {\bibinfo
   {journal} {\prd}\ }\textbf {\bibinfo {volume} {100}},\ \bibinfo {eid}
  {104036} (\bibinfo {year} {2019}{\natexlab{a}})}\BibitemShut {NoStop}%
\bibitem [{\citenamefont {{Isi}}\ \emph {et~al.}(2019)\citenamefont {{Isi}},
  \citenamefont {{Giesler}}, \citenamefont {{Farr}}, \citenamefont {{Scheel}},\
  and\ \citenamefont {{Teukolsky}}}]{Ilsi2019}%
  \BibitemOpen
  \bibfield  {author} {\bibinfo {author} {\bibfnamefont {M.}~\bibnamefont
  {{Isi}}}, \bibinfo {author} {\bibfnamefont {M.}~\bibnamefont {{Giesler}}},
  \bibinfo {author} {\bibfnamefont {W.~M.}\ \bibnamefont {{Farr}}}, \bibinfo
  {author} {\bibfnamefont {M.~A.}\ \bibnamefont {{Scheel}}},\ and\ \bibinfo
  {author} {\bibfnamefont {S.~A.}\ \bibnamefont {{Teukolsky}}},\ }\href
  {https://doi.org/10.1103/PhysRevLett.123.111102} {\bibfield  {journal}
  {\bibinfo  {journal} {\prl}\ }\textbf {\bibinfo {volume} {123}},\ \bibinfo
  {eid} {111102} (\bibinfo {year} {2019})}\BibitemShut {NoStop}%
\bibitem [{\citenamefont {{Abbott}}\ \emph
  {et~al.}(2020{\natexlab{a}})\citenamefont {{Abbott}}, \citenamefont
  {{Abbott}}, \citenamefont {{Abraham}}, \citenamefont {{Acernese}} \emph
  {et~al.}}]{GWTC2:GR}%
  \BibitemOpen
  \bibfield  {author} {\bibinfo {author} {\bibfnamefont {R.}~\bibnamefont
  {{Abbott}}}, \bibinfo {author} {\bibfnamefont {T.~D.}\ \bibnamefont
  {{Abbott}}}, \bibinfo {author} {\bibfnamefont {S.}~\bibnamefont {{Abraham}}},
  \bibinfo {author} {\bibfnamefont {F.}~\bibnamefont {{Acernese}}}, \emph
  {et~al.},\ }\href@noop {} {\bibfield  {journal} {\bibinfo  {journal} {arXiv
  e-prints}\ ,\ \bibinfo {eid} {arXiv:2010.14529}} (\bibinfo {year}
  {2020}{\natexlab{a}})},\ \Eprint {https://arxiv.org/abs/2010.14529}
  {arXiv:2010.14529 [gr-qc]} \BibitemShut {NoStop}%
\bibitem [{\citenamefont {{Ferreira}}(2019)}]{Ferreira2019}%
  \BibitemOpen
  \bibfield  {author} {\bibinfo {author} {\bibfnamefont {P.~G.}\ \bibnamefont
  {{Ferreira}}},\ }\href {https://doi.org/10.1146/annurev-astro-091918-104423}
  {\bibfield  {journal} {\bibinfo  {journal} {\araa}\ }\textbf {\bibinfo
  {volume} {57}},\ \bibinfo {pages} {335} (\bibinfo {year} {2019})}\BibitemShut
  {NoStop}%
\bibitem [{\citenamefont {{Wagoner}}(1970)}]{Wagoner1970}%
  \BibitemOpen
  \bibfield  {author} {\bibinfo {author} {\bibfnamefont {R.~V.}\ \bibnamefont
  {{Wagoner}}},\ }\href {https://doi.org/10.1103/PhysRevD.1.3209} {\bibfield
  {journal} {\bibinfo  {journal} {\prd}\ }\textbf {\bibinfo {volume} {1}},\
  \bibinfo {pages} {3209} (\bibinfo {year} {1970})}\BibitemShut {NoStop}%
\bibitem [{\citenamefont {{Damour}}\ and\ \citenamefont
  {{Esposito-Farese}}(1993)}]{Damour1993}%
  \BibitemOpen
  \bibfield  {author} {\bibinfo {author} {\bibfnamefont {T.}~\bibnamefont
  {{Damour}}}\ and\ \bibinfo {author} {\bibfnamefont {G.}~\bibnamefont
  {{Esposito-Farese}}},\ }\href {https://doi.org/10.1103/PhysRevLett.70.2220}
  {\bibfield  {journal} {\bibinfo  {journal} {\prl}\ }\textbf {\bibinfo
  {volume} {70}},\ \bibinfo {pages} {2220} (\bibinfo {year}
  {1993})}\BibitemShut {NoStop}%
\bibitem [{\citenamefont {{Damour}}\ and\ \citenamefont
  {{Esposito-Far{\`e}se}}(1996)}]{Damour1996}%
  \BibitemOpen
  \bibfield  {author} {\bibinfo {author} {\bibfnamefont {T.}~\bibnamefont
  {{Damour}}}\ and\ \bibinfo {author} {\bibfnamefont {G.}~\bibnamefont
  {{Esposito-Far{\`e}se}}},\ }\href {https://doi.org/10.1103/PhysRevD.54.1474}
  {\bibfield  {journal} {\bibinfo  {journal} {\prd}\ }\textbf {\bibinfo
  {volume} {54}},\ \bibinfo {pages} {1474} (\bibinfo {year}
  {1996})}\BibitemShut {NoStop}%
\bibitem [{Note1()}]{Note1}%
  \BibitemOpen
  \bibinfo {note} {Here, $M$ denotes the mass scale of the test, $R$ the length
  scale, $G$ is the gravitational constant, and $c$ is the speed of
  light.}\BibitemShut {Stop}%
\bibitem [{\citenamefont {{Baker}}\ \emph {et~al.}(2015)\citenamefont
  {{Baker}}, \citenamefont {{Psaltis}},\ and\ \citenamefont
  {{Skordis}}}]{Baker2015}%
  \BibitemOpen
  \bibfield  {author} {\bibinfo {author} {\bibfnamefont {T.}~\bibnamefont
  {{Baker}}}, \bibinfo {author} {\bibfnamefont {D.}~\bibnamefont {{Psaltis}}},\
  and\ \bibinfo {author} {\bibfnamefont {C.}~\bibnamefont {{Skordis}}},\ }\href
  {https://doi.org/10.1088/0004-637X/802/1/63} {\bibfield  {journal} {\bibinfo
  {journal} {\apj}\ }\textbf {\bibinfo {volume} {802}},\ \bibinfo {eid} {63}
  (\bibinfo {year} {2015})}\BibitemShut {NoStop}%
\bibitem [{Note2()}]{Note2}%
  \BibitemOpen
  \bibinfo {note} {Using the gravitational-wave events as sirens, additional
  tests have been performed on the propagation of gravitational waves across
  the large distances from the sources to the Earth, in order to constrain the
  mass of the graviton~\cite {Abbott2016} or the speed of gravitational waves
  that is allowed to be different from the speed of light in some modified
  gravity theories inspired by cosmological observations~\cite
  {Baker2017}}\BibitemShut {NoStop}%
\bibitem [{\citenamefont {{Gravity Collaboration}}(2019)}]{Gravity2019}%
  \BibitemOpen
  \bibfield  {author} {\bibinfo {author} {\bibnamefont {{Gravity
  Collaboration}}},\ }\href {https://doi.org/10.1103/PhysRevLett.122.101102}
  {\bibfield  {journal} {\bibinfo  {journal} {\prl}\ }\textbf {\bibinfo
  {volume} {122}},\ \bibinfo {eid} {101102} (\bibinfo {year}
  {2019})}\BibitemShut {NoStop}%
\bibitem [{\citenamefont {{Do}}\ and\ \citenamefont {{et al.}}(2019)}]{Do2019}%
  \BibitemOpen
  \bibfield  {author} {\bibinfo {author} {\bibfnamefont {T.}~\bibnamefont
  {{Do}}}\ and\ \bibinfo {author} {\bibnamefont {{et al.}}},\ }\href
  {https://doi.org/10.1126/science.aav8137} {\bibfield  {journal} {\bibinfo
  {journal} {Science}\ }\textbf {\bibinfo {volume} {365}},\ \bibinfo {pages}
  {664} (\bibinfo {year} {2019})}\BibitemShut {NoStop}%
\bibitem [{\citenamefont {{Gravity Collaboration}}(2020)}]{Gravity2020}%
  \BibitemOpen
  \bibfield  {author} {\bibinfo {author} {\bibnamefont {{Gravity
  Collaboration}}},\ }\href {https://doi.org/10.1051/0004-6361/202037813}
  {\bibfield  {journal} {\bibinfo  {journal} {\aap}\ }\textbf {\bibinfo
  {volume} {636}},\ \bibinfo {eid} {L5} (\bibinfo {year} {2020})}\BibitemShut
  {NoStop}%
\bibitem [{\citenamefont {{Hees}}\ \emph {et~al.}(2020)\citenamefont {{Hees}},
  \citenamefont {{Do}}, \citenamefont {{Roberts}}, \citenamefont {{Ghez}},
  \citenamefont {{Nishiyama}}, \citenamefont {{Bentley}}, \citenamefont
  {{Gautam}}, \citenamefont {{Jia}}, \citenamefont {{Kara}}, \citenamefont
  {{Lu}}, \citenamefont {{Saida}}, \citenamefont {{Sakai}}, \citenamefont
  {{Takahashi}},\ and\ \citenamefont {{Takamori}}}]{Hees2020}%
  \BibitemOpen
  \bibfield  {author} {\bibinfo {author} {\bibfnamefont {A.}~\bibnamefont
  {{Hees}}}, \bibinfo {author} {\bibfnamefont {T.}~\bibnamefont {{Do}}},
  \bibinfo {author} {\bibfnamefont {B.~M.}\ \bibnamefont {{Roberts}}}, \bibinfo
  {author} {\bibfnamefont {A.~M.}\ \bibnamefont {{Ghez}}}, \bibinfo {author}
  {\bibfnamefont {S.}~\bibnamefont {{Nishiyama}}}, \bibinfo {author}
  {\bibfnamefont {R.~O.}\ \bibnamefont {{Bentley}}}, \bibinfo {author}
  {\bibfnamefont {A.~K.}\ \bibnamefont {{Gautam}}}, \bibinfo {author}
  {\bibfnamefont {S.}~\bibnamefont {{Jia}}}, \bibinfo {author} {\bibfnamefont
  {T.}~\bibnamefont {{Kara}}}, \bibinfo {author} {\bibfnamefont {J.~R.}\
  \bibnamefont {{Lu}}}, \bibinfo {author} {\bibfnamefont {H.}~\bibnamefont
  {{Saida}}}, \bibinfo {author} {\bibfnamefont {S.}~\bibnamefont {{Sakai}}},
  \bibinfo {author} {\bibfnamefont {M.}~\bibnamefont {{Takahashi}}},\ and\
  \bibinfo {author} {\bibfnamefont {Y.}~\bibnamefont {{Takamori}}},\ }\href
  {https://doi.org/10.1103/PhysRevLett.124.081101} {\bibfield  {journal}
  {\bibinfo  {journal} {\prl}\ }\textbf {\bibinfo {volume} {124}},\ \bibinfo
  {eid} {081101} (\bibinfo {year} {2020})}\BibitemShut {NoStop}%
\bibitem [{\citenamefont {{Hees}}\ \emph {et~al.}(2017)\citenamefont {{Hees}},
  \citenamefont {{Do}}, \citenamefont {{Ghez}}, \citenamefont {{Martinez}},
  \citenamefont {{Naoz}}, \citenamefont {{Becklin}}, \citenamefont {{Boehle}},
  \citenamefont {{Chappell}}, \citenamefont {{Chu}}, \citenamefont
  {{Dehghanfar}}, \citenamefont {{Kosmo}}, \citenamefont {{Lu}}, \citenamefont
  {{Matthews}}, \citenamefont {{Morris}}, \citenamefont {{Sakai}},
  \citenamefont {{Sch{\"o}del}},\ and\ \citenamefont {{Witzel}}}]{Hees2017}%
  \BibitemOpen
  \bibfield  {author} {\bibinfo {author} {\bibfnamefont {A.}~\bibnamefont
  {{Hees}}}, \bibinfo {author} {\bibfnamefont {T.}~\bibnamefont {{Do}}},
  \bibinfo {author} {\bibfnamefont {A.~M.}\ \bibnamefont {{Ghez}}}, \bibinfo
  {author} {\bibfnamefont {G.~D.}\ \bibnamefont {{Martinez}}}, \bibinfo
  {author} {\bibfnamefont {S.}~\bibnamefont {{Naoz}}}, \bibinfo {author}
  {\bibfnamefont {E.~E.}\ \bibnamefont {{Becklin}}}, \bibinfo {author}
  {\bibfnamefont {A.}~\bibnamefont {{Boehle}}}, \bibinfo {author}
  {\bibfnamefont {S.}~\bibnamefont {{Chappell}}}, \bibinfo {author}
  {\bibfnamefont {D.}~\bibnamefont {{Chu}}}, \bibinfo {author} {\bibfnamefont
  {A.}~\bibnamefont {{Dehghanfar}}}, \bibinfo {author} {\bibfnamefont
  {K.}~\bibnamefont {{Kosmo}}}, \bibinfo {author} {\bibfnamefont {J.~R.}\
  \bibnamefont {{Lu}}}, \bibinfo {author} {\bibfnamefont {K.}~\bibnamefont
  {{Matthews}}}, \bibinfo {author} {\bibfnamefont {M.~R.}\ \bibnamefont
  {{Morris}}}, \bibinfo {author} {\bibfnamefont {S.}~\bibnamefont {{Sakai}}},
  \bibinfo {author} {\bibfnamefont {R.}~\bibnamefont {{Sch{\"o}del}}},\ and\
  \bibinfo {author} {\bibfnamefont {G.}~\bibnamefont {{Witzel}}},\ }\href
  {https://doi.org/10.1103/PhysRevLett.118.211101} {\bibfield  {journal}
  {\bibinfo  {journal} {\prl}\ }\textbf {\bibinfo {volume} {118}},\ \bibinfo
  {eid} {211101} (\bibinfo {year} {2017})}\BibitemShut {NoStop}%
\bibitem [{\citenamefont {{Event Horizon Telescope
  Collaboration}}(2019{\natexlab{a}})}]{PaperI}%
  \BibitemOpen
  \bibfield  {author} {\bibinfo {author} {\bibnamefont {{Event Horizon
  Telescope Collaboration}}},\ }\href
  {https://doi.org/10.3847/2041-8213/ab0ec7} {\bibfield  {journal} {\bibinfo
  {journal} {\apjl}\ }\textbf {\bibinfo {volume} {875}},\ \bibinfo {eid} {L1}
  (\bibinfo {year} {2019}{\natexlab{a}})}\BibitemShut {NoStop}%
\bibitem [{\citenamefont {{Event Horizon Telescope
  Collaboration}}(2019{\natexlab{b}})}]{PaperVI}%
  \BibitemOpen
  \bibfield  {author} {\bibinfo {author} {\bibnamefont {{Event Horizon
  Telescope Collaboration}}},\ }\href
  {https://doi.org/10.3847/2041-8213/ab1141} {\bibfield  {journal} {\bibinfo
  {journal} {\apjl}\ }\textbf {\bibinfo {volume} {875}},\ \bibinfo {eid} {L6}
  (\bibinfo {year} {2019}{\natexlab{b}})}\BibitemShut {NoStop}%
\bibitem [{\citenamefont {{Psaltis}}\ \emph {et~al.}(2020)\citenamefont
  {{Psaltis}}, \citenamefont {{et al}},\ and\ \citenamefont {{the EHT
  Collaboration}}}]{Psaltis2020}%
  \BibitemOpen
  \bibfield  {author} {\bibinfo {author} {\bibfnamefont {D.}~\bibnamefont
  {{Psaltis}}}, \bibinfo {author} {\bibnamefont {{et al}}},\ and\ \bibinfo
  {author} {\bibnamefont {{the EHT Collaboration}}},\ }\href@noop {} {\bibfield
   {journal} {\bibinfo  {journal} {\prl}\ } (\bibinfo {year}
  {2020})}\BibitemShut {NoStop}%
\bibitem [{\citenamefont {{Johannsen}}\ and\ \citenamefont
  {{Psaltis}}(2011)}]{Johannsen2011}%
  \BibitemOpen
  \bibfield  {author} {\bibinfo {author} {\bibfnamefont {T.}~\bibnamefont
  {{Johannsen}}}\ and\ \bibinfo {author} {\bibfnamefont {D.}~\bibnamefont
  {{Psaltis}}},\ }\href {https://doi.org/10.1103/PhysRevD.83.124015} {\bibfield
   {journal} {\bibinfo  {journal} {\prd}\ }\textbf {\bibinfo {volume} {83}},\
  \bibinfo {eid} {124015} (\bibinfo {year} {2011})}\BibitemShut {NoStop}%
\bibitem [{\citenamefont {{Vigeland}}\ \emph {et~al.}(2011)\citenamefont
  {{Vigeland}}, \citenamefont {{Yunes}},\ and\ \citenamefont
  {{Stein}}}]{Vigeland2011}%
  \BibitemOpen
  \bibfield  {author} {\bibinfo {author} {\bibfnamefont {S.}~\bibnamefont
  {{Vigeland}}}, \bibinfo {author} {\bibfnamefont {N.}~\bibnamefont
  {{Yunes}}},\ and\ \bibinfo {author} {\bibfnamefont {L.~C.}\ \bibnamefont
  {{Stein}}},\ }\href {https://doi.org/10.1103/PhysRevD.83.104027} {\bibfield
  {journal} {\bibinfo  {journal} {\prd}\ }\textbf {\bibinfo {volume} {83}},\
  \bibinfo {eid} {104027} (\bibinfo {year} {2011})}\BibitemShut {NoStop}%
\bibitem [{\citenamefont {{Johannsen}}(2013{\natexlab{a}})}]{Johannsen2013}%
  \BibitemOpen
  \bibfield  {author} {\bibinfo {author} {\bibfnamefont {T.}~\bibnamefont
  {{Johannsen}}},\ }\href {https://doi.org/10.1103/PhysRevD.87.124017}
  {\bibfield  {journal} {\bibinfo  {journal} {\prd}\ }\textbf {\bibinfo
  {volume} {87}},\ \bibinfo {eid} {124017} (\bibinfo {year}
  {2013}{\natexlab{a}})}\BibitemShut {NoStop}%
\bibitem [{\citenamefont {{Johannsen}}(2013{\natexlab{b}})}]{Johannsen2013b}%
  \BibitemOpen
  \bibfield  {author} {\bibinfo {author} {\bibfnamefont {T.}~\bibnamefont
  {{Johannsen}}},\ }\href {https://doi.org/10.1103/PhysRevD.88.044002}
  {\bibfield  {journal} {\bibinfo  {journal} {\prd}\ }\textbf {\bibinfo
  {volume} {88}},\ \bibinfo {eid} {044002} (\bibinfo {year}
  {2013}{\natexlab{b}})}\BibitemShut {NoStop}%
\bibitem [{\citenamefont {{Johannsen}}\ and\ \citenamefont
  {{Psaltis}}(2010{\natexlab{a}})}]{Johannsen2010b}%
  \BibitemOpen
  \bibfield  {author} {\bibinfo {author} {\bibfnamefont {T.}~\bibnamefont
  {{Johannsen}}}\ and\ \bibinfo {author} {\bibfnamefont {D.}~\bibnamefont
  {{Psaltis}}},\ }\href {https://doi.org/10.1088/0004-637X/718/1/446}
  {\bibfield  {journal} {\bibinfo  {journal} {\apj}\ }\textbf {\bibinfo
  {volume} {718}},\ \bibinfo {pages} {446} (\bibinfo {year}
  {2010}{\natexlab{a}})}\BibitemShut {NoStop}%
\bibitem [{\citenamefont {{Johannsen}}(2013{\natexlab{c}})}]{Johannsen2013c}%
  \BibitemOpen
  \bibfield  {author} {\bibinfo {author} {\bibfnamefont {T.}~\bibnamefont
  {{Johannsen}}},\ }\href {https://doi.org/10.1088/0004-637X/777/2/170}
  {\bibfield  {journal} {\bibinfo  {journal} {\apj}\ }\textbf {\bibinfo
  {volume} {777}},\ \bibinfo {eid} {170} (\bibinfo {year}
  {2013}{\natexlab{c}})}\BibitemShut {NoStop}%
\bibitem [{\citenamefont {{Medeiros}}\ \emph {et~al.}(2019)\citenamefont
  {{Medeiros}}, \citenamefont {{Psaltis}},\ and\ \citenamefont
  {{{\"O}zel}}}]{Medeiros2020}%
  \BibitemOpen
  \bibfield  {author} {\bibinfo {author} {\bibfnamefont {L.}~\bibnamefont
  {{Medeiros}}}, \bibinfo {author} {\bibfnamefont {D.}~\bibnamefont
  {{Psaltis}}},\ and\ \bibinfo {author} {\bibfnamefont {F.}~\bibnamefont
  {{{\"O}zel}}},\ }\href@noop {} {\bibfield  {journal} {\bibinfo  {journal}
  {arXiv e-prints}\ ,\ \bibinfo {pages} {arXiv:1907.12575}} (\bibinfo {year}
  {2019})}\BibitemShut {NoStop}%
\bibitem [{\citenamefont {{Khan}}\ \emph {et~al.}(2016)\citenamefont {{Khan}},
  \citenamefont {{Husa}}, \citenamefont {{Hannam}}, \citenamefont {{Ohme}},
  \citenamefont {{P{\"u}rrer}}, \citenamefont {{Forteza}},\ and\ \citenamefont
  {{Boh{\'e}}}}]{Khan2016}%
  \BibitemOpen
  \bibfield  {author} {\bibinfo {author} {\bibfnamefont {S.}~\bibnamefont
  {{Khan}}}, \bibinfo {author} {\bibfnamefont {S.}~\bibnamefont {{Husa}}},
  \bibinfo {author} {\bibfnamefont {M.}~\bibnamefont {{Hannam}}}, \bibinfo
  {author} {\bibfnamefont {F.}~\bibnamefont {{Ohme}}}, \bibinfo {author}
  {\bibfnamefont {M.}~\bibnamefont {{P{\"u}rrer}}}, \bibinfo {author}
  {\bibfnamefont {X.~J.}\ \bibnamefont {{Forteza}}},\ and\ \bibinfo {author}
  {\bibfnamefont {A.}~\bibnamefont {{Boh{\'e}}}},\ }\href
  {https://doi.org/10.1103/PhysRevD.93.044007} {\bibfield  {journal} {\bibinfo
  {journal} {\prd}\ }\textbf {\bibinfo {volume} {93}},\ \bibinfo {eid} {044007}
  (\bibinfo {year} {2016})}\BibitemShut {NoStop}%
\bibitem [{\citenamefont {{Abbott}}\ \emph
  {et~al.}(2020{\natexlab{b}})\citenamefont {{Abbott}}, \citenamefont
  {{Abbott}}, \citenamefont {{Abraham}}, \citenamefont {{Acernese}} \emph
  {et~al.}}]{GWTC2}%
  \BibitemOpen
  \bibfield  {author} {\bibinfo {author} {\bibfnamefont {R.}~\bibnamefont
  {{Abbott}}}, \bibinfo {author} {\bibfnamefont {T.~D.}\ \bibnamefont
  {{Abbott}}}, \bibinfo {author} {\bibfnamefont {S.}~\bibnamefont {{Abraham}}},
  \bibinfo {author} {\bibfnamefont {F.}~\bibnamefont {{Acernese}}}, \emph
  {et~al.},\ }\href@noop {} {\bibfield  {journal} {\bibinfo  {journal} {arXiv
  e-prints}\ ,\ \bibinfo {eid} {arXiv:2010.14527}} (\bibinfo {year}
  {2020}{\natexlab{b}})},\ \Eprint {https://arxiv.org/abs/2010.14527}
  {arXiv:2010.14527 [gr-qc]} \BibitemShut {NoStop}%
\bibitem [{\citenamefont {{Abbott}}\ \emph
  {et~al.}(2020{\natexlab{c}})\citenamefont {{Abbott}}, \citenamefont
  {{Abbott}}, \citenamefont {{Abraham}}, \citenamefont {{Acernese}} \emph
  {et~al.}}]{GWTC2:pop}%
  \BibitemOpen
  \bibfield  {author} {\bibinfo {author} {\bibfnamefont {R.}~\bibnamefont
  {{Abbott}}}, \bibinfo {author} {\bibfnamefont {T.~D.}\ \bibnamefont
  {{Abbott}}}, \bibinfo {author} {\bibfnamefont {S.}~\bibnamefont {{Abraham}}},
  \bibinfo {author} {\bibfnamefont {F.}~\bibnamefont {{Acernese}}}, \emph
  {et~al.},\ }\href@noop {} {\bibfield  {journal} {\bibinfo  {journal} {arXiv
  e-prints}\ ,\ \bibinfo {eid} {arXiv:2010.14533}} (\bibinfo {year}
  {2020}{\natexlab{c}})},\ \Eprint {https://arxiv.org/abs/2010.14533}
  {arXiv:2010.14533 [astro-ph.HE]} \BibitemShut {NoStop}%
\bibitem [{\citenamefont {{Joyce}}\ \emph {et~al.}(2015)\citenamefont
  {{Joyce}}, \citenamefont {{Jain}}, \citenamefont {{Khoury}},\ and\
  \citenamefont {{Trodden}}}]{Joyce2015}%
  \BibitemOpen
  \bibfield  {author} {\bibinfo {author} {\bibfnamefont {A.}~\bibnamefont
  {{Joyce}}}, \bibinfo {author} {\bibfnamefont {B.}~\bibnamefont {{Jain}}},
  \bibinfo {author} {\bibfnamefont {J.}~\bibnamefont {{Khoury}}},\ and\
  \bibinfo {author} {\bibfnamefont {M.}~\bibnamefont {{Trodden}}},\ }\href
  {https://doi.org/10.1016/j.physrep.2014.12.002} {\bibfield  {journal}
  {\bibinfo  {journal} {Phys.\ Rep.}\ }\textbf {\bibinfo {volume} {568}},\
  \bibinfo {pages} {1} (\bibinfo {year} {2015})}\BibitemShut {NoStop}%
\bibitem [{\citenamefont {{Burrage}}\ and\ \citenamefont
  {{Sakstein}}(2018)}]{Burrage2018}%
  \BibitemOpen
  \bibfield  {author} {\bibinfo {author} {\bibfnamefont {C.}~\bibnamefont
  {{Burrage}}}\ and\ \bibinfo {author} {\bibfnamefont {J.}~\bibnamefont
  {{Sakstein}}},\ }\href {https://doi.org/10.1007/s41114-018-0011-x} {\bibfield
   {journal} {\bibinfo  {journal} {Living Reviews in Relativity}\ }\textbf
  {\bibinfo {volume} {21}},\ \bibinfo {eid} {1} (\bibinfo {year}
  {2018})}\BibitemShut {NoStop}%
\bibitem [{\citenamefont {{Thorne}}\ and\ \citenamefont
  {{Dykla}}(1971)}]{Thorne1971}%
  \BibitemOpen
  \bibfield  {author} {\bibinfo {author} {\bibfnamefont {K.~S.}\ \bibnamefont
  {{Thorne}}}\ and\ \bibinfo {author} {\bibfnamefont {J.~J.}\ \bibnamefont
  {{Dykla}}},\ }\href {https://doi.org/10.1086/180734} {\bibfield  {journal}
  {\bibinfo  {journal} {\apjl}\ }\textbf {\bibinfo {volume} {166}},\ \bibinfo
  {pages} {L35} (\bibinfo {year} {1971})}\BibitemShut {NoStop}%
\bibitem [{\citenamefont {{Scheel}}\ \emph {et~al.}(1995)\citenamefont
  {{Scheel}}, \citenamefont {{Shapiro}},\ and\ \citenamefont
  {{Teukolsky}}}]{Scheel1995}%
  \BibitemOpen
  \bibfield  {author} {\bibinfo {author} {\bibfnamefont {M.~A.}\ \bibnamefont
  {{Scheel}}}, \bibinfo {author} {\bibfnamefont {S.~L.}\ \bibnamefont
  {{Shapiro}}},\ and\ \bibinfo {author} {\bibfnamefont {S.~A.}\ \bibnamefont
  {{Teukolsky}}},\ }\href {https://doi.org/10.1103/PhysRevD.51.4236} {\bibfield
   {journal} {\bibinfo  {journal} {\prd}\ }\textbf {\bibinfo {volume} {51}},\
  \bibinfo {pages} {4236} (\bibinfo {year} {1995})}\BibitemShut {NoStop}%
\bibitem [{\citenamefont {{Psaltis}}\ \emph {et~al.}(2008)\citenamefont
  {{Psaltis}}, \citenamefont {{Perrodin}}, \citenamefont {{Dienes}},\ and\
  \citenamefont {{Mocioiu}}}]{Psaltis2008b}%
  \BibitemOpen
  \bibfield  {author} {\bibinfo {author} {\bibfnamefont {D.}~\bibnamefont
  {{Psaltis}}}, \bibinfo {author} {\bibfnamefont {D.}~\bibnamefont
  {{Perrodin}}}, \bibinfo {author} {\bibfnamefont {K.~R.}\ \bibnamefont
  {{Dienes}}},\ and\ \bibinfo {author} {\bibfnamefont {I.}~\bibnamefont
  {{Mocioiu}}},\ }\href {https://doi.org/10.1103/PhysRevLett.100.091101}
  {\bibfield  {journal} {\bibinfo  {journal} {\prl}\ }\textbf {\bibinfo
  {volume} {100}},\ \bibinfo {eid} {091101} (\bibinfo {year}
  {2008})}\BibitemShut {NoStop}%
\bibitem [{\citenamefont {{Sotiriou}}\ and\ \citenamefont
  {{Faraoni}}(2012)}]{Sotiriou2012}%
  \BibitemOpen
  \bibfield  {author} {\bibinfo {author} {\bibfnamefont {T.~P.}\ \bibnamefont
  {{Sotiriou}}}\ and\ \bibinfo {author} {\bibfnamefont {V.}~\bibnamefont
  {{Faraoni}}},\ }\href {https://doi.org/10.1103/PhysRevLett.108.081103}
  {\bibfield  {journal} {\bibinfo  {journal} {\prl}\ }\textbf {\bibinfo
  {volume} {108}},\ \bibinfo {eid} {081103} (\bibinfo {year}
  {2012})}\BibitemShut {NoStop}%
\bibitem [{\citenamefont {{Carson}}\ and\ \citenamefont
  {{Yagi}}(2020{\natexlab{a}})}]{Carson2020}%
  \BibitemOpen
  \bibfield  {author} {\bibinfo {author} {\bibfnamefont {Z.}~\bibnamefont
  {{Carson}}}\ and\ \bibinfo {author} {\bibfnamefont {K.}~\bibnamefont
  {{Yagi}}},\ }\href {https://doi.org/10.1103/PhysRevD.101.084050} {\bibfield
  {journal} {\bibinfo  {journal} {\prd}\ }\textbf {\bibinfo {volume} {101}},\
  \bibinfo {eid} {084050} (\bibinfo {year} {2020}{\natexlab{a}})},\ \Eprint
  {https://arxiv.org/abs/2003.02374} {arXiv:2003.02374 [gr-qc]} \BibitemShut
  {NoStop}%
\bibitem [{\citenamefont {{Will}}(2011)}]{Will2011}%
  \BibitemOpen
  \bibfield  {author} {\bibinfo {author} {\bibfnamefont {C.~M.}\ \bibnamefont
  {{Will}}},\ }\href {https://doi.org/10.1073/pnas.1103127108} {\bibfield
  {journal} {\bibinfo  {journal} {Proceedings of the National Academy of
  Science}\ }\textbf {\bibinfo {volume} {108}},\ \bibinfo {pages} {5938}
  (\bibinfo {year} {2011})},\ \Eprint {https://arxiv.org/abs/1102.5192}
  {arXiv:1102.5192 [gr-qc]} \BibitemShut {NoStop}%
\bibitem [{\citenamefont {{Kanti}}\ \emph {et~al.}(1996)\citenamefont
  {{Kanti}}, \citenamefont {{Mavromatos}}, \citenamefont {{Rizos}},
  \citenamefont {{Tamvakis}},\ and\ \citenamefont {{Winstanley}}}]{Kanti1996}%
  \BibitemOpen
  \bibfield  {author} {\bibinfo {author} {\bibfnamefont {P.}~\bibnamefont
  {{Kanti}}}, \bibinfo {author} {\bibfnamefont {N.~E.}\ \bibnamefont
  {{Mavromatos}}}, \bibinfo {author} {\bibfnamefont {J.}~\bibnamefont
  {{Rizos}}}, \bibinfo {author} {\bibfnamefont {K.}~\bibnamefont
  {{Tamvakis}}},\ and\ \bibinfo {author} {\bibfnamefont {E.}~\bibnamefont
  {{Winstanley}}},\ }\href {https://doi.org/10.1103/PhysRevD.54.5049}
  {\bibfield  {journal} {\bibinfo  {journal} {\prd}\ }\textbf {\bibinfo
  {volume} {54}},\ \bibinfo {pages} {5049} (\bibinfo {year}
  {1996})}\BibitemShut {NoStop}%
\bibitem [{\citenamefont {{Kanti}}\ \emph {et~al.}(1998)\citenamefont
  {{Kanti}}, \citenamefont {{Mavromatos}}, \citenamefont {{Rizos}},
  \citenamefont {{Tamvakis}},\ and\ \citenamefont {{Winstanley}}}]{Kanti1998}%
  \BibitemOpen
  \bibfield  {author} {\bibinfo {author} {\bibfnamefont {P.}~\bibnamefont
  {{Kanti}}}, \bibinfo {author} {\bibfnamefont {N.~E.}\ \bibnamefont
  {{Mavromatos}}}, \bibinfo {author} {\bibfnamefont {J.}~\bibnamefont
  {{Rizos}}}, \bibinfo {author} {\bibfnamefont {K.}~\bibnamefont
  {{Tamvakis}}},\ and\ \bibinfo {author} {\bibfnamefont {E.}~\bibnamefont
  {{Winstanley}}},\ }\href {https://doi.org/10.1103/PhysRevD.57.6255}
  {\bibfield  {journal} {\bibinfo  {journal} {\prd}\ }\textbf {\bibinfo
  {volume} {57}},\ \bibinfo {pages} {6255} (\bibinfo {year}
  {1998})}\BibitemShut {NoStop}%
\bibitem [{\citenamefont {{Yunes}}\ and\ \citenamefont
  {{Stein}}(2011)}]{Yunes2011}%
  \BibitemOpen
  \bibfield  {author} {\bibinfo {author} {\bibfnamefont {N.}~\bibnamefont
  {{Yunes}}}\ and\ \bibinfo {author} {\bibfnamefont {L.~C.}\ \bibnamefont
  {{Stein}}},\ }\href {https://doi.org/10.1103/PhysRevD.83.104002} {\bibfield
  {journal} {\bibinfo  {journal} {\prd}\ }\textbf {\bibinfo {volume} {83}},\
  \bibinfo {eid} {104002} (\bibinfo {year} {2011})}\BibitemShut {NoStop}%
\bibitem [{\citenamefont {{Yagi}}\ \emph {et~al.}(2012)\citenamefont {{Yagi}},
  \citenamefont {{Yunes}},\ and\ \citenamefont {{Tanaka}}}]{Yagi2012}%
  \BibitemOpen
  \bibfield  {author} {\bibinfo {author} {\bibfnamefont {K.}~\bibnamefont
  {{Yagi}}}, \bibinfo {author} {\bibfnamefont {N.}~\bibnamefont {{Yunes}}},\
  and\ \bibinfo {author} {\bibfnamefont {T.}~\bibnamefont {{Tanaka}}},\ }\href
  {https://doi.org/10.1103/PhysRevD.86.044037} {\bibfield  {journal} {\bibinfo
  {journal} {\prd}\ }\textbf {\bibinfo {volume} {86}},\ \bibinfo {eid} {044037}
  (\bibinfo {year} {2012})}\BibitemShut {NoStop}%
\bibitem [{\citenamefont {{Ayzenberg}}\ and\ \citenamefont
  {{Yunes}}(2014)}]{Ayzenberg2014}%
  \BibitemOpen
  \bibfield  {author} {\bibinfo {author} {\bibfnamefont {D.}~\bibnamefont
  {{Ayzenberg}}}\ and\ \bibinfo {author} {\bibfnamefont {N.}~\bibnamefont
  {{Yunes}}},\ }\href {https://doi.org/10.1103/PhysRevD.90.044066} {\bibfield
  {journal} {\bibinfo  {journal} {\prd}\ }\textbf {\bibinfo {volume} {90}},\
  \bibinfo {eid} {044066} (\bibinfo {year} {2014})}\BibitemShut {NoStop}%
\bibitem [{\citenamefont {{McNees}}\ \emph {et~al.}(2016)\citenamefont
  {{McNees}}, \citenamefont {{Stein}},\ and\ \citenamefont
  {{Yunes}}}]{McNees2016}%
  \BibitemOpen
  \bibfield  {author} {\bibinfo {author} {\bibfnamefont {R.}~\bibnamefont
  {{McNees}}}, \bibinfo {author} {\bibfnamefont {L.~C.}\ \bibnamefont
  {{Stein}}},\ and\ \bibinfo {author} {\bibfnamefont {N.}~\bibnamefont
  {{Yunes}}},\ }\href {https://doi.org/10.1088/0264-9381/33/23/235013}
  {\bibfield  {journal} {\bibinfo  {journal} {Classical and Quantum Gravity}\
  }\textbf {\bibinfo {volume} {33}},\ \bibinfo {eid} {235013} (\bibinfo {year}
  {2016})}\BibitemShut {NoStop}%
\bibitem [{\citenamefont {{Antoniou}}\ \emph
  {et~al.}(2018{\natexlab{a}})\citenamefont {{Antoniou}}, \citenamefont
  {{Bakopoulos}},\ and\ \citenamefont {{Kanti}}}]{Antoniou2018a}%
  \BibitemOpen
  \bibfield  {author} {\bibinfo {author} {\bibfnamefont {G.}~\bibnamefont
  {{Antoniou}}}, \bibinfo {author} {\bibfnamefont {A.}~\bibnamefont
  {{Bakopoulos}}},\ and\ \bibinfo {author} {\bibfnamefont {P.}~\bibnamefont
  {{Kanti}}},\ }\href {https://doi.org/10.1103/PhysRevLett.120.131102}
  {\bibfield  {journal} {\bibinfo  {journal} {\prl}\ }\textbf {\bibinfo
  {volume} {120}},\ \bibinfo {eid} {131102} (\bibinfo {year}
  {2018}{\natexlab{a}})}\BibitemShut {NoStop}%
\bibitem [{\citenamefont {{Antoniou}}\ \emph
  {et~al.}(2018{\natexlab{b}})\citenamefont {{Antoniou}}, \citenamefont
  {{Bakopoulos}},\ and\ \citenamefont {{Kanti}}}]{Antoniou2018b}%
  \BibitemOpen
  \bibfield  {author} {\bibinfo {author} {\bibfnamefont {G.}~\bibnamefont
  {{Antoniou}}}, \bibinfo {author} {\bibfnamefont {A.}~\bibnamefont
  {{Bakopoulos}}},\ and\ \bibinfo {author} {\bibfnamefont {P.}~\bibnamefont
  {{Kanti}}},\ }\href {https://doi.org/10.1103/PhysRevD.97.084037} {\bibfield
  {journal} {\bibinfo  {journal} {\prd}\ }\textbf {\bibinfo {volume} {97}},\
  \bibinfo {eid} {084037} (\bibinfo {year} {2018}{\natexlab{b}})}\BibitemShut
  {NoStop}%
\bibitem [{\citenamefont {{Silva}}\ \emph {et~al.}(2018)\citenamefont
  {{Silva}}, \citenamefont {{Sakstein}}, \citenamefont {{Gualtieri}},
  \citenamefont {{Sotiriou}},\ and\ \citenamefont {{Berti}}}]{Silva2018}%
  \BibitemOpen
  \bibfield  {author} {\bibinfo {author} {\bibfnamefont {H.~O.}\ \bibnamefont
  {{Silva}}}, \bibinfo {author} {\bibfnamefont {J.}~\bibnamefont {{Sakstein}}},
  \bibinfo {author} {\bibfnamefont {L.}~\bibnamefont {{Gualtieri}}}, \bibinfo
  {author} {\bibfnamefont {T.~P.}\ \bibnamefont {{Sotiriou}}},\ and\ \bibinfo
  {author} {\bibfnamefont {E.}~\bibnamefont {{Berti}}},\ }\href
  {https://doi.org/10.1103/PhysRevLett.120.131104} {\bibfield  {journal}
  {\bibinfo  {journal} {\prl}\ }\textbf {\bibinfo {volume} {120}},\ \bibinfo
  {eid} {131104} (\bibinfo {year} {2018})}\BibitemShut {NoStop}%
\bibitem [{\citenamefont {{Doneva}}\ and\ \citenamefont
  {{Yazadjiev}}(2018)}]{Doneva2018}%
  \BibitemOpen
  \bibfield  {author} {\bibinfo {author} {\bibfnamefont {D.~D.}\ \bibnamefont
  {{Doneva}}}\ and\ \bibinfo {author} {\bibfnamefont {S.~S.}\ \bibnamefont
  {{Yazadjiev}}},\ }\href {https://doi.org/10.1103/PhysRevLett.120.131103}
  {\bibfield  {journal} {\bibinfo  {journal} {\prl}\ }\textbf {\bibinfo
  {volume} {120}},\ \bibinfo {eid} {131103} (\bibinfo {year}
  {2018})}\BibitemShut {NoStop}%
\bibitem [{\citenamefont {{Stein}}\ and\ \citenamefont
  {{Yagi}}(2014)}]{Stein2014}%
  \BibitemOpen
  \bibfield  {author} {\bibinfo {author} {\bibfnamefont {L.~C.}\ \bibnamefont
  {{Stein}}}\ and\ \bibinfo {author} {\bibfnamefont {K.}~\bibnamefont
  {{Yagi}}},\ }\href {https://doi.org/10.1103/PhysRevD.89.044026} {\bibfield
  {journal} {\bibinfo  {journal} {\prd}\ }\textbf {\bibinfo {volume} {89}},\
  \bibinfo {eid} {044026} (\bibinfo {year} {2014})},\ \Eprint
  {https://arxiv.org/abs/1310.6743} {arXiv:1310.6743 [gr-qc]} \BibitemShut
  {NoStop}%
\bibitem [{\citenamefont {{Giddings}}(2016{\natexlab{a}})}]{Giddings2016a}%
  \BibitemOpen
  \bibfield  {author} {\bibinfo {author} {\bibfnamefont {S.~B.}\ \bibnamefont
  {{Giddings}}},\ }\href {https://doi.org/10.1142/S0218271816440144} {\bibfield
   {journal} {\bibinfo  {journal} {International Journal of Modern Physics D}\
  }\textbf {\bibinfo {volume} {25}},\ \bibinfo {eid} {1644014} (\bibinfo {year}
  {2016}{\natexlab{a}})}\BibitemShut {NoStop}%
\bibitem [{\citenamefont {{Giddings}}(2016{\natexlab{b}})}]{Giddings2016b}%
  \BibitemOpen
  \bibfield  {author} {\bibinfo {author} {\bibfnamefont {S.~B.}\ \bibnamefont
  {{Giddings}}},\ }\href {https://doi.org/10.1088/0264-9381/33/23/235010}
  {\bibfield  {journal} {\bibinfo  {journal} {Classical and Quantum Gravity}\
  }\textbf {\bibinfo {volume} {33}},\ \bibinfo {eid} {235010} (\bibinfo {year}
  {2016}{\natexlab{b}})},\ \Eprint {https://arxiv.org/abs/1602.03622}
  {arXiv:1602.03622 [gr-qc]} \BibitemShut {NoStop}%
\bibitem [{\citenamefont {{Giddings}}\ and\ \citenamefont
  {{Psaltis}}(2018)}]{Giddings2018}%
  \BibitemOpen
  \bibfield  {author} {\bibinfo {author} {\bibfnamefont {S.~B.}\ \bibnamefont
  {{Giddings}}}\ and\ \bibinfo {author} {\bibfnamefont {D.}~\bibnamefont
  {{Psaltis}}},\ }\href {https://doi.org/10.1103/PhysRevD.97.084035} {\bibfield
   {journal} {\bibinfo  {journal} {\prd}\ }\textbf {\bibinfo {volume} {97}},\
  \bibinfo {eid} {084035} (\bibinfo {year} {2018})}\BibitemShut {NoStop}%
\bibitem [{\citenamefont {{Event Horizon Telescope
  Collaboration}}(2019{\natexlab{c}})}]{PaperIV}%
  \BibitemOpen
  \bibfield  {author} {\bibinfo {author} {\bibnamefont {{Event Horizon
  Telescope Collaboration}}},\ }\href
  {https://doi.org/10.3847/2041-8213/ab0e85} {\bibfield  {journal} {\bibinfo
  {journal} {\apjl}\ }\textbf {\bibinfo {volume} {875}},\ \bibinfo {eid} {L4}
  (\bibinfo {year} {2019}{\natexlab{c}})}\BibitemShut {NoStop}%
\bibitem [{\citenamefont {{Gebhardt}}\ \emph {et~al.}(2011)\citenamefont
  {{Gebhardt}}, \citenamefont {{Adams}}, \citenamefont {{Richstone}},
  \citenamefont {{Lauer}}, \citenamefont {{Faber}}, \citenamefont
  {{G{\"u}ltekin}}, \citenamefont {{Murphy}},\ and\ \citenamefont
  {{Tremaine}}}]{Gebhardt2011}%
  \BibitemOpen
  \bibfield  {author} {\bibinfo {author} {\bibfnamefont {K.}~\bibnamefont
  {{Gebhardt}}}, \bibinfo {author} {\bibfnamefont {J.}~\bibnamefont {{Adams}}},
  \bibinfo {author} {\bibfnamefont {D.}~\bibnamefont {{Richstone}}}, \bibinfo
  {author} {\bibfnamefont {T.~R.}\ \bibnamefont {{Lauer}}}, \bibinfo {author}
  {\bibfnamefont {S.~M.}\ \bibnamefont {{Faber}}}, \bibinfo {author}
  {\bibfnamefont {K.}~\bibnamefont {{G{\"u}ltekin}}}, \bibinfo {author}
  {\bibfnamefont {J.}~\bibnamefont {{Murphy}}},\ and\ \bibinfo {author}
  {\bibfnamefont {S.}~\bibnamefont {{Tremaine}}},\ }\href
  {https://doi.org/10.1088/0004-637X/729/2/119} {\bibfield  {journal} {\bibinfo
   {journal} {\apj}\ }\textbf {\bibinfo {volume} {729}},\ \bibinfo {eid} {119}
  (\bibinfo {year} {2011})}\BibitemShut {NoStop}%
\bibitem [{\citenamefont {{Psaltis}}\ \emph {et~al.}(2015)\citenamefont
  {{Psaltis}}, \citenamefont {{{\"O}zel}}, \citenamefont {{Chan}},\ and\
  \citenamefont {{Marrone}}}]{Psaltis2015}%
  \BibitemOpen
  \bibfield  {author} {\bibinfo {author} {\bibfnamefont {D.}~\bibnamefont
  {{Psaltis}}}, \bibinfo {author} {\bibfnamefont {F.}~\bibnamefont
  {{{\"O}zel}}}, \bibinfo {author} {\bibfnamefont {C.-K.}\ \bibnamefont
  {{Chan}}},\ and\ \bibinfo {author} {\bibfnamefont {D.~P.}\ \bibnamefont
  {{Marrone}}},\ }\href {https://doi.org/10.1088/0004-637X/814/2/115}
  {\bibfield  {journal} {\bibinfo  {journal} {\apj}\ }\textbf {\bibinfo
  {volume} {814}},\ \bibinfo {pages} {115} (\bibinfo {year}
  {2015})}\BibitemShut {NoStop}%
\bibitem [{Note3()}]{Note3}%
  \BibitemOpen
  \bibinfo {note} {Had we instead used the mass smaller black-hole mass
  measurements based on gas dynamics~\cite {Walsh2013}, we would have concluded
  that the black hole spacetime is not described by the Kerr metric, a
  situation that was assigned a negligible prior.}\BibitemShut {Stop}%
\bibitem [{\citenamefont {{Bardeen}}(1973)}]{Bardeen1973}%
  \BibitemOpen
  \bibfield  {author} {\bibinfo {author} {\bibfnamefont {J.~M.}\ \bibnamefont
  {{Bardeen}}},\ }in\ \href@noop {} {\emph {\bibinfo {booktitle} {Black Holes
  (Les Astres Occlus)}}},\ \bibinfo {editor} {edited by\ \bibinfo {editor}
  {\bibfnamefont {C.}~\bibnamefont {{Dewitt}}}\ and\ \bibinfo {editor}
  {\bibfnamefont {B.~S.}\ \bibnamefont {{Dewitt}}}}\ (\bibinfo {year} {1973})\
  pp.\ \bibinfo {pages} {215--239}\BibitemShut {NoStop}%
\bibitem [{\citenamefont {{Chandrasekhar}}(1983)}]{Chandra1983}%
  \BibitemOpen
  \bibfield  {author} {\bibinfo {author} {\bibfnamefont {S.}~\bibnamefont
  {{Chandrasekhar}}},\ }\href@noop {} {\emph {\bibinfo {title} {{The
  mathematical theory of black holes}}}}\ (\bibinfo {year} {1983})\BibitemShut
  {NoStop}%
\bibitem [{\citenamefont {{Teo}}(2003)}]{Tao2003}%
  \BibitemOpen
  \bibfield  {author} {\bibinfo {author} {\bibfnamefont {E.}~\bibnamefont
  {{Teo}}},\ }\href {https://doi.org/10.1023/A:1026286607562} {\bibfield
  {journal} {\bibinfo  {journal} {GRG}\ }\textbf {\bibinfo {volume} {35}},\
  \bibinfo {pages} {1909} (\bibinfo {year} {2003})}\BibitemShut {NoStop}%
\bibitem [{\citenamefont {{Takahashi}}(2004)}]{Takahashi2004}%
  \BibitemOpen
  \bibfield  {author} {\bibinfo {author} {\bibfnamefont {R.}~\bibnamefont
  {{Takahashi}}},\ }\href {https://doi.org/10.1086/422403} {\bibfield
  {journal} {\bibinfo  {journal} {\apj}\ }\textbf {\bibinfo {volume} {611}},\
  \bibinfo {pages} {996} (\bibinfo {year} {2004})}\BibitemShut {NoStop}%
\bibitem [{\citenamefont {{Bambi}}\ and\ \citenamefont
  {{Freese}}(2009)}]{Bambi2009}%
  \BibitemOpen
  \bibfield  {author} {\bibinfo {author} {\bibfnamefont {C.}~\bibnamefont
  {{Bambi}}}\ and\ \bibinfo {author} {\bibfnamefont {K.}~\bibnamefont
  {{Freese}}},\ }\href {https://doi.org/10.1103/PhysRevD.79.043002} {\bibfield
  {journal} {\bibinfo  {journal} {\prd}\ }\textbf {\bibinfo {volume} {79}},\
  \bibinfo {eid} {043002} (\bibinfo {year} {2009})}\BibitemShut {NoStop}%
\bibitem [{\citenamefont {{Israel}}(1967)}]{Israel1967}%
  \BibitemOpen
  \bibfield  {author} {\bibinfo {author} {\bibfnamefont {W.}~\bibnamefont
  {{Israel}}},\ }\href {https://doi.org/10.1103/PhysRev.164.1776} {\bibfield
  {journal} {\bibinfo  {journal} {Physical Review}\ }\textbf {\bibinfo {volume}
  {164}},\ \bibinfo {pages} {1776} (\bibinfo {year} {1967})}\BibitemShut
  {NoStop}%
\bibitem [{\citenamefont {{Israel}}(1968)}]{Israel1968}%
  \BibitemOpen
  \bibfield  {author} {\bibinfo {author} {\bibfnamefont {W.}~\bibnamefont
  {{Israel}}},\ }\href {https://doi.org/10.1007/BF01645859} {\bibfield
  {journal} {\bibinfo  {journal} {Communications in Mathematical Physics}\
  }\textbf {\bibinfo {volume} {8}},\ \bibinfo {pages} {245} (\bibinfo {year}
  {1968})}\BibitemShut {NoStop}%
\bibitem [{\citenamefont {{Carter}}(1971)}]{Carter1971}%
  \BibitemOpen
  \bibfield  {author} {\bibinfo {author} {\bibfnamefont {B.}~\bibnamefont
  {{Carter}}},\ }\href {https://doi.org/10.1103/PhysRevLett.26.331} {\bibfield
  {journal} {\bibinfo  {journal} {\prl}\ }\textbf {\bibinfo {volume} {26}},\
  \bibinfo {pages} {331} (\bibinfo {year} {1971})}\BibitemShut {NoStop}%
\bibitem [{\citenamefont {{Hawking}}(1972)}]{Hawking1972}%
  \BibitemOpen
  \bibfield  {author} {\bibinfo {author} {\bibfnamefont {S.~W.}\ \bibnamefont
  {{Hawking}}},\ }\href {https://doi.org/10.1007/BF01877517} {\bibfield
  {journal} {\bibinfo  {journal} {Communications in Mathematical Physics}\
  }\textbf {\bibinfo {volume} {25}},\ \bibinfo {pages} {152} (\bibinfo {year}
  {1972})}\BibitemShut {NoStop}%
\bibitem [{\citenamefont {{Robinson}}(1975)}]{Robinson1975}%
  \BibitemOpen
  \bibfield  {author} {\bibinfo {author} {\bibfnamefont {D.~C.}\ \bibnamefont
  {{Robinson}}},\ }\href {https://doi.org/10.1103/PhysRevLett.34.905}
  {\bibfield  {journal} {\bibinfo  {journal} {\prl}\ }\textbf {\bibinfo
  {volume} {34}},\ \bibinfo {pages} {905} (\bibinfo {year} {1975})}\BibitemShut
  {NoStop}%
\bibitem [{\citenamefont {{Johannsen}}\ and\ \citenamefont
  {{Psaltis}}(2010{\natexlab{b}})}]{Johannsen2010a}%
  \BibitemOpen
  \bibfield  {author} {\bibinfo {author} {\bibfnamefont {T.}~\bibnamefont
  {{Johannsen}}}\ and\ \bibinfo {author} {\bibfnamefont {D.}~\bibnamefont
  {{Psaltis}}},\ }\href {https://doi.org/10.1088/0004-637X/716/1/187}
  {\bibfield  {journal} {\bibinfo  {journal} {\apj}\ }\textbf {\bibinfo
  {volume} {716}},\ \bibinfo {pages} {187} (\bibinfo {year}
  {2010}{\natexlab{b}})}\BibitemShut {NoStop}%
\bibitem [{\citenamefont {{Cardoso}}\ \emph {et~al.}(2014)\citenamefont
  {{Cardoso}}, \citenamefont {{Pani}},\ and\ \citenamefont
  {{Rico}}}]{Cardoso2014}%
  \BibitemOpen
  \bibfield  {author} {\bibinfo {author} {\bibfnamefont {V.}~\bibnamefont
  {{Cardoso}}}, \bibinfo {author} {\bibfnamefont {P.}~\bibnamefont {{Pani}}},\
  and\ \bibinfo {author} {\bibfnamefont {J.}~\bibnamefont {{Rico}}},\ }\href
  {https://doi.org/10.1103/PhysRevD.89.064007} {\bibfield  {journal} {\bibinfo
  {journal} {\prd}\ }\textbf {\bibinfo {volume} {89}},\ \bibinfo {eid} {064007}
  (\bibinfo {year} {2014})}\BibitemShut {NoStop}%
\bibitem [{\citenamefont {{Carson}}\ and\ \citenamefont
  {{Yagi}}(2020{\natexlab{b}})}]{Carson2020b}%
  \BibitemOpen
  \bibfield  {author} {\bibinfo {author} {\bibfnamefont {Z.}~\bibnamefont
  {{Carson}}}\ and\ \bibinfo {author} {\bibfnamefont {K.}~\bibnamefont
  {{Yagi}}},\ }\href {https://doi.org/10.1103/PhysRevD.101.084030} {\bibfield
  {journal} {\bibinfo  {journal} {\prd}\ }\textbf {\bibinfo {volume} {101}},\
  \bibinfo {eid} {084030} (\bibinfo {year} {2020}{\natexlab{b}})},\ \Eprint
  {https://arxiv.org/abs/2002.01028} {arXiv:2002.01028 [gr-qc]} \BibitemShut
  {NoStop}%
\bibitem [{Note4()}]{Note4}%
  \BibitemOpen
  \bibinfo {note} {Albeit useful in numerical studies of known metrics, such
  parametrizations do not guarantee the absence of pathologies when the various
  coefficients are chosen outside the discrete sets that are known to describe
  regular metrics}\BibitemShut {NoStop}%
\bibitem [{\citenamefont {{Rezzolla}}\ and\ \citenamefont
  {{Zhidenko}}(2014)}]{Rezzolla2014}%
  \BibitemOpen
  \bibfield  {author} {\bibinfo {author} {\bibfnamefont {L.}~\bibnamefont
  {{Rezzolla}}}\ and\ \bibinfo {author} {\bibfnamefont {A.}~\bibnamefont
  {{Zhidenko}}},\ }\href {https://doi.org/10.1103/PhysRevD.90.084009}
  {\bibfield  {journal} {\bibinfo  {journal} {\prd}\ }\textbf {\bibinfo
  {volume} {90}},\ \bibinfo {eid} {084009} (\bibinfo {year}
  {2014})}\BibitemShut {NoStop}%
\bibitem [{\citenamefont {{Konoplya}}\ \emph {et~al.}(2016)\citenamefont
  {{Konoplya}}, \citenamefont {{Rezzolla}},\ and\ \citenamefont
  {{Zhidenko}}}]{Konoplya2016}%
  \BibitemOpen
  \bibfield  {author} {\bibinfo {author} {\bibfnamefont {R.}~\bibnamefont
  {{Konoplya}}}, \bibinfo {author} {\bibfnamefont {L.}~\bibnamefont
  {{Rezzolla}}},\ and\ \bibinfo {author} {\bibfnamefont {A.}~\bibnamefont
  {{Zhidenko}}},\ }\href {https://doi.org/10.1103/PhysRevD.93.064015}
  {\bibfield  {journal} {\bibinfo  {journal} {\prd}\ }\textbf {\bibinfo
  {volume} {93}},\ \bibinfo {eid} {064015} (\bibinfo {year}
  {2016})}\BibitemShut {NoStop}%
\bibitem [{\citenamefont {{Younsi}}\ \emph {et~al.}(2016)\citenamefont
  {{Younsi}}, \citenamefont {{Zhidenko}}, \citenamefont {{Rezzolla}},
  \citenamefont {{Konoplya}},\ and\ \citenamefont {{Mizuno}}}]{Younsi2016}%
  \BibitemOpen
  \bibfield  {author} {\bibinfo {author} {\bibfnamefont {Z.}~\bibnamefont
  {{Younsi}}}, \bibinfo {author} {\bibfnamefont {A.}~\bibnamefont
  {{Zhidenko}}}, \bibinfo {author} {\bibfnamefont {L.}~\bibnamefont
  {{Rezzolla}}}, \bibinfo {author} {\bibfnamefont {R.}~\bibnamefont
  {{Konoplya}}},\ and\ \bibinfo {author} {\bibfnamefont {Y.}~\bibnamefont
  {{Mizuno}}},\ }\href {https://doi.org/10.1103/PhysRevD.94.084025} {\bibfield
  {journal} {\bibinfo  {journal} {\prd}\ }\textbf {\bibinfo {volume} {94}},\
  \bibinfo {eid} {084025} (\bibinfo {year} {2016})}\BibitemShut {NoStop}%
\bibitem [{\citenamefont {{Mizuno}}\ \emph {et~al.}(2018)\citenamefont
  {{Mizuno}}, \citenamefont {{Younsi}}, \citenamefont {{Fromm}}, \citenamefont
  {{Porth}}, \citenamefont {{De Laurentis}}, \citenamefont {{Olivares}},
  \citenamefont {{Falcke}}, \citenamefont {{Kramer}},\ and\ \citenamefont
  {{Rezzolla}}}]{Mizuno2018}%
  \BibitemOpen
  \bibfield  {author} {\bibinfo {author} {\bibfnamefont {Y.}~\bibnamefont
  {{Mizuno}}}, \bibinfo {author} {\bibfnamefont {Z.}~\bibnamefont {{Younsi}}},
  \bibinfo {author} {\bibfnamefont {C.~M.}\ \bibnamefont {{Fromm}}}, \bibinfo
  {author} {\bibfnamefont {O.}~\bibnamefont {{Porth}}}, \bibinfo {author}
  {\bibfnamefont {M.}~\bibnamefont {{De Laurentis}}}, \bibinfo {author}
  {\bibfnamefont {H.}~\bibnamefont {{Olivares}}}, \bibinfo {author}
  {\bibfnamefont {H.}~\bibnamefont {{Falcke}}}, \bibinfo {author}
  {\bibfnamefont {M.}~\bibnamefont {{Kramer}}},\ and\ \bibinfo {author}
  {\bibfnamefont {L.}~\bibnamefont {{Rezzolla}}},\ }\href
  {https://doi.org/10.1038/s41550-018-0449-5} {\bibfield  {journal} {\bibinfo
  {journal} {Nature Astronomy}\ }\textbf {\bibinfo {volume} {2}},\ \bibinfo
  {pages} {585} (\bibinfo {year} {2018})}\BibitemShut {NoStop}%
\bibitem [{\citenamefont {{LIGO Scientific Collaboration}}\ and\ \citenamefont
  {{Aasi}}(2015)}]{AdLIGO}%
  \BibitemOpen
  \bibfield  {author} {\bibinfo {author} {\bibnamefont {{LIGO Scientific
  Collaboration}}}\ and\ \bibinfo {author} {\bibfnamefont {J.~o.}\ \bibnamefont
  {{Aasi}}},\ }\href {https://doi.org/10.1088/0264-9381/32/7/074001} {\bibfield
   {journal} {\bibinfo  {journal} {Classical and Quantum Gravity}\ }\textbf
  {\bibinfo {volume} {32}},\ \bibinfo {eid} {074001} (\bibinfo {year}
  {2015})},\ \Eprint {https://arxiv.org/abs/1411.4547} {arXiv:1411.4547
  [gr-qc]} \BibitemShut {NoStop}%
\bibitem [{\citenamefont {{Acernese}}\ \emph {et~al.}(2015)\citenamefont
  {{Acernese}} \emph {et~al.}}]{AdVirgo}%
  \BibitemOpen
  \bibfield  {author} {\bibinfo {author} {\bibfnamefont {F.}~\bibnamefont
  {{Acernese}}} \emph {et~al.},\ }\href
  {https://doi.org/10.1088/0264-9381/32/2/024001} {\bibfield  {journal}
  {\bibinfo  {journal} {Classical and Quantum Gravity}\ }\textbf {\bibinfo
  {volume} {32}},\ \bibinfo {eid} {024001} (\bibinfo {year} {2015})},\ \Eprint
  {https://arxiv.org/abs/1408.3978} {arXiv:1408.3978 [gr-qc]} \BibitemShut
  {NoStop}%
\bibitem [{\citenamefont {{Hannam}}\ \emph {et~al.}(2014)\citenamefont
  {{Hannam}}, \citenamefont {{Schmidt}}, \citenamefont {{Boh{\'e}}},
  \citenamefont {{Haegel}}, \citenamefont {{Husa}}, \citenamefont {{Ohme}},
  \citenamefont {{Pratten}},\ and\ \citenamefont
  {{P{\"u}rrer}}}]{IMRPhenomPv2}%
  \BibitemOpen
  \bibfield  {author} {\bibinfo {author} {\bibfnamefont {M.}~\bibnamefont
  {{Hannam}}}, \bibinfo {author} {\bibfnamefont {P.}~\bibnamefont {{Schmidt}}},
  \bibinfo {author} {\bibfnamefont {A.}~\bibnamefont {{Boh{\'e}}}}, \bibinfo
  {author} {\bibfnamefont {L.}~\bibnamefont {{Haegel}}}, \bibinfo {author}
  {\bibfnamefont {S.}~\bibnamefont {{Husa}}}, \bibinfo {author} {\bibfnamefont
  {F.}~\bibnamefont {{Ohme}}}, \bibinfo {author} {\bibfnamefont
  {G.}~\bibnamefont {{Pratten}}},\ and\ \bibinfo {author} {\bibfnamefont
  {M.}~\bibnamefont {{P{\"u}rrer}}},\ }\href
  {https://doi.org/10.1103/PhysRevLett.113.151101} {\bibfield  {journal}
  {\bibinfo  {journal} {\prl}\ }\textbf {\bibinfo {volume} {113}},\ \bibinfo
  {eid} {151101} (\bibinfo {year} {2014})}\BibitemShut {NoStop}%
\bibitem [{\citenamefont {{Agathos}}\ \emph {et~al.}(2014)\citenamefont
  {{Agathos}}, \citenamefont {{Del Pozzo}}, \citenamefont {{Li}}, \citenamefont
  {{Van Den Broeck}}, \citenamefont {{Veitch}},\ and\ \citenamefont
  {{Vitale}}}]{Agathos2014}%
  \BibitemOpen
  \bibfield  {author} {\bibinfo {author} {\bibfnamefont {M.}~\bibnamefont
  {{Agathos}}}, \bibinfo {author} {\bibfnamefont {W.}~\bibnamefont {{Del
  Pozzo}}}, \bibinfo {author} {\bibfnamefont {T.~G.~F.}\ \bibnamefont {{Li}}},
  \bibinfo {author} {\bibfnamefont {C.}~\bibnamefont {{Van Den Broeck}}},
  \bibinfo {author} {\bibfnamefont {J.}~\bibnamefont {{Veitch}}},\ and\
  \bibinfo {author} {\bibfnamefont {S.}~\bibnamefont {{Vitale}}},\ }\href
  {https://doi.org/10.1103/PhysRevD.89.082001} {\bibfield  {journal} {\bibinfo
  {journal} {\prd}\ }\textbf {\bibinfo {volume} {89}},\ \bibinfo {eid} {082001}
  (\bibinfo {year} {2014})},\ \Eprint {https://arxiv.org/abs/1311.0420}
  {arXiv:1311.0420 [gr-qc]} \BibitemShut {NoStop}%
\bibitem [{\citenamefont {{Barausse}}\ and\ \citenamefont
  {{Sotiriou}}(2008)}]{Barausse2008}%
  \BibitemOpen
  \bibfield  {author} {\bibinfo {author} {\bibfnamefont {E.}~\bibnamefont
  {{Barausse}}}\ and\ \bibinfo {author} {\bibfnamefont {T.~P.}\ \bibnamefont
  {{Sotiriou}}},\ }\href {https://doi.org/10.1103/PhysRevLett.101.099001}
  {\bibfield  {journal} {\bibinfo  {journal} {\prl}\ }\textbf {\bibinfo
  {volume} {101}},\ \bibinfo {eid} {099001} (\bibinfo {year}
  {2008})}\BibitemShut {NoStop}%
\bibitem [{Note5()}]{Note5}%
  \BibitemOpen
  \bibinfo {note} {We do not explore the possibility that back-scattering of
  gravitational waves off the background -- so-called tails of tails terms --
  could contribute to higher non-integer deviations in the gravitational-wave
  signature: the 2.5PN term corresponds to a fixed phase offset that is
  marginalized over, and we restrict our analysis to terms at or below
  3PN.}\BibitemShut {Stop}%
\bibitem [{\citenamefont {{C{\'a}rdenas-Avenda{\~n}o}}\ \emph
  {et~al.}(2020)\citenamefont {{C{\'a}rdenas-Avenda{\~n}o}}, \citenamefont
  {{Nampalliwar}},\ and\ \citenamefont {{Yunes}}}]{Cardenas2020}%
  \BibitemOpen
  \bibfield  {author} {\bibinfo {author} {\bibfnamefont {A.}~\bibnamefont
  {{C{\'a}rdenas-Avenda{\~n}o}}}, \bibinfo {author} {\bibfnamefont
  {S.}~\bibnamefont {{Nampalliwar}}},\ and\ \bibinfo {author} {\bibfnamefont
  {N.}~\bibnamefont {{Yunes}}},\ }\href
  {https://doi.org/10.1088/1361-6382/ab8f64} {\bibfield  {journal} {\bibinfo
  {journal} {Classical and Quantum Gravity}\ }\textbf {\bibinfo {volume}
  {37}},\ \bibinfo {eid} {135008} (\bibinfo {year} {2020})},\ \Eprint
  {https://arxiv.org/abs/1912.08062} {arXiv:1912.08062 [gr-qc]} \BibitemShut
  {NoStop}%
\bibitem [{\citenamefont {{Tahura}}\ and\ \citenamefont
  {{Yagi}}(2018)}]{Tahura2018}%
  \BibitemOpen
  \bibfield  {author} {\bibinfo {author} {\bibfnamefont {S.}~\bibnamefont
  {{Tahura}}}\ and\ \bibinfo {author} {\bibfnamefont {K.}~\bibnamefont
  {{Yagi}}},\ }\href {https://doi.org/10.1103/PhysRevD.98.084042} {\bibfield
  {journal} {\bibinfo  {journal} {\prd}\ }\textbf {\bibinfo {volume} {98}},\
  \bibinfo {eid} {084042} (\bibinfo {year} {2018})},\ \Eprint
  {https://arxiv.org/abs/1809.00259} {arXiv:1809.00259 [gr-qc]} \BibitemShut
  {NoStop}%
\bibitem [{\citenamefont {{Yunes}}\ and\ \citenamefont
  {{Pretorius}}(2009)}]{Yunes2009}%
  \BibitemOpen
  \bibfield  {author} {\bibinfo {author} {\bibfnamefont {N.}~\bibnamefont
  {{Yunes}}}\ and\ \bibinfo {author} {\bibfnamefont {F.}~\bibnamefont
  {{Pretorius}}},\ }\href {https://doi.org/10.1103/PhysRevD.79.084043}
  {\bibfield  {journal} {\bibinfo  {journal} {\prd}\ }\textbf {\bibinfo
  {volume} {79}},\ \bibinfo {eid} {084043} (\bibinfo {year}
  {2009})}\BibitemShut {NoStop}%
\bibitem [{\citenamefont {{Buonanno}}\ and\ \citenamefont
  {{Damour}}(1999)}]{Buonanno1999}%
  \BibitemOpen
  \bibfield  {author} {\bibinfo {author} {\bibfnamefont {A.}~\bibnamefont
  {{Buonanno}}}\ and\ \bibinfo {author} {\bibfnamefont {T.}~\bibnamefont
  {{Damour}}},\ }\href {https://doi.org/10.1103/PhysRevD.59.084006} {\bibfield
  {journal} {\bibinfo  {journal} {\prd}\ }\textbf {\bibinfo {volume} {59}},\
  \bibinfo {eid} {084006} (\bibinfo {year} {1999})},\ \Eprint
  {https://arxiv.org/abs/gr-qc/9811091} {arXiv:gr-qc/9811091 [gr-qc]}
  \BibitemShut {NoStop}%
\bibitem [{Note6()}]{Note6}%
  \BibitemOpen
  \bibinfo {note} {Even though the effective one-body approach was proven to be
  valid only for the radiative properties of GR, this is consistent with the
  approach in this work of allowing only for the equilibrium spacetimes to be
  different than GR}\BibitemShut {NoStop}%
\bibitem [{\citenamefont {{Foreman-Mackey}}\ \emph {et~al.}(2013)\citenamefont
  {{Foreman-Mackey}}, \citenamefont {{Hogg}}, \citenamefont {{Lang}},\ and\
  \citenamefont {{Goodman}}}]{ForemanMackey2013}%
  \BibitemOpen
  \bibfield  {author} {\bibinfo {author} {\bibfnamefont {D.}~\bibnamefont
  {{Foreman-Mackey}}}, \bibinfo {author} {\bibfnamefont {D.~W.}\ \bibnamefont
  {{Hogg}}}, \bibinfo {author} {\bibfnamefont {D.}~\bibnamefont {{Lang}}},\
  and\ \bibinfo {author} {\bibfnamefont {J.}~\bibnamefont {{Goodman}}},\ }\href
  {https://doi.org/10.1086/670067} {\bibfield  {journal} {\bibinfo  {journal}
  {PASP}\ }\textbf {\bibinfo {volume} {125}},\ \bibinfo {pages} {306} (\bibinfo
  {year} {2013})},\ \Eprint {https://arxiv.org/abs/1202.3665} {arXiv:1202.3665
  [astro-ph.IM]} \BibitemShut {NoStop}%
\bibitem [{\citenamefont {{Ashton}}\ \emph {et~al.}(2019)\citenamefont
  {{Ashton}}, \citenamefont {{H{\"u}bner}}, \citenamefont {{Lasky}},
  \citenamefont {{Talbot}}, \citenamefont {{Ackley}}, \citenamefont
  {{Biscoveanu}} \emph {et~al.}}]{Ashton2019}%
  \BibitemOpen
  \bibfield  {author} {\bibinfo {author} {\bibfnamefont {G.}~\bibnamefont
  {{Ashton}}}, \bibinfo {author} {\bibfnamefont {M.}~\bibnamefont
  {{H{\"u}bner}}}, \bibinfo {author} {\bibfnamefont {P.~D.}\ \bibnamefont
  {{Lasky}}}, \bibinfo {author} {\bibfnamefont {C.}~\bibnamefont {{Talbot}}},
  \bibinfo {author} {\bibfnamefont {K.}~\bibnamefont {{Ackley}}}, \bibinfo
  {author} {\bibfnamefont {S.}~\bibnamefont {{Biscoveanu}}}, \emph {et~al.},\
  }\href {https://doi.org/10.3847/1538-4365/ab06fc} {\bibfield  {journal}
  {\bibinfo  {journal} {\apjs}\ }\textbf {\bibinfo {volume} {241}},\ \bibinfo
  {eid} {27} (\bibinfo {year} {2019})},\ \Eprint
  {https://arxiv.org/abs/1811.02042} {arXiv:1811.02042 [astro-ph.IM]}
  \BibitemShut {NoStop}%
\bibitem [{\citenamefont {{Romero-Shaw}}\ \emph {et~al.}(2020)\citenamefont
  {{Romero-Shaw}}, \citenamefont {{Talbot}}, \citenamefont {{Biscoveanu}},
  \citenamefont {{D'Emilio}}, \citenamefont {{Ashton}}, \citenamefont {{Berry}}
  \emph {et~al.}}]{RomeroShaw2020}%
  \BibitemOpen
  \bibfield  {author} {\bibinfo {author} {\bibfnamefont {I.~M.}\ \bibnamefont
  {{Romero-Shaw}}}, \bibinfo {author} {\bibfnamefont {C.}~\bibnamefont
  {{Talbot}}}, \bibinfo {author} {\bibfnamefont {S.}~\bibnamefont
  {{Biscoveanu}}}, \bibinfo {author} {\bibfnamefont {V.}~\bibnamefont
  {{D'Emilio}}}, \bibinfo {author} {\bibfnamefont {G.}~\bibnamefont
  {{Ashton}}}, \bibinfo {author} {\bibfnamefont {C.~P.~L.}\ \bibnamefont
  {{Berry}}}, \emph {et~al.},\ }\bibfield  {journal} {\bibinfo  {journal}
  {\mnras}\ }\href {https://doi.org/10.1093/mnras/staa2850}
  {10.1093/mnras/staa2850} (\bibinfo {year} {2020}),\ \Eprint
  {https://arxiv.org/abs/2006.00714} {arXiv:2006.00714 [astro-ph.IM]}
  \BibitemShut {NoStop}%
\bibitem [{\citenamefont {{Payne}}\ \emph {et~al.}(2020)\citenamefont
  {{Payne}}, \citenamefont {{Talbot}}, \citenamefont {{Lasky}}, \citenamefont
  {{Thrane}},\ and\ \citenamefont {{Kissel}}}]{Payne2020}%
  \BibitemOpen
  \bibfield  {author} {\bibinfo {author} {\bibfnamefont {E.}~\bibnamefont
  {{Payne}}}, \bibinfo {author} {\bibfnamefont {C.}~\bibnamefont {{Talbot}}},
  \bibinfo {author} {\bibfnamefont {P.~D.}\ \bibnamefont {{Lasky}}}, \bibinfo
  {author} {\bibfnamefont {E.}~\bibnamefont {{Thrane}}},\ and\ \bibinfo
  {author} {\bibfnamefont {J.~S.}\ \bibnamefont {{Kissel}}},\ }\href@noop {}
  {\bibfield  {journal} {\bibinfo  {journal} {arXiv e-prints}\ ,\ \bibinfo
  {eid} {arXiv:2009.10193}} (\bibinfo {year} {2020})},\ \Eprint
  {https://arxiv.org/abs/2009.10193} {arXiv:2009.10193 [astro-ph.IM]}
  \BibitemShut {NoStop}%
\bibitem [{\citenamefont {{Abbott}}\ \emph
  {et~al.}(2019{\natexlab{b}})\citenamefont {{Abbott}}, \citenamefont {{et
  al}}, \citenamefont {{LIGO Scientific}},\ and\ \citenamefont {{Virgo
  Collaborations}}}]{Abbott2019b}%
  \BibitemOpen
  \bibfield  {author} {\bibinfo {author} {\bibfnamefont {B.~P.}\ \bibnamefont
  {{Abbott}}}, \bibinfo {author} {\bibnamefont {{et al}}}, \bibinfo {author}
  {\bibnamefont {{LIGO Scientific}}},\ and\ \bibinfo {author} {\bibnamefont
  {{Virgo Collaborations}}},\ }\href
  {https://doi.org/10.1103/PhysRevX.9.031040} {\bibfield  {journal} {\bibinfo
  {journal} {Physical Review X}\ }\textbf {\bibinfo {volume} {9}},\ \bibinfo
  {eid} {031040} (\bibinfo {year} {2019}{\natexlab{b}})},\ \Eprint
  {https://arxiv.org/abs/1811.12907} {arXiv:1811.12907 [astro-ph.HE]}
  \BibitemShut {NoStop}%
\bibitem [{\citenamefont {{Littenberg}}\ and\ \citenamefont
  {{Cornish}}(2015)}]{Littenberg2015}%
  \BibitemOpen
  \bibfield  {author} {\bibinfo {author} {\bibfnamefont {T.~B.}\ \bibnamefont
  {{Littenberg}}}\ and\ \bibinfo {author} {\bibfnamefont {N.~J.}\ \bibnamefont
  {{Cornish}}},\ }\href {https://doi.org/10.1103/PhysRevD.91.084034} {\bibfield
   {journal} {\bibinfo  {journal} {\prd}\ }\textbf {\bibinfo {volume} {91}},\
  \bibinfo {eid} {084034} (\bibinfo {year} {2015})},\ \Eprint
  {https://arxiv.org/abs/1410.3852} {arXiv:1410.3852 [gr-qc]} \BibitemShut
  {NoStop}%
\bibitem [{\citenamefont {Vallisneri}\ \emph {et~al.}(2015)\citenamefont
  {Vallisneri}, \citenamefont {Kanner}, \citenamefont {Williams}, \citenamefont
  {Weinstein},\ and\ \citenamefont {Stephens}}]{Vallisneri2015}%
  \BibitemOpen
  \bibfield  {author} {\bibinfo {author} {\bibfnamefont {M.}~\bibnamefont
  {Vallisneri}}, \bibinfo {author} {\bibfnamefont {J.}~\bibnamefont {Kanner}},
  \bibinfo {author} {\bibfnamefont {R.}~\bibnamefont {Williams}}, \bibinfo
  {author} {\bibfnamefont {A.}~\bibnamefont {Weinstein}},\ and\ \bibinfo
  {author} {\bibfnamefont {B.}~\bibnamefont {Stephens}},\ }\href
  {https://doi.org/10.1088/1742-6596/610/1/012021} {\bibfield  {journal}
  {\bibinfo  {journal} {J. Phys. Conf. Ser.}\ }\textbf {\bibinfo {volume}
  {610}},\ \bibinfo {pages} {012021} (\bibinfo {year} {2015})},\ \Eprint
  {https://arxiv.org/abs/1410.4839} {arXiv:1410.4839} \BibitemShut {NoStop}%
\bibitem [{\citenamefont {{Abbott}}\ \emph
  {et~al.}(2019{\natexlab{c}})\citenamefont {{Abbott}}, \citenamefont
  {{Abbott}}, \citenamefont {{Abraham}}, \citenamefont {{Acernese}},
  \citenamefont {{Ackley}}, \citenamefont {{Adams}}, \citenamefont
  {{Adhikari}}, \citenamefont {{Adya}}, \citenamefont {{Affeldt}},
  \citenamefont {{Agathos}}, \citenamefont {{Agatsuma}}, \citenamefont
  {{Aggarwal}} \emph {et~al.}}]{OpenData}%
  \BibitemOpen
  \bibfield  {author} {\bibinfo {author} {\bibfnamefont {R.}~\bibnamefont
  {{Abbott}}}, \bibinfo {author} {\bibfnamefont {T.~D.}\ \bibnamefont
  {{Abbott}}}, \bibinfo {author} {\bibfnamefont {S.}~\bibnamefont {{Abraham}}},
  \bibinfo {author} {\bibfnamefont {F.}~\bibnamefont {{Acernese}}}, \bibinfo
  {author} {\bibfnamefont {K.}~\bibnamefont {{Ackley}}}, \bibinfo {author}
  {\bibfnamefont {C.}~\bibnamefont {{Adams}}}, \bibinfo {author} {\bibfnamefont
  {R.~X.}\ \bibnamefont {{Adhikari}}}, \bibinfo {author} {\bibfnamefont
  {V.~B.}\ \bibnamefont {{Adya}}}, \bibinfo {author} {\bibfnamefont
  {C.}~\bibnamefont {{Affeldt}}}, \bibinfo {author} {\bibfnamefont
  {M.}~\bibnamefont {{Agathos}}}, \bibinfo {author} {\bibfnamefont
  {K.}~\bibnamefont {{Agatsuma}}}, \bibinfo {author} {\bibfnamefont
  {N.}~\bibnamefont {{Aggarwal}}}, \emph {et~al.},\ }\href@noop {} {\bibfield
  {journal} {\bibinfo  {journal} {arXiv e-prints}\ ,\ \bibinfo {eid}
  {arXiv:1912.11716}} (\bibinfo {year} {2019}{\natexlab{c}})},\ \Eprint
  {https://arxiv.org/abs/1912.11716} {arXiv:1912.11716 [gr-qc]} \BibitemShut
  {NoStop}%
\bibitem [{\citenamefont {{Gravitational Wave Open Science
  Center}}(2018)}]{GWTC1Data}%
  \BibitemOpen
  \bibfield  {author} {\bibinfo {author} {\bibnamefont {{Gravitational Wave
  Open Science Center}}},\ }\href {https://doi.org/10.7935/82H3-HH23} {\bibinfo
  {title} {Strain data release for gwtc-1: A gravitational-wave transient
  catalog of compact binary mergers observed by ligo and virgo during the first
  and second observing runs}} (\bibinfo {year} {2018})\BibitemShut {NoStop}%
\bibitem [{\citenamefont {{LIGO Scientific Collaboration}}\ and\ \citenamefont
  {{Virgo Collaboration}}(2020)}]{GWTC2Data}%
  \BibitemOpen
  \bibfield  {author} {\bibinfo {author} {\bibnamefont {{LIGO Scientific
  Collaboration}}}\ and\ \bibinfo {author} {\bibnamefont {{Virgo
  Collaboration}}},\ }\href {https://doi.org/10.7935/99GF-AX93} {\bibinfo
  {title} {Ligo virgo strain data from gwtc-2 catalog}} (\bibinfo {year}
  {2020})\BibitemShut {NoStop}%
\bibitem [{\citenamefont {{Baker}}\ \emph {et~al.}(2017)\citenamefont
  {{Baker}}, \citenamefont {{Bellini}}, \citenamefont {{Ferreira}},
  \citenamefont {{Lagos}}, \citenamefont {{Noller}},\ and\ \citenamefont
  {{Sawicki}}}]{Baker2017}%
  \BibitemOpen
  \bibfield  {author} {\bibinfo {author} {\bibfnamefont {T.}~\bibnamefont
  {{Baker}}}, \bibinfo {author} {\bibfnamefont {E.}~\bibnamefont {{Bellini}}},
  \bibinfo {author} {\bibfnamefont {P.~G.}\ \bibnamefont {{Ferreira}}},
  \bibinfo {author} {\bibfnamefont {M.}~\bibnamefont {{Lagos}}}, \bibinfo
  {author} {\bibfnamefont {J.}~\bibnamefont {{Noller}}},\ and\ \bibinfo
  {author} {\bibfnamefont {I.}~\bibnamefont {{Sawicki}}},\ }\href
  {https://doi.org/10.1103/PhysRevLett.119.251301} {\bibfield  {journal}
  {\bibinfo  {journal} {\prl}\ }\textbf {\bibinfo {volume} {119}},\ \bibinfo
  {eid} {251301} (\bibinfo {year} {2017})}\BibitemShut {NoStop}%
\bibitem [{\citenamefont {{Walsh}}\ \emph {et~al.}(2013)\citenamefont
  {{Walsh}}, \citenamefont {{Barth}}, \citenamefont {{Ho}},\ and\ \citenamefont
  {{Sarzi}}}]{Walsh2013}%
  \BibitemOpen
  \bibfield  {author} {\bibinfo {author} {\bibfnamefont {J.~L.}\ \bibnamefont
  {{Walsh}}}, \bibinfo {author} {\bibfnamefont {A.~J.}\ \bibnamefont
  {{Barth}}}, \bibinfo {author} {\bibfnamefont {L.~C.}\ \bibnamefont {{Ho}}},\
  and\ \bibinfo {author} {\bibfnamefont {M.}~\bibnamefont {{Sarzi}}},\ }\href
  {https://doi.org/10.1088/0004-637X/770/2/86} {\bibfield  {journal} {\bibinfo
  {journal} {\apj}\ }\textbf {\bibinfo {volume} {770}},\ \bibinfo {eid} {86}
  (\bibinfo {year} {2013})}\BibitemShut {NoStop}%
\end{thebibliography}%
\end{document}